\documentclass[12pt, twoside, onecolumn, reqno, notitlepage, centertags]{amsart} 
\usepackage[tmargin=30mm,bmargin=25mm,lmargin=20mm,rmargin=20mm]{geometry}
\usepackage{rotate,graphicx,url,color}
\usepackage{rotating}
\usepackage{enumitem}
\usepackage{textcomp}
\usepackage{placeins}
\usepackage[utf8]{inputenc}
\usepackage[round]{natbib}
\usepackage[brazil,english]{babel} 
\usepackage{amssymb,amsthm,amsmath,amsaddr}		
\usepackage{booktabs} 
\usepackage{multirow} 			
\usepackage{diagbox}
\usepackage{bm}
\usepackage{subfig}

\def\bSig\mathbf{\Sigma}

\usepackage{url}
\usepackage{soul} 

\def\corr{\mbox{{\rm corr\,}}}

\newcommand{\vbeta}{\pmb{\beta}}
\newcommand{\vxi}{\pmb{\xi}}
\newcommand{\vepsilon}{\pmb{\epsilon}}

\newcommand{\vD}{{\textbf D}}
\newcommand{\vd}{{\textbf d}}

\newcommand{\vR}{{\textbf R}}

\newcommand{\vX}{{\textbf X}}

\newcommand{\vT}{{\textbf T}}
\newcommand{\vt}{{\textbf t}}

\allowdisplaybreaks[4]

\title{Wavelet Spatio-Temporal Change Detection on multi-temporal PolSAR images}
\author{Rodney Fonseca$^{\star}$, Alu\'{i}sio Pinheiro$^{\star}$ and Abdourrahmane Atto$^\diamond$}

\address{$^{\star}$ Department of Statistics, University of Campinas, Campinas-SP, Brazil\\
$^\diamond$ LISTIC, Universit\'e Savoie Mont Blanc, Annecy-le-Vieux, France
}

\begin{document}

\begin{abstract}
We introduce WECS (Wavelet Energies Correlation Sreening), an unsupervised sparse procedure to detect spatio-temporal change points on multi-temporal SAR (POLSAR) images or even on sequences of very high resolution images. The procedure is based on wavelet approximation for the multi-temporal images, wavelet energy apportionment, and ultra-high dimensional correlation screening for the wavelet coefficients. We present two complimentary wavelet measures in order to detect sudden and/or cumulative changes, as well as for the case of stationary or non-stationary multi-temporal images. We show WECS performance on synthetic multi-temporal image data. We also apply the proposed method to a time series of 85 satellite images in the border region of Brazil and the French Guiana. The images were captured from November 08, 2015 to December 09 2017.
\end{abstract}

\maketitle

\section{Introduction}

We discuss here a novel method for unsupervised spatio-temporal change detection in multi-temporal SAR/POLSAR images. WECS is based upon correlation screening for energy apportionment on wavelet approximations. The spatial character of the change detection is attained on pixel level. The method is fast, scalable, linearly updatable, and the resulting measures are sparse. 

A review for change detection in multi-temporal remote sensing is given by \cite{ban2016change}. 
Different proposals for this purpose may be found  in the literature. They vary in their motivations as well as in their applicability.  Change detection in multi-temporal hyperspectral images is discussed in \cite{bovolo2015time},  \cite{liu2019review}, and \cite{matsunaga2017current}.  \cite{jia2018novel} pursue change detection techniques via non-local means and principal component analysis. Compressed projection and image fusion are employed by \cite{hou2014unsupervised}. Deep learning by slow feature analysis for change detection is the subject of \cite{du2019unsupervised}. \cite{chen2020change} proposes a change detection method driven by adaptive parameter estimation.

Besides different methodological paradigms, several areas of application receive special attention.  For instance, urban change detection applications via polarimetric SAR images are discussed in \cite{ansari2020urban}. \cite{song2018multi} discusses land cover change detection in mountainous terrain via multi-temporal and multi-sensor remote sensing images. \cite{ru2021multi} studies multi-temporal scene classification and scene change detection. Deforestation change detection is discussed by \cite{barreto2016deforestation}. 

Wavelet methods present many advantages for a plethora of applications \citep{vidakovic1999statistical}. Their  computational efficiency and sparseness are specially relevant for large images and other high-dimensional data \citep{morettin2017wavelets}. \cite{atto2012multidate}, \cite{bouhlel2015multivariate}, \cite{celik2009multiscale}, \cite{cui2012statistical} use different wavelet methods for change detection in satellite images. 

The motivation for our proposed method is multi-fold. We aim a fast and accurate method. We would also like this method to be easily updatable when a new observation is captured. Finally, escalability was a concern as well. We propose a wavelet-based procedure for change detection in multi-temporal remore sensing images (WECS). It is unsupervised and built on ultra-high dimensional correlation screening \citep{fan2020statistical} for the wavelet coefficients.  We present two complimentary wavelet measures in order to detect sudden and/or cumulative changes, as well as for the case of stationary or non-stationary multi-temporal images. The procedure presents some advantages. It is unsupervised, fast and updatable, thus allowing for real-time change detection. Moreover, it is sparse and scalable. 

The rest of the text goes as follows. Section \ref{section_method} introduces the problem and presents the proposed method. We show WECS performance on synthetic multi-temporal image data in Section \ref{section_validation}. In Section \ref{section_realdata} we apply the proposed method to a time series of 85 satellite images in the border region of Brazil and the French Guiana, for  images captured from November 08, 2015 to December 09 2017.  Section \ref{section_discussion} concludes the paper with a discussion.
   
\section{Methodology}\label{section_method}

Let $\mathcal{I}(1),\ldots,\mathcal{I}(m)$ be a set of matrices representing the log-images of some region of interest. These images may be relative to one satellite channel or a combination of channels; this will be specified when appropriate. Our goal is twofold: to find possible points in time where some relevant change might have taken place at the region represented in $\mathcal{I}(m)$, $m=1,\ldots,n$, and to find which regions are closely associated to the observed changes along time. We shall address these tasks by analyzing the bidimensional discrete wavelet decomposition of $\mathcal{I}(m)$ at the $J$-th approximation level, i.e., 
such that
\begin{equation} 
\mathcal{I}(m)=\mathcal{I}_J(m)+\vepsilon_J(m),
\end{equation}
where $\mathcal{I}_J(m)$ is the wavelet approximation of log-image $\mathcal{I}(m)$ at level $J$.We denote $\mathcal{I}_J(m)$'s approximation coefficients matrix by $\vX(m)$ \citep{morettin2017wavelets,vidakovic1999statistical}. The advantages of doing so is that we linearize possible speckle (multiplicative) noise in the original images by taking logarithms to obtain $\mathcal{I}(m)$ and later perform a smoothing to obtain $\mathcal{I}_J(m)$, whose approximation coefficients matrix is $\vX(m)$.

We can then consider further apportioning the total $\mathbb{L}_2$ energy of $\left\{\mathcal I(m)\right\}$ as
\begin{equation} 
\sum_{m=1}^n\|\mathcal I_J(m)\|^2_2=n\|\bar{\mathcal I}\|^2_2+\sum_{m=1}^n\|\mathcal I_J(m)-\bar{\mathcal I}\|^2_2,\label{eq_ANOVAlogimage}
\end{equation}
where $\bar{\mathcal{I}}=n^{-1}\sum_{m=1}^n\mathcal{I}_J(m)$. The same can be done on the wavelet domain with $\vX(m)$, which is the way we shall proceed in the following steps.

We discuss here two procedures which are similar, but may yield different results depending on the nature of relevant changes. The first procedure makes use of (\ref{eq_ANOVAlogimage}), so that we establish an {\it average wavelet-approximated image} and detect time points for which images differ from the {\it characteristic} image. Changes are detected by absolute correlations between individual wavelet approximation coefficients time series and the time series of overall wavelet energy. The second procedure also uses absolute correlations but instead of an average wavelet image energy, we compute  the overall energy differences for subsequent wavelet-smoothed images and their approximation coefficients. The former should help us detect image changes from an overall behavior over time whilst the latter should also detect changes on non-stationary set-ups.  In practive, both procedures shall be performed in the wavelet domain by using $\{\vX(m)\}$ instead of $\{\mathcal{I}_J(m)\}$.

The average image $\bar{\mathcal{I}}=n^{-1}\sum_{m=1}^n\mathcal{I}_J(m)$ has approximation coefficients matrix given by $\bar{\vX} = n^{-1}\sum_{m=1}^n \vX(m)$. Let $X_{k,l}(m)$ and $\bar{X}_{k,l}$ be the entry $(k,l)$ of the matrices $\vX(m)$ and $\bar{\vX}$, respectively. We take the matrix $\vD(m) = [D_{k.l}(m)]$, where $D_{k,l}(m)=(X_{k,l}(m)-\bar{X}_{k,l})^2$. We then 
analyze the time series given by
\begin{equation} 
\vd(m)= \sum_{k,l}D_{k,l}(m) = \sum_{k,l}(X_{k,l}(m) - \bar{X}_{k,l})^2, \quad m=1,\ldots,n,
\label{eq_defdm}
\end{equation}
which displays the temporal variation with respect to $\bar{\vX}$ of spatial energies.

The time points with highest values of $d(m)$ represent the images for which the most expressive changes take place, where changes here are measured through $\mathbb{L}_2$ energy. Define the  $n\times p$ matrix
\begin{equation}
 \vD=\left(
 \begin{array}{c}
 vec(\vD(1))^T\\
 \vdots\\
 vec(\vD(n))^T\\
 \end{array}
 \right),
\label{eq_defmatrixD}
\end{equation}
where $vec(\vD(m))$ is the $p\times 1$ vector of wavelet coefficients for time $m$ and $p=\#\{k,l\}$ is total number of locations represented by $\vX(m)$, $m=1,\ldots,n$. Sparsity \citep{johnstone2009statistical} on the wavelet coefficients plays a special role here.  We suppose a handful of coefficients drive the changes given by $\vd$, so that the effective dimension of $\vD$ (number of locations where relevant changes occur), say $e_d$, is such that $e_d<<p$. This can be represented as the following linear model
\begin{equation} 
\vd=\vD\vbeta^{(d)}+\vxi^{(d)}
\label{sparsemodel_d}
\end{equation}
where $\vbeta^{(d)}$ is sparse, i.e., it has $p-e_d$ null elements, and $\vxi^{(d)}$ is some $n\times 1$ random vector of errors.

 In order to identify spatio-temporal changes we employ the idea of ultra-high dimensional correlation screening \citep{fan2020statistical} as follows. For each squared mean-corrected approximation coefficient time series, given by $\vD(m)$, consider its Pearson correlation with the mean-corrected total approximation energy, given by $\vd$:
\begin{equation}
R_{k,l}^{(d)}= \corr\left( \vD_{k,l}, \vd\right),
\label{def_Rkld}
\end{equation}
where $\vD_{k,l}=(D_{k,l}(1),\ldots,D_{k,l}(n))^T$ is the time series of squared mean deviations of wavelet coefficients for the two-dimensional index $\{k,l\}$.

We have a matrix $\vR^{(d)}=[R_{k,l}^{(d)}]$ of correlations of ultra-high dimension. Define the set of 
{\it important} indices for changes in images with respect to $\bar{\mathcal I}$ as 
$\mathcal{M}^{*d}=\{(k,l):\mbox{ Change in }\mathcal{I}(m)\mbox{ with respect}$ to $\bar{\mathcal{I}}$  
$\mbox{ are caused by changes in approximation coefficients of index }(k,l)\}$.  This set coincides with the non-zero vectorized one-dimensional indices for the sparse representation of $\vbeta^{(d)}$ in (\ref{sparsemodel_d}). 
We build the empirical set of selected indices by
\begin{equation}
\mathcal{M}_{\tau}^{(d)}=\{(k,l):|R_{k,l}^{(d)}|>\tau_d\},
\label{def_Mtaud}
\end{equation}
where $\tau_d>0$ is a convenient threshold value, function of $n$ and $J$. Under some regularity conditions,
\[
P(\mathcal{M}_{\tau}^{(d)}\supset\mathcal{M}^{*d})\rightarrow 1,
\]
as $n\rightarrow\infty$ \citep{fan2020statistical}.

Analogously we take $\vT(m) = [T_{k,l}(m)]$, where $T_{k,l}(m)=(X_{k,l}(m+1)-X_{k,l}(m))^2$, for $m=1,\ldots,n-1$. We then analyze the time series given by
\begin{equation}
t(m)= \sum_{k,l}T_{k,l}(m) = \sum_{k,l}(X_{k,l}(m+1) - X_{k,l}(m))^2, \quad m=1,\ldots,n-1.
\label{eq_deftm}
\end{equation}

The time points with highest values of $t(m)$ represent the images for which the most expressive changes take place, where changes here are measured through $\mathbb{L}_2$ energy. Define the  $(n-1)\times p$ matrix
\begin{equation}
 \vT=\left(
 \begin{array}{c}
 vec(\vT(1))^T\\
 \vdots\\
 vec(\vT(n))^T\\
 \end{array}
 \right),
\label{eq_defmatrixT}
\end{equation}
where $vec(\vT(m))$ is the $p\times 1$ vector of wavelet coefficients for time $m=1,\ldots,n-1$. 

The highest values of $\{t(m)\}$ represent the images for which the most expressive changes take place between time points $m$ and $m+1$, where changes here are measured through $\mathbb{L}_2$ energy. For each squared mean-corrected approximation coefficient time series, given by $\vT(m) = [T_{k,l}(m)]$, consider its Pearson correlation with the mean-corrected total approximation energy, given by $\vt$:
\begin{equation}
R_{k,l}^{(t)}= \corr\left( \vT_{k,l}, \vt\right).
\label{def_Rklt}
\end{equation}

We again suppose a handful of coefficients drive the changes given by $\vt$, so that the effective dimension, say $e_t$, is such that $e_t<<p$. This can be represented as the following linear model
\begin{equation} 
\vt=\vT\vbeta^{(t)}+\vxi^{(t)}
\label{sparsemodel_t}
\end{equation}
where $\vbeta^{(t)}$ is sparse, i.e. $\vbeta^{(t)}$ has $p-e_t$ null elements, and $\vxi^{(t)}$ is some $(n-1)\times 1$ random vector of errors.

We have a matrix $\vR^{(t)}=[R_{k,l}^{(t)}]$ of correlations of ultra-high dimension. Define the set of 
{\it important} indices for changes in subsequent images as $\mathcal{M}^{*t}=\{(k,l):\mbox{ Change in }\mathcal{I}(m+1)\mbox{ with respect to\hfill }\\ \mathcal{I}(m)  \mbox{  for some }m=1,\ldots,n-1\mbox{ are caused by the approximation coefficients of index }(k,l)\}$. We build a set of selected indices by
\[
\mathcal{M}_{\tau}^{(t)}=\{(k,l):|R_{k,l}^{(t)}|>\tau_t\},
\] 
where $\tau_t>0$ is a convenient threshold value, function of $n$ and $J$. Under some regularity conditions,
\[
P(\mathcal{M}_{\tau}^{(t)}\supset\mathcal{M}^{*t})\rightarrow 1,
\]
as $n\rightarrow\infty$ \citep{fan2020statistical}.

Therefore, if we define
\begin{eqnarray}
\mathcal{M}_{\tau}&=&\mathcal{M}_{\tau}^{(t)}\cup\mathcal{M}_{\tau}^{(d)},\label{def_Mpop}\\
\mathcal{M}^{*}&=& \mathcal{M}^{*d}\cup\mathcal{M}^{*t}\label{def_Memp},
\end{eqnarray}
we have the following consistency property:
\[
P(\mathcal{M}_{\tau}\supset\mathcal{M}^{*})\rightarrow 1.
\]
as $n\rightarrow\infty$.

Thence, if we compute $\{d(m)\}$ and $\{t(m)\}$, and build $\mathcal{M}_{\tau}=\mathcal{M}_{\tau}^{(t)}\cup\mathcal{M}_{\tau}^{(d)}$, the consistency of the screening methods above guarantees asymptotic coverage of all approximation coefficients strongly associated with changes with respect to the average image as well as immediate previous image with a high probability, as long as the required regularity conditions hold. 

Further geometrical motivation for our proposal is given as follows.  We argue the case of $\vd$ but the same may be written regarding $\vt$ as well. As defined by (\ref{eq_defdm}), we expect $\vd$ to be a vector with some few high values, say $s_d$, and $n-s_d$ smaller values. This segregates the multi-temporal images, since the former time points identify the images in which significant changes occur, while the latter indices identify time points with no major changes. Consider $U>L>0$ such that the $s_d$ highest values of $\vd$ are larger then $U$, and the $n-s_d$ smallest values of $\vd$ are smaller then $L$. We also take $\delta=U-L$. The indices defined by  (\ref{def_Mtaud}) are such that   
\[
\frac{<\vD_{k,l},\vd>}{\|\vD_{k,l}\|_2\|\vd\|_2}>\tau_d,
\]
i.e., such that  $\sum_{m=1}^nD_{k,l}(m)d(m)>\tau_d\|\vD_{k,l}\|_2\|\vd\|_2$. This can be rewritten as
\[
\left|\sum_{m:d(m)>U} D_{k,l}(m)\right|-\left|\sum_{m:d(m)<L} D_{k,l}(m)\right|>\Delta,
\]
for some arbitrary $\Delta>>0$ (which can be a function of $n$ and $J$). Thence, when we employ correlation screening we select the two-dimensional wavelet indices which have the closest empirical directions to the vector of image temporal changes. Thus we are performing a truly spatio-temporal change detection in a single procedure.

\section{Validation on synthetic data}\label{section_validation}

\begin{figure}[htb!]
\centering
\includegraphics[width=10pc]{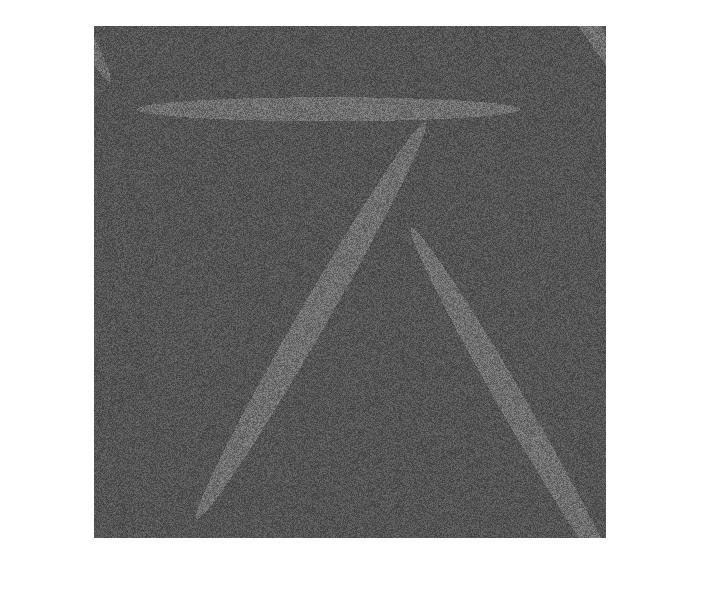}
\includegraphics[width=10pc]{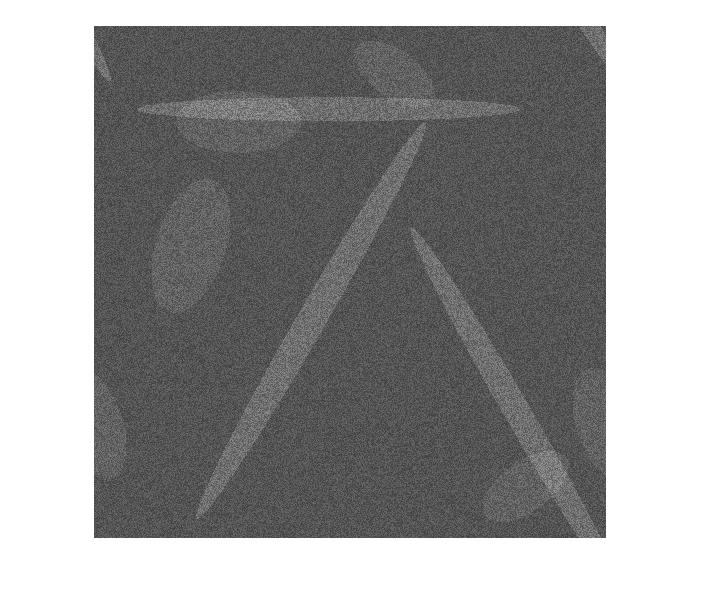}
\includegraphics[width=10pc]{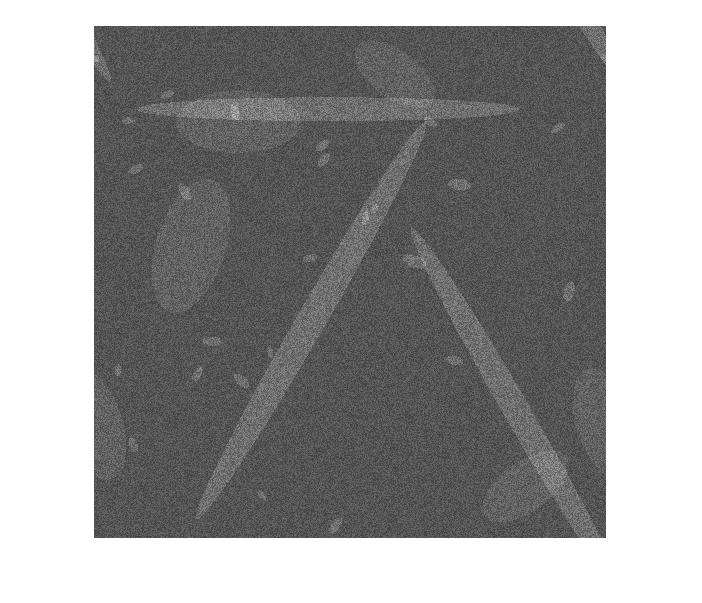}
\includegraphics[width=10pc]{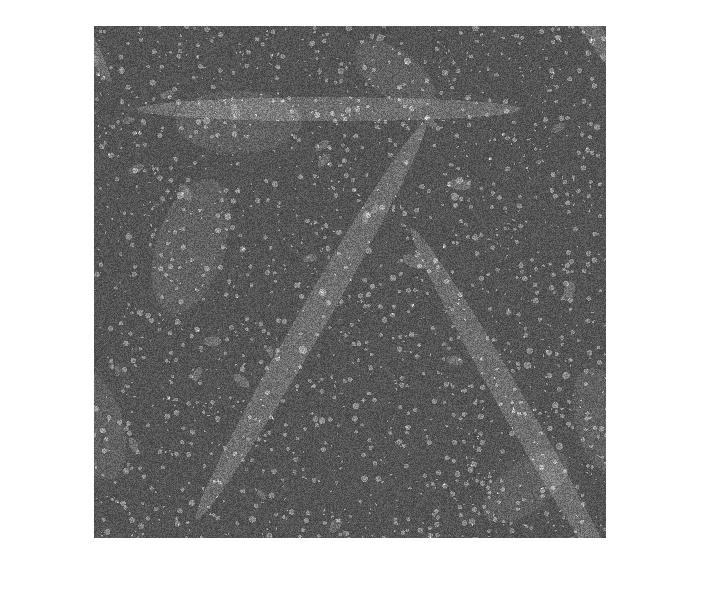}
\caption{Synthetic multi-temporal ($n=4$) images. Features and changes come as ellipses and dots.}
\label{F:EllipsoidChanges}
\end{figure}

In this section we apply the change detection methods above on synthetic data of multi-temporal images. 
The synthetic multi-temporal images ($n=4$) are shown in Figure \ref{F:EllipsoidChanges}. The first image, $I(1)$, presents three elongated ellipses. Changes consist of three different types of ellipses that are successively added to the original image $I(1)$. The second image, $I(2)$, has new large ellipses added. Smaller ellipses are then added to form $I(3)$ and small dots are added to form $I(4)$. 

\begin{figure}[htp!]
\noindent\includegraphics[width=12pc,height=10pc]{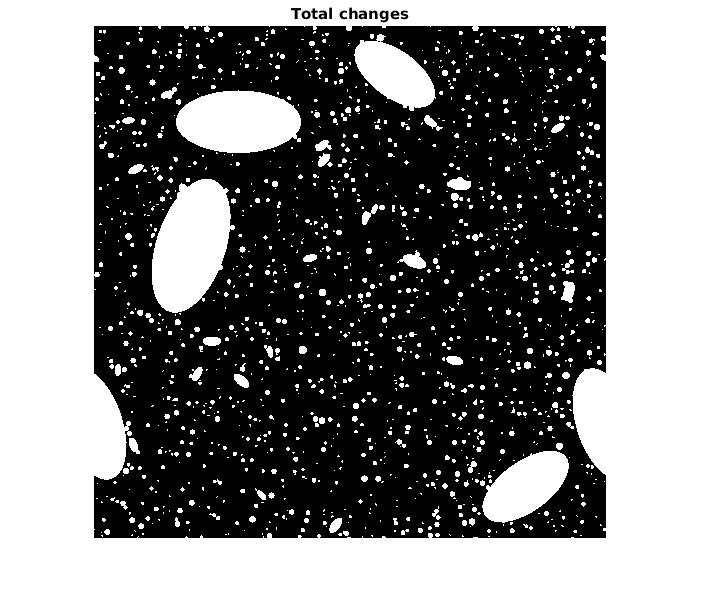}(a)
\includegraphics[width=12pc,height=10pc]{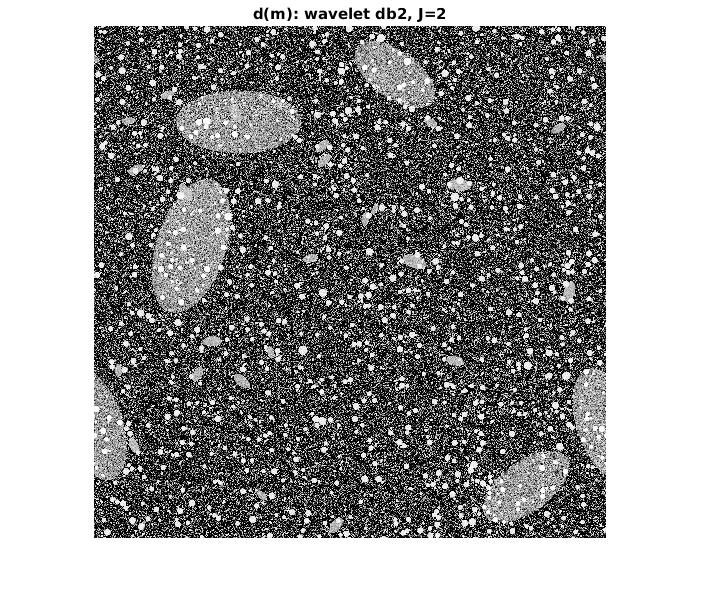}(b)
\includegraphics[width=12pc,height=10pc]{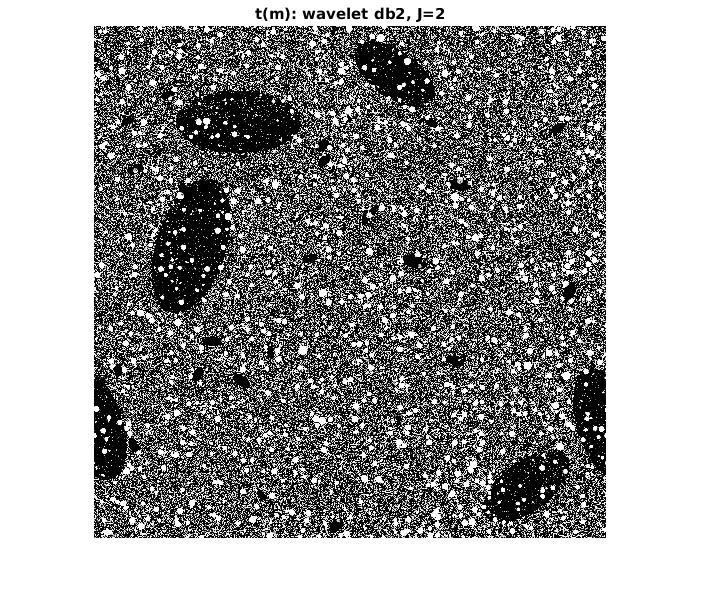}(c)

\includegraphics[width=12pc,height=10pc]{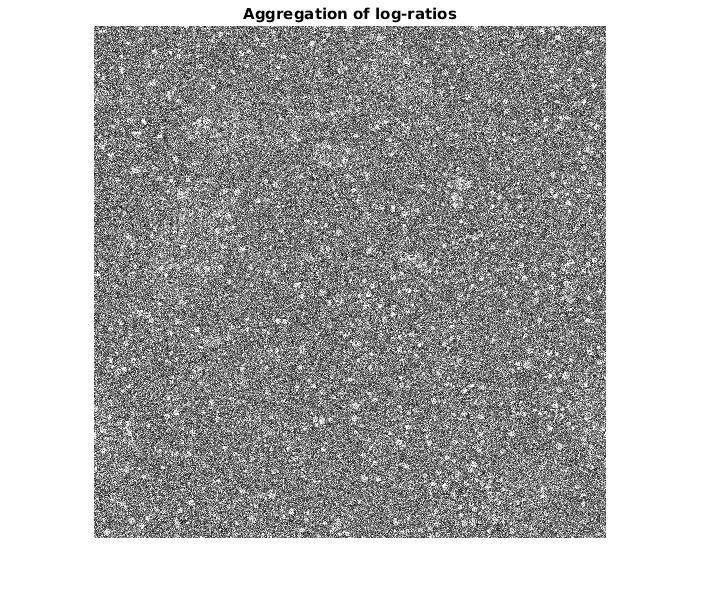}(d)
\includegraphics[width=12pc,height=10pc]{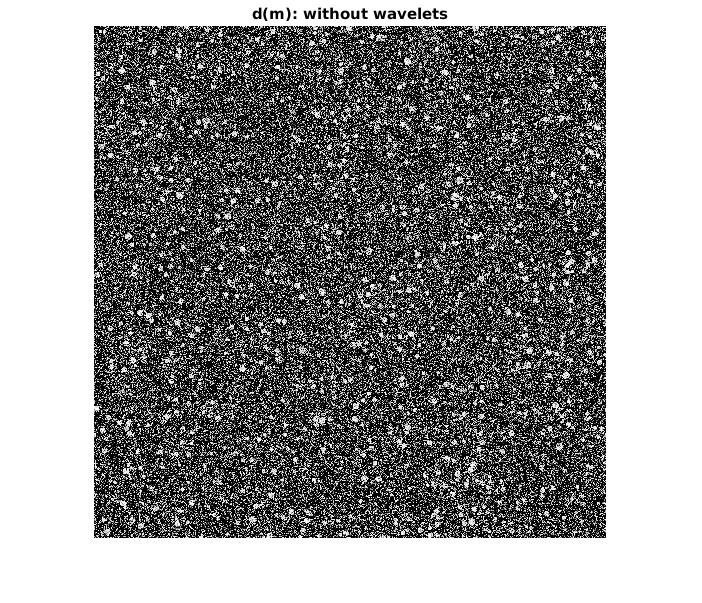}(e)
\includegraphics[width=12pc,height=10pc]{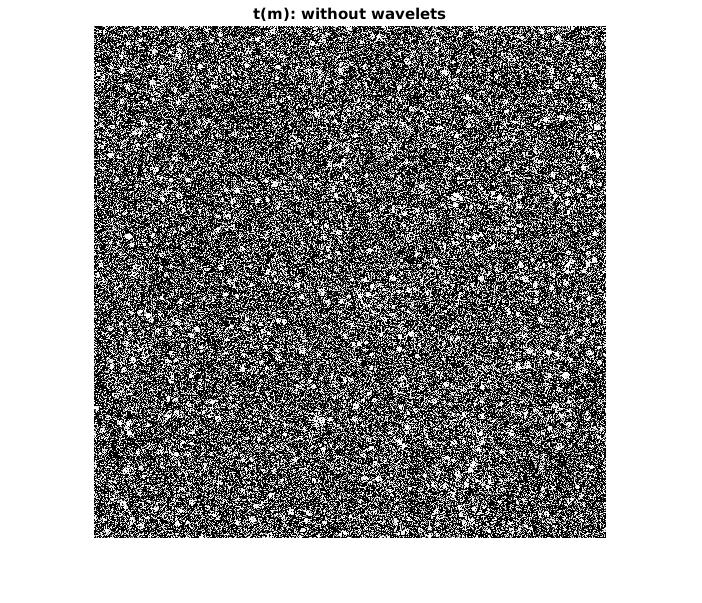}(f)
\caption{Synthetic images with changing ellipses. (a) Image composed by the total changes over time. 
(b) Proposed db2 wavelet $\vd(m)$ with $J=2$; (c) Proposed db2 wavelet $\vt(m)$ with $J=2$.
(d) Aggregation of log-ratios. (e) $\vd(m)$ without wavelets. (f) $\vt(m)$  without wavelets. 
}
\label{F:Changes_methods_images}
\end{figure}
Figure \ref{F:Changes_methods_images} illustrates the simulated synthetic images, the proposed wavelet detection methods, and three classic detection methods, as well. Panel (a) presents the total changes with respect to image $I(1)$. Panels (b) and (c) show the results by the proposed wavelet methods using Daubechies db2 and $J=2$ by $\vd(m)$ and $\vt(m)$, respectively. Panel (d) presents the results by the aggregated log-ratios. Finally, in Panels (e) and (f)  we can see the results if $\vd(m)$ and $\vt(m)$ are performed purely on the spatial domain, without wavelets. The spatio-temporal advantages of the proposed wavelet  $\vd(m)$ and $\vt(m)$ are clear in Figure \ref{F:Changes_methods_images}. A slight advantage for the detection of dots is attained by $\vd(m)$ over $\vt(m)$.

We compute ROC curves to compare the detection performance of different methods. For this we simulate noisy versions of the synthetic images illustrated by Figure \ref{F:EllipsoidChanges}. The change detection methods are then employed. Each method generates a correlation matrix between the real image of total changes and the estimated one.

Each ROC curve presents how close the magnitude variation of change measures is to the variation of the image of total changes in the following way:
\begin{enumerate}
\item Let $R$ be the matrix of change measures. Compute the range $[r_{\min},r_{\max}]$ of the values in $R$;
\item Let $(r_{(1)},\ldots,r_{(100)})$ be equally space values between $r_{\min}$ and $r_{\max}$;
\item For each $k=1,\ldots,n$, check how many pixels are such that $R_{i,j}>r_{(k)}$ coincide with the pixels $(i,j)$ where a change really occurs on the image of total changes. Dividing this number by the total number of changes gives the true positive rate.
\item For each $k=1,\ldots,n$, check how many pixels are such that $R_{i,j}>r_{(k)}$  do not coincide with the pixels $(i,j)$ where a change really occurs. Dividing this number by the total number of pixels where changes do not occur gives the false positive rate.
\end{enumerate}

\begin{figure}[htp!]
\includegraphics[width=12pc,height=10pc]{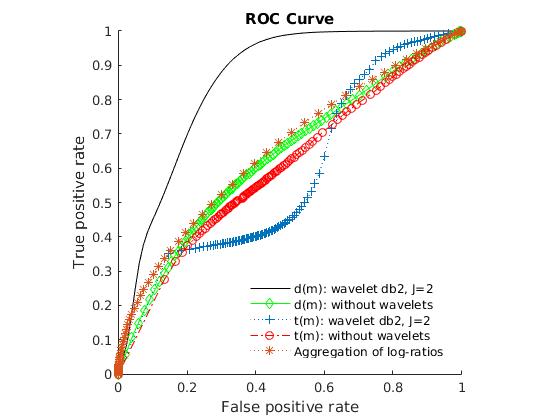}(a) 
\includegraphics[width=12pc,height=10pc]{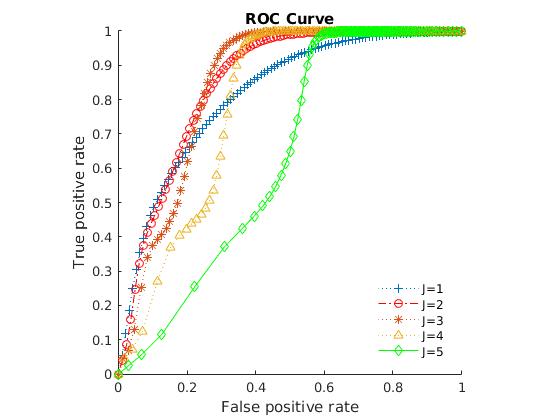}(b)
\noindent\includegraphics[width=12pc,height=10pc]{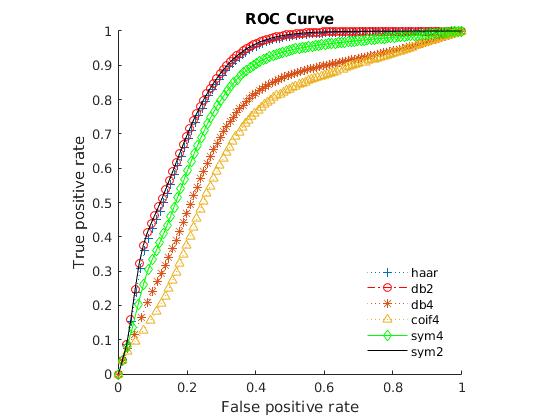}(c)

\includegraphics[width=12pc,height=10pc]{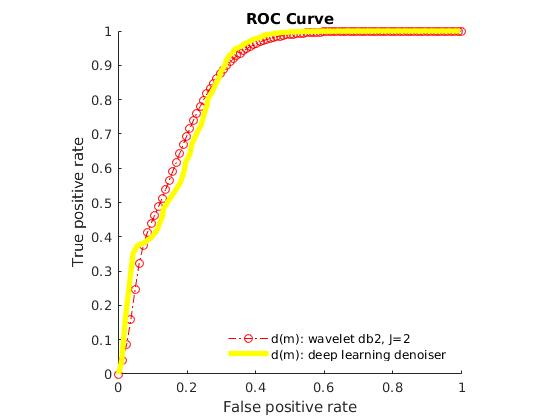}(d) 
\includegraphics[width=12pc,height=10pc]{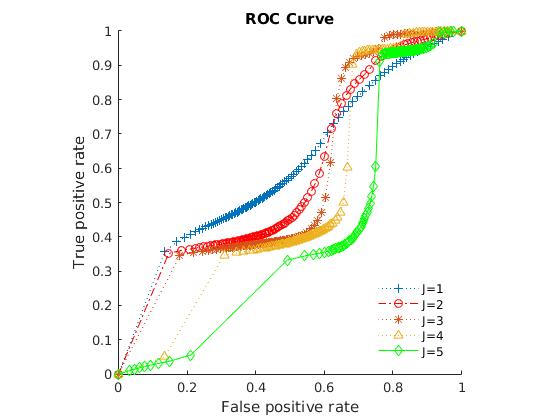}(e)
\includegraphics[width=12pc,height=10pc]{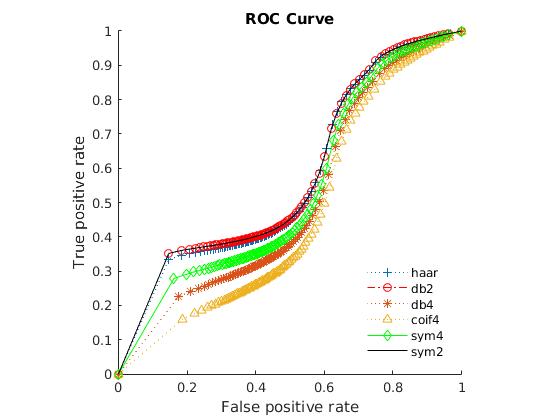}(f)

\caption{ROC curves for detection of changing ellipses in synthetic images and different methods. (a) The proposed methods in black (db2 wavelet $\vd(m)$) and green (db2 wavelet  $\vt(m)$) vs three non-wavelet methods: aggregated log-ratios (red stars); $d(m)$ (blue); and $\vt(m)$ (red circles). 
(b) db2 $\vd(m)$ with different levels.
(c) $\vd(m)$ with different wavelet bases and J=2;
(d) The proposed db2 wavelet $\vd(m)$ with (yellow) and without (red) deep-learning pre-treatment.  
(e) db2 $\vt(m)$ with different levels. (f) $\vt(m)$ with different wavelet bases and J=2. }
\label{F:EllipsoidChanges_details}
\end{figure}

Figure \ref{F:EllipsoidChanges_details} presents the different ROC curves for change detection methods applied to the synthetic data as follows. The effects of wavelet bases, level of decomposition, image pre-treatment and $\vd(m)$/$\vt(m)$ usage are shown on the ROC curves. We employ the following wavelet bases: Haar;  Daubechies db2; Daubechies db4; Coiflets coif4; Symlets sym2 ; and Symlets sym4. Panels (c) and (f) present the ROC curves for the proposed methods under the aforementioned bases for $\vd(m)$ and $\vt(m)$, respectively. On both instances $J=2$ is employed. The results for $\vd(m)$ are much more robust to basis variation then the ones for $\vt(m)$. Combining the ROC curves's comparison from both panels, Daubechies db2 is the best choice.  Panels (b) and (e) present the ROC curves for different levels of decomposition under the aforementioned bases for $\vd(m)$ and $\vt(m)$, respectively. On both instances db2 is employed, and five levels are considered: $J=1,2,3,4,5$. Levels $J=2,3$ have a clear better performance for $\vd(m)$ (with a slight advantage to $J=2$), whilst $J=1$ is competitive for $\vt(m)$. The overall performance  of $J=2$ warrants its use for the rest of the comparisons. Panel (d) shows how the proposed method performs with or without images' deep learning pre-treatment. The change detection method is the proposed db2 $\vd(m)$ with $J=2$. We can see that the ROC curves for treated or untreated images are almost identical. The proposed untreated method runs in 3.35s, while the combined deep-learning/db2 $\vd(m)$ with $J=2$ runs in 559.12s on a notebook. The configuration of the notebook is: OS - Ubuntu 18.04.5 LTS; RAM 7.7 GB; Intel\textregistered Core\texttrademark $ $i7-7500U CPU @ 2.70GHz x 4; graphics - Intel\textregistered HD Graphics 620 (KBL GT2); GNOME - 3.28.2; OS type - 64-bit. We finally have in Panel (a) the proposed db2 $\vd(m)$ with $J=2$ and db2 $\vt(m)$ with $J=2$  compared to three other non-wavelet methods. These are $\vd(m)$ and $\vt(m)$ where wavelet decomposition is not performed, i.e., the squared deviations are computed using $\{\mathcal{I}(m)\}$ instead of $\{\vX(m)\}$, and the classic method of analyzing aggregated log-ratios of $\{\mathcal{I}(m)\}$. The ROC curves in Panel (a) clearly show that  the proposed db2 $\vd(m)$ with $J=2$ outperforms the rest. 

We may summarize these results as: the proposed wavelet $\vd(m)$ method presents a superior performance. It is also equipped with the following nice properties: (i) it is scalable; (ii) it is sparse; (iii) it is parsimonious; (iv) it performs equally well with or without image denoising pre-treatment; iv) it can be easily adapted to be linearly updated when a new image is acquired; and (vi) it is fast. Thence, the proposed wavelet change-detection procedure can be used as a real-time change detection tool for long time series of large images.

\FloatBarrier

\section{Real Data Results}\label{section_realdata}

We employed the proposed change detection method on a series of 85 multi-date satellite images. The images were taken on a forest region at the border of Brazil and the French Guiana from November 08, 2015 to December 09, 2017. Each image has two channels and 1200 by 1000 pixels. We perform three change detection wavelet analyses: VV Polarization Channel;  VH Polarization Channel; and the Combined Image by Euclidean norm. 

A multi-resolution analysis (MRA) based on a Symlet basis with filter of length 16 (symlet 8) is built.  The log-images are approximated at levels $J=1,2,3,4$. Table \ref{T:approxenergy} shows the 85 images' average energy for each approximation level. We notice that roughly 99\% of the energy is recovered with $J=1$, and more than  90\% with $J=2$. The VV channel shows better overall energy recovery than the VH channel. For $J=4$ and $J=3$, the Euclidean combination of the polarization channels increases the energy representation percentage.  For $J=2$, VV channel and combined channels are equivalent. VH results in 4\% less energy than VV for $J=3$, and $-10\%$,  for $J=4$.

\begin{table}[h!]
\caption{Wavelet Approximation Mean Energy Percentage for log-images. Forest region at the border of Brazil and the French Guiana from November 08, 2015 to December 09, 2017. $n=85$ multi-date satellite images. Each image has two channels and 1200 by 1000 pixels. VV Polarization Channel;  VH Polarization Channel; and the Combined Image by Euclidean norm. Approximation $J=1,2,3,4$.}
\begin{tabular}{cccc|cccc|cccc}
\hline
\multicolumn{12}{c}{\sc Mean Approximated Energy Percentage}\\
\hline
\multicolumn{4}{c|}{VV Channel}&\multicolumn{4}{c|}{VH Channel}&\multicolumn{4}{c}{Combined Channels}\\
\hline
$J=4$&$J=3$&$J=2$&$J=1$&$J=4$&$J=3$&$J=2$&$J=1$&$J=4$&$J=3$&$J=2$&$J=1$\\
\hline
0.803&0.847&0.924&0.990&0.763&0.814&0.908&0.988&0.816&0.858&0.931&0.991\\
\hline
\end{tabular}\label{T:approxenergy}
\end{table}

\begin{figure}[htp!]
\noindent\includegraphics[width=18pc,height=4pc]{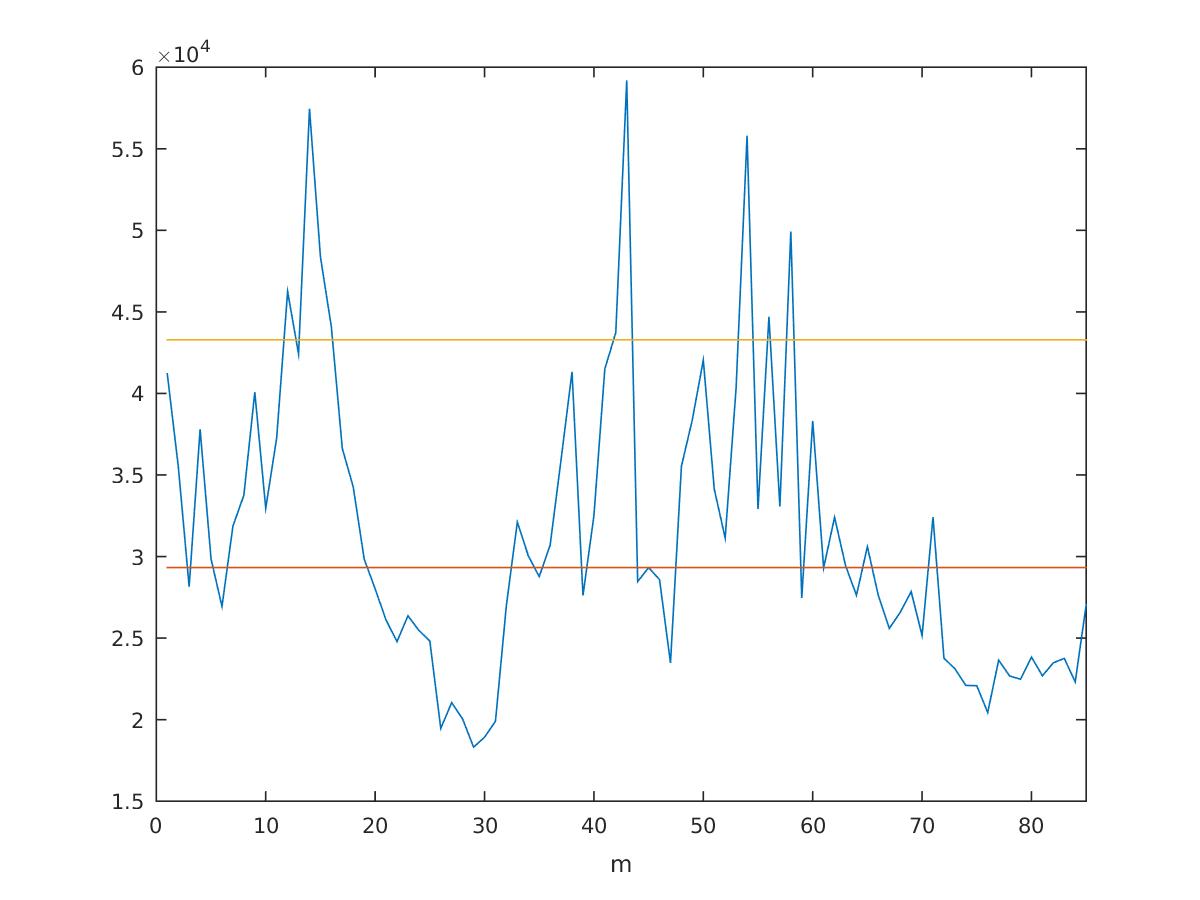}(a)
\includegraphics[width=18pc,height=4pc]{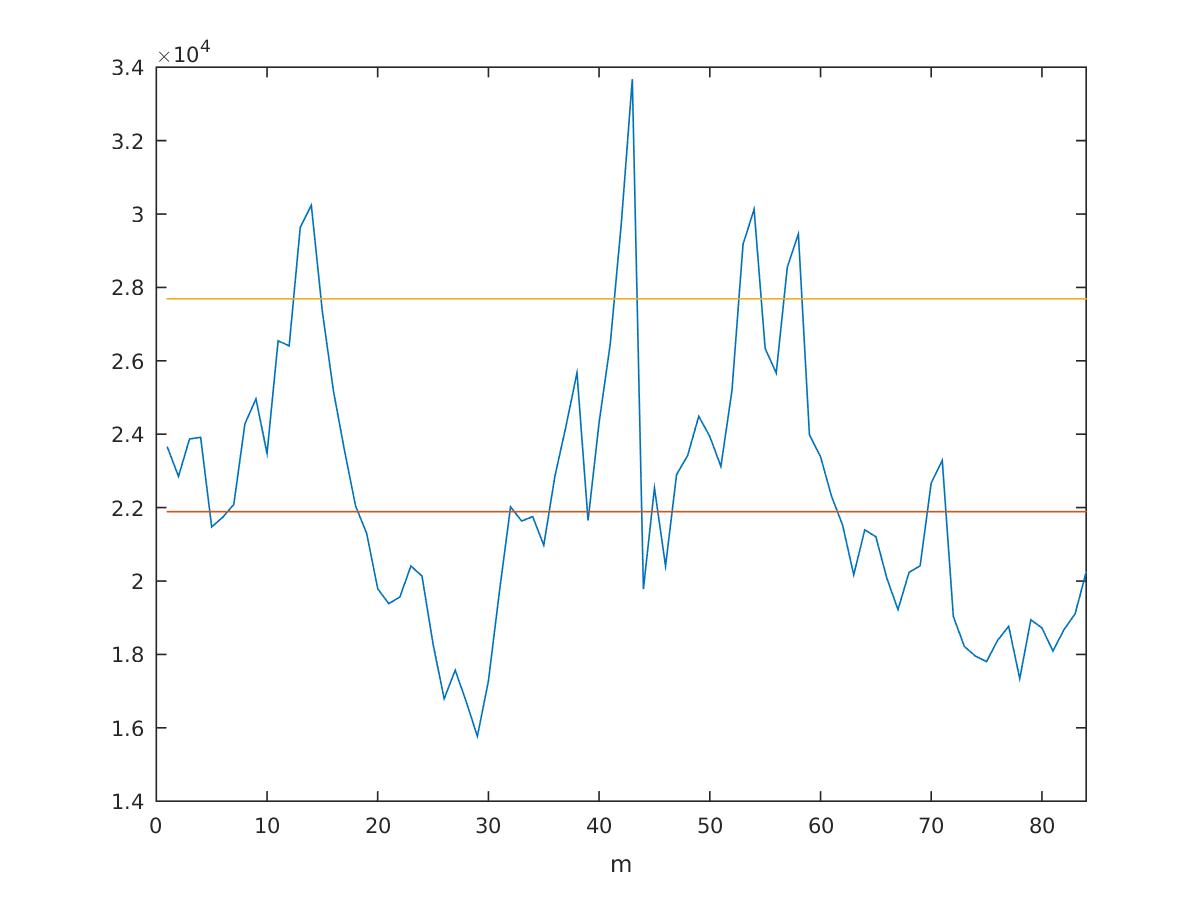}(b)

\includegraphics[width=18pc,height=4pc]{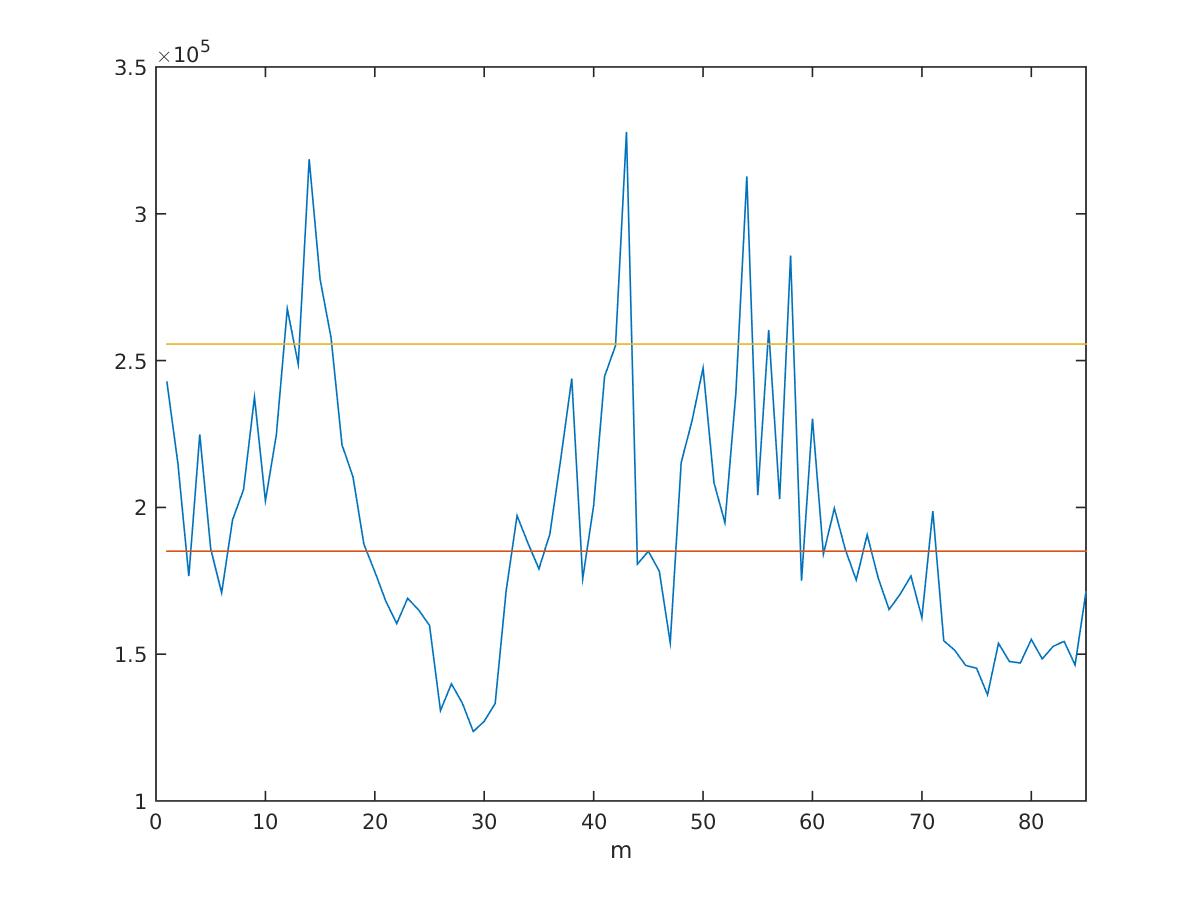}(c)
\includegraphics[width=18pc,height=4pc]{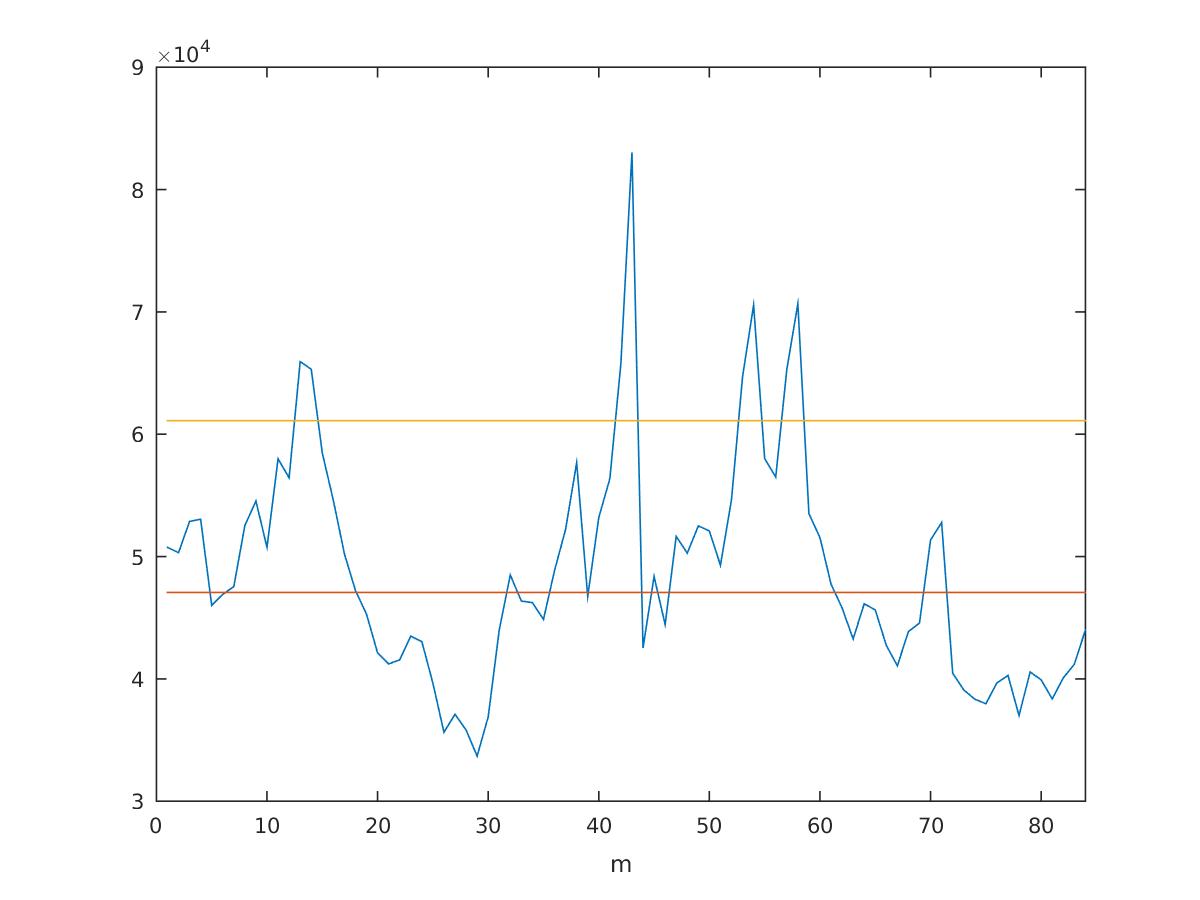}(d)

\includegraphics[width=18pc,height=4pc]{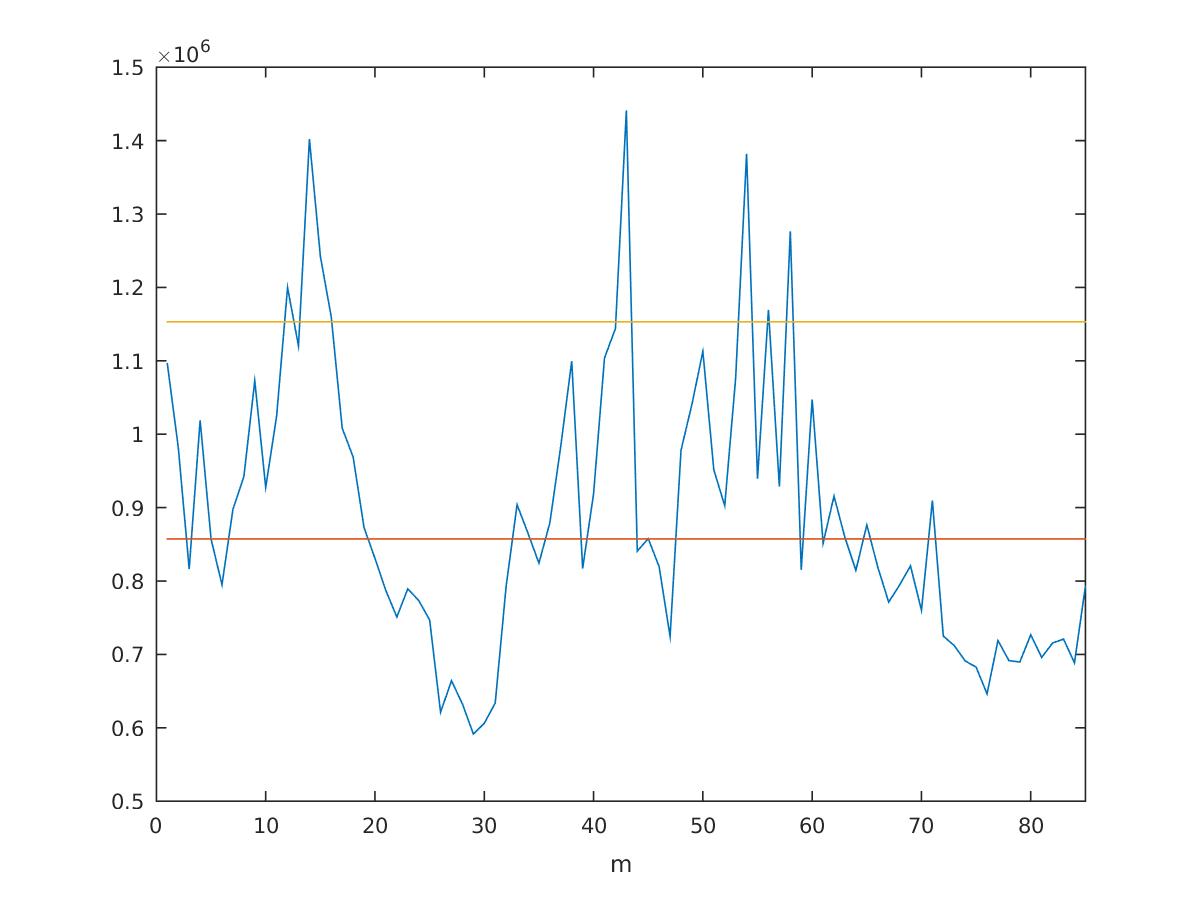}(e)
\includegraphics[width=18pc,height=4pc]{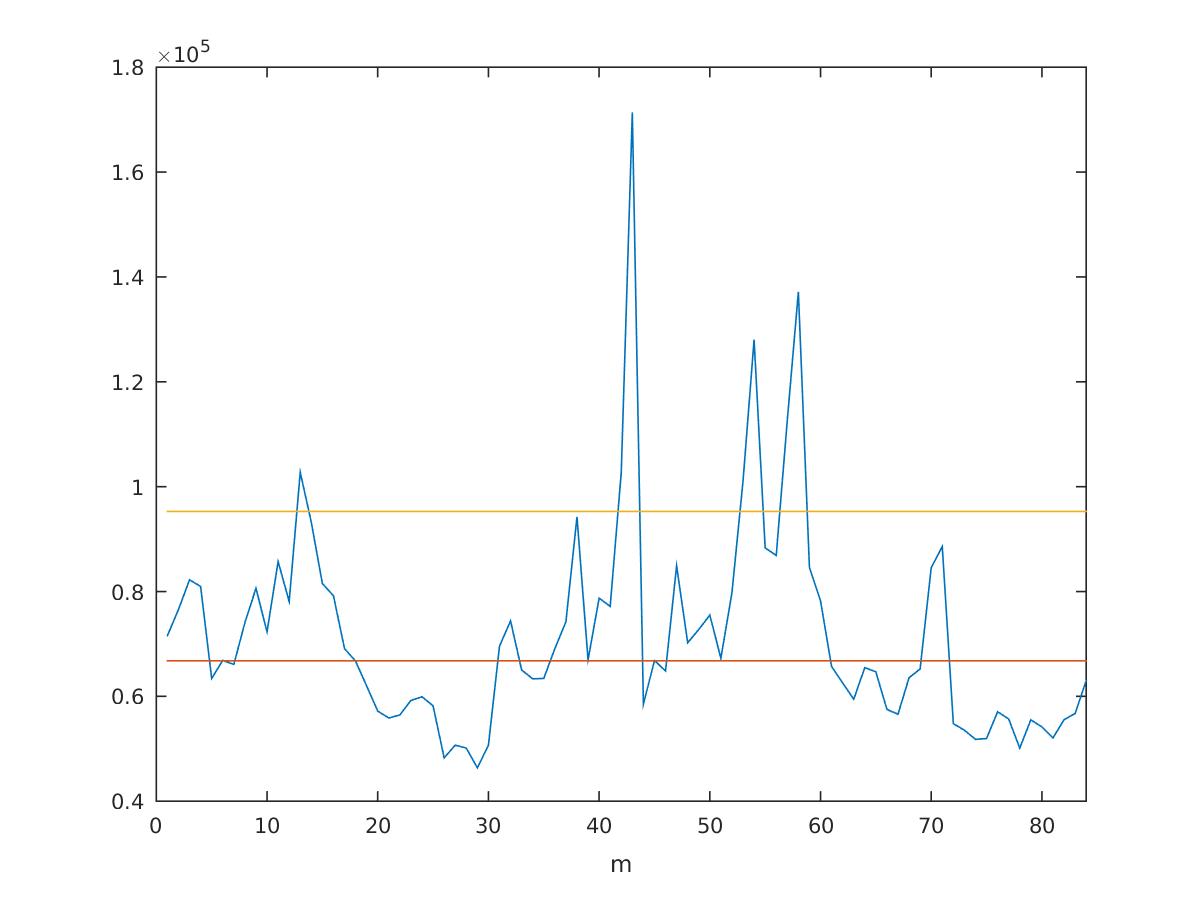}(f)

\includegraphics[width=18pc,height=4pc]{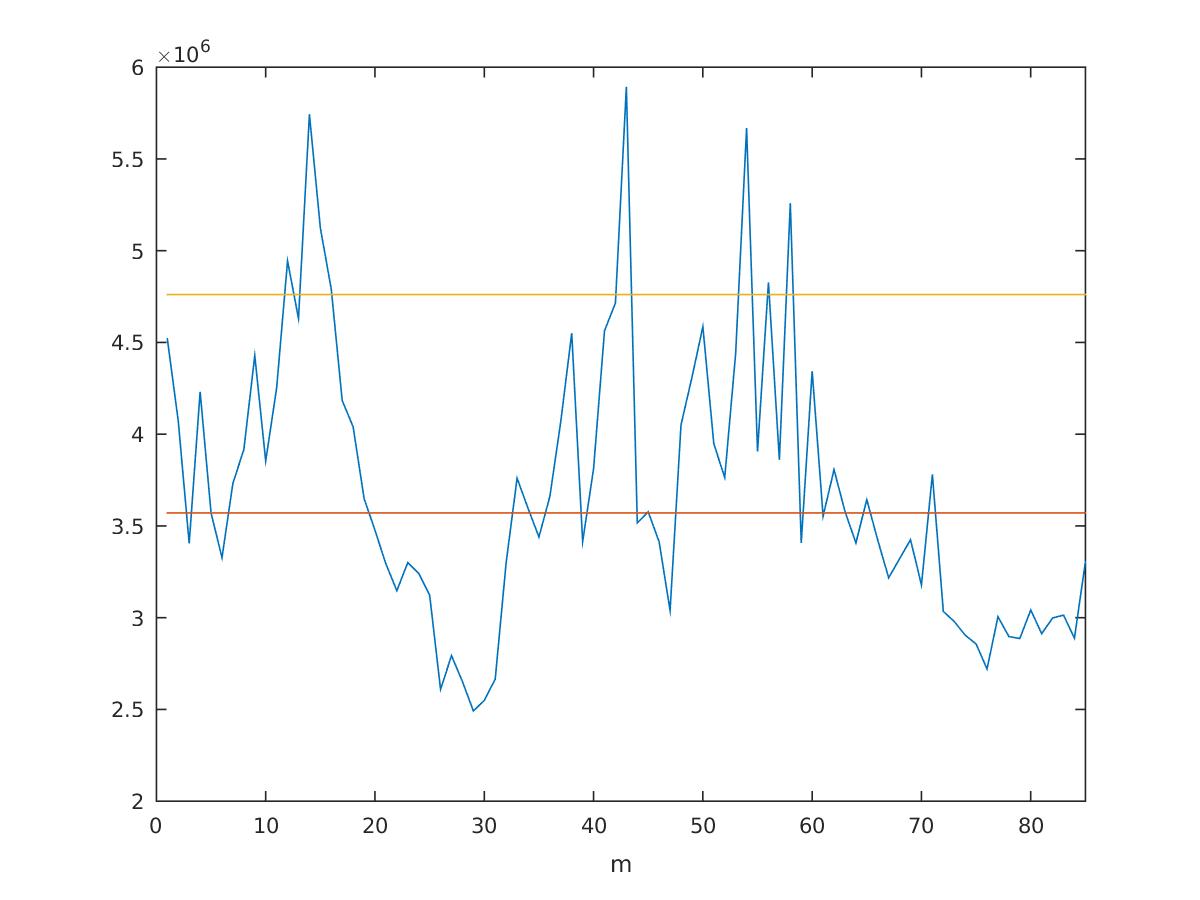}(g)
\includegraphics[width=18pc,height=4pc]{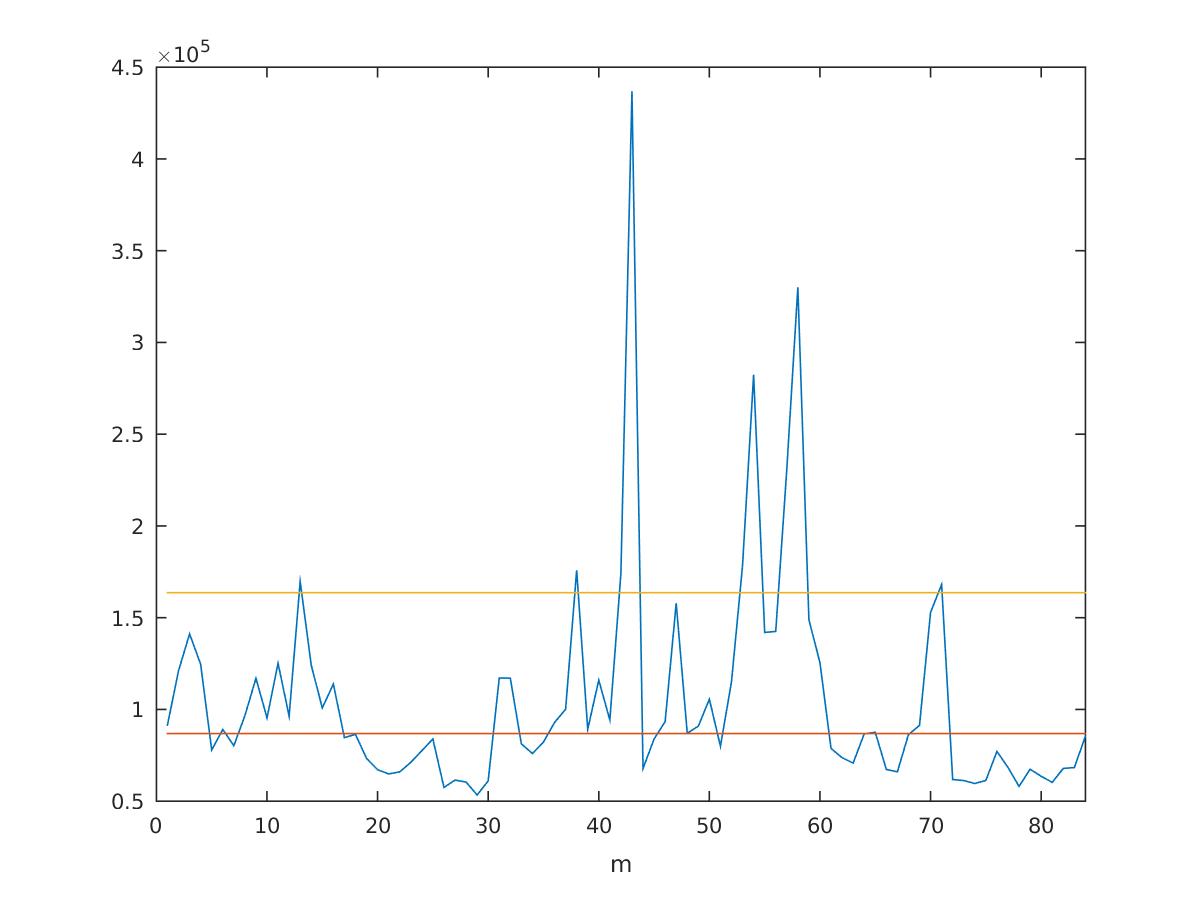}(h)

\caption{{\sc VV Polarization Channel} Series of squared deviations $d(m)$ and $t(m)$. The red horizontal line represents the median value and the yellow horizontal line represents their median plus two times their absolute median deviation.  $d(m)$ - Approximation Levels:  (a) $J=1$; (c) $J=2$; (e) $J=3$; (g) $J=4$. $t(m)$ - Approximation Levels:  (b) $J=1$; (d) $J=2$; (f) $J=3$; (h) $J=4$. 
}
\label{F:squared_J1-4_VV}
\end{figure}

\begin{figure}[htp!]
\noindent\includegraphics[width=18pc,height=4pc]{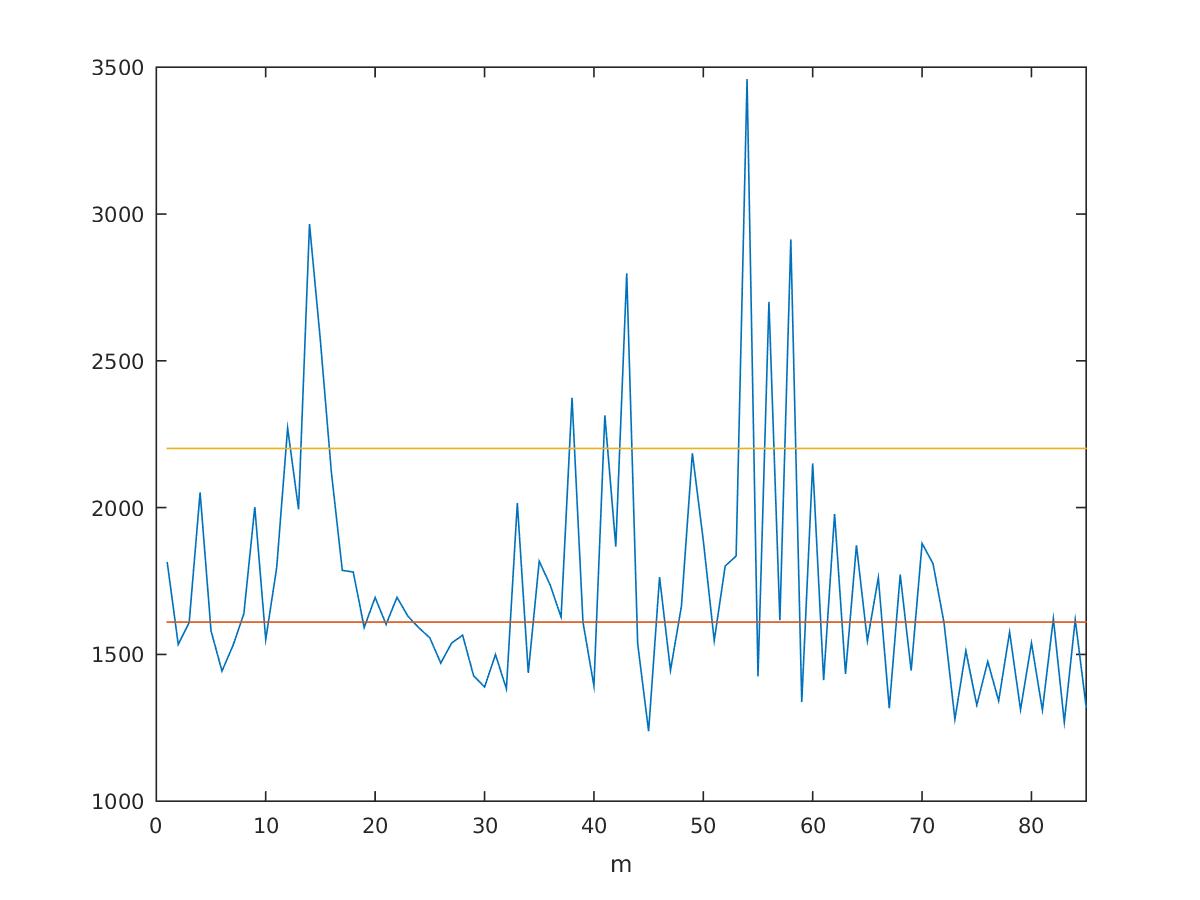}(a)
\includegraphics[width=18pc,height=4pc]{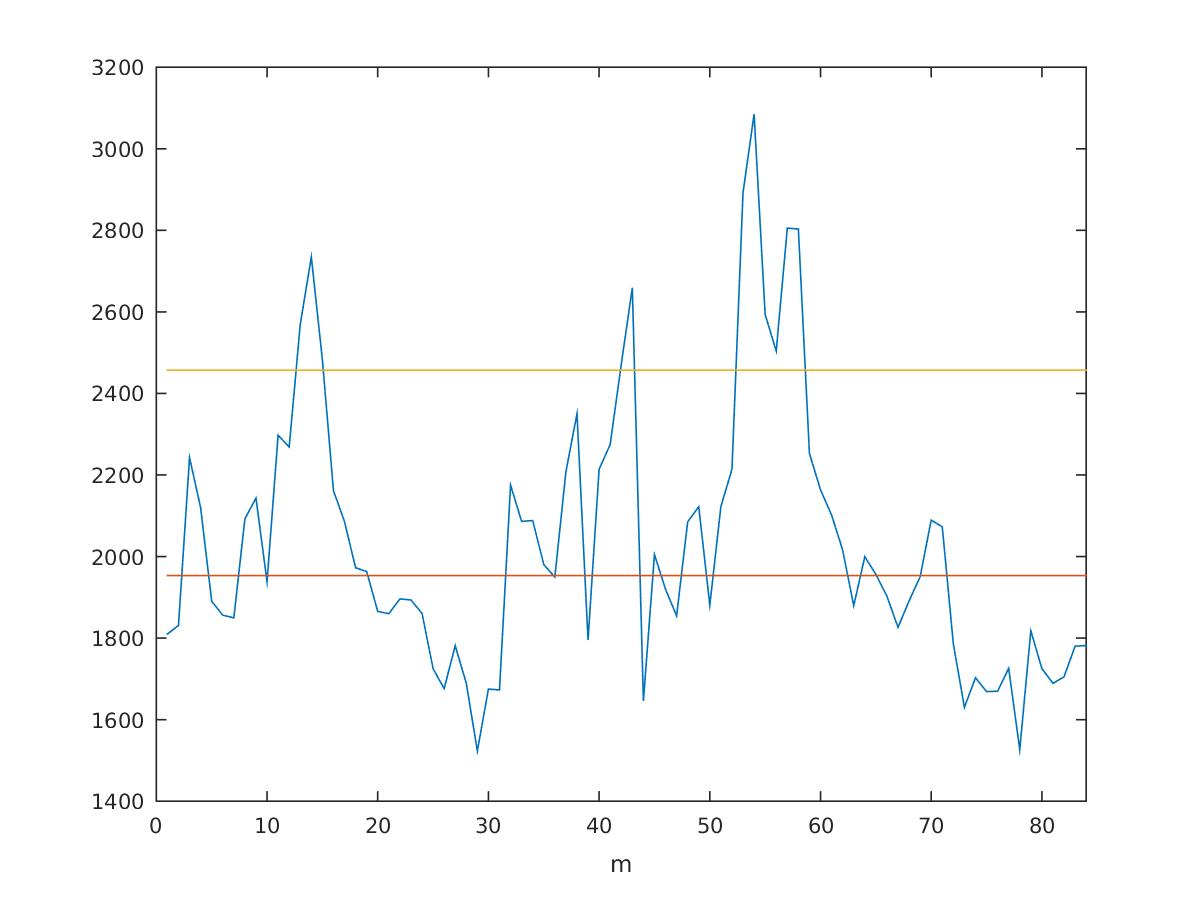}(b)

\includegraphics[width=18pc,height=4pc]{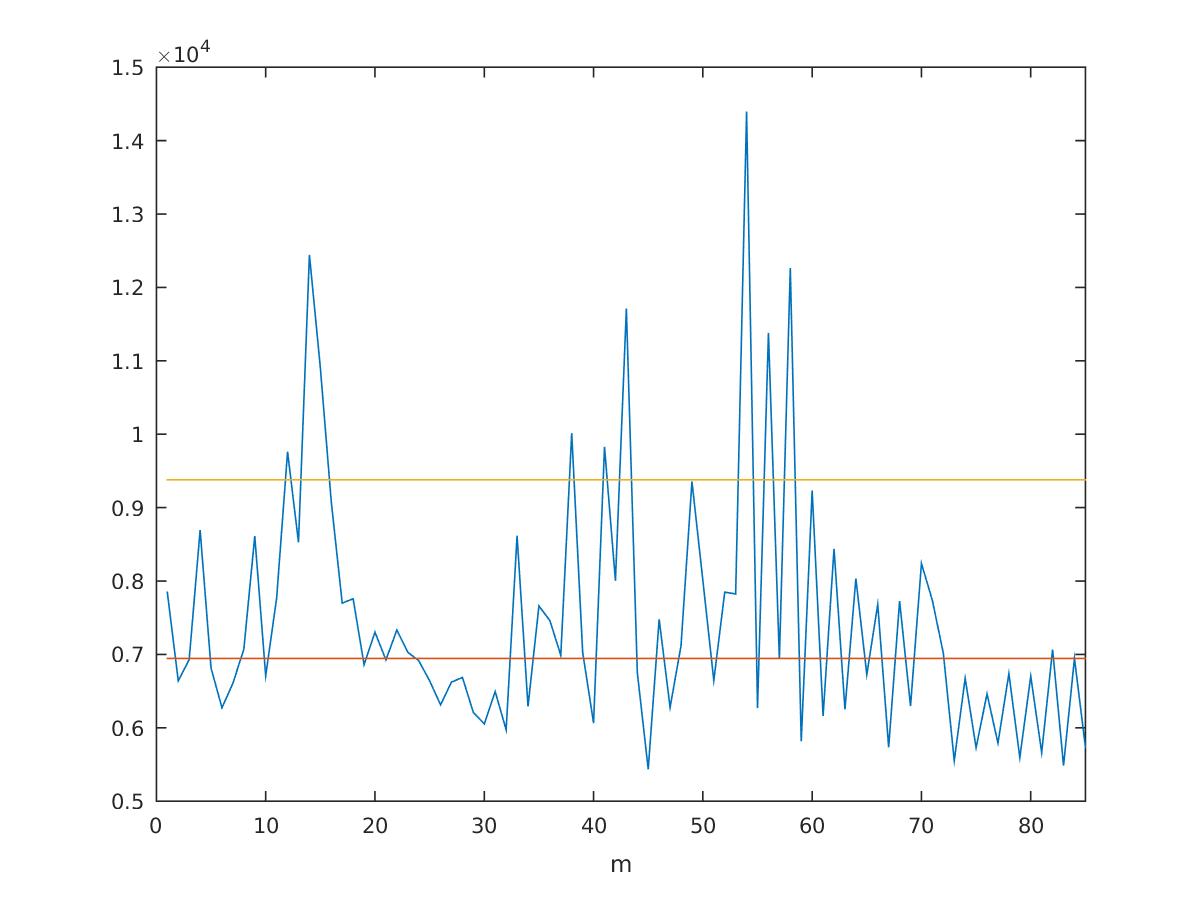}(c)
\includegraphics[width=18pc,height=4pc]{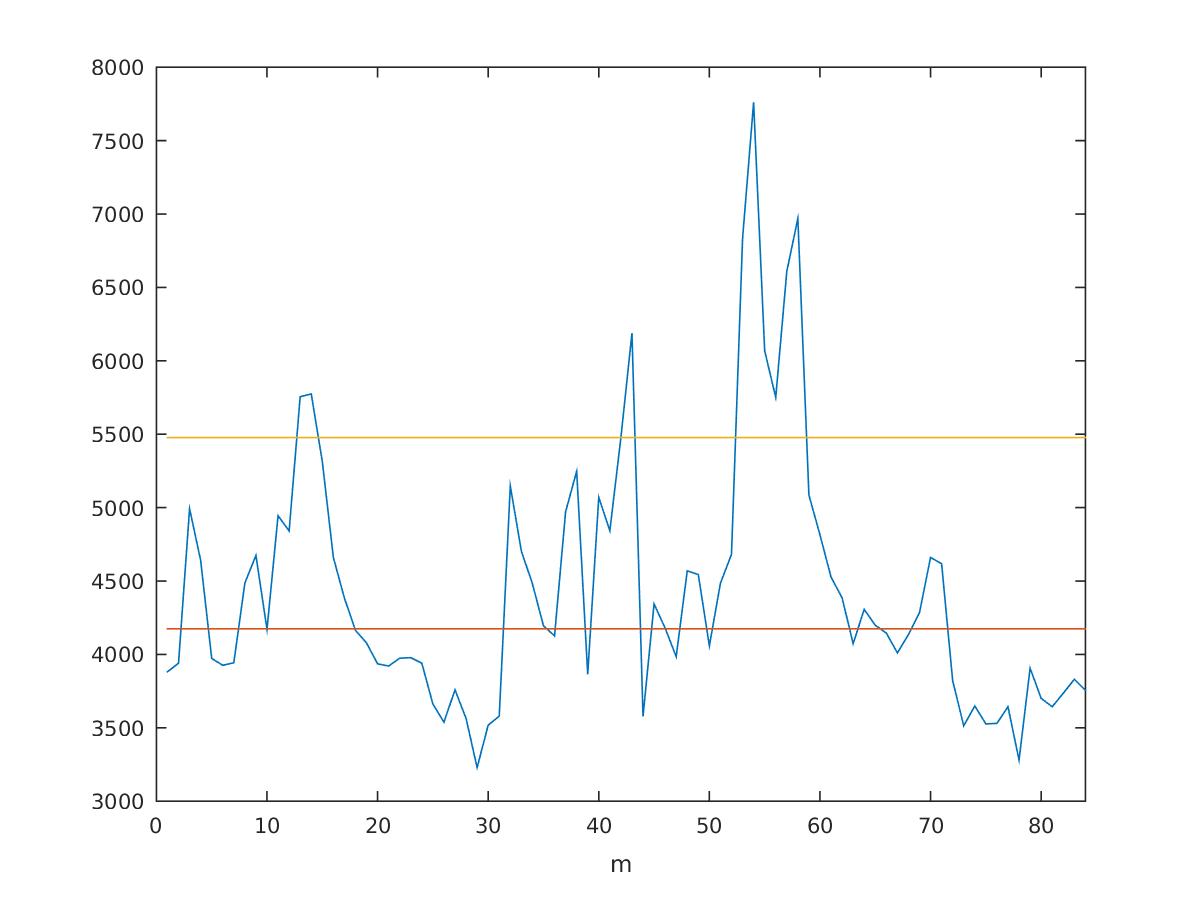}(d)

\includegraphics[width=18pc,height=4pc]{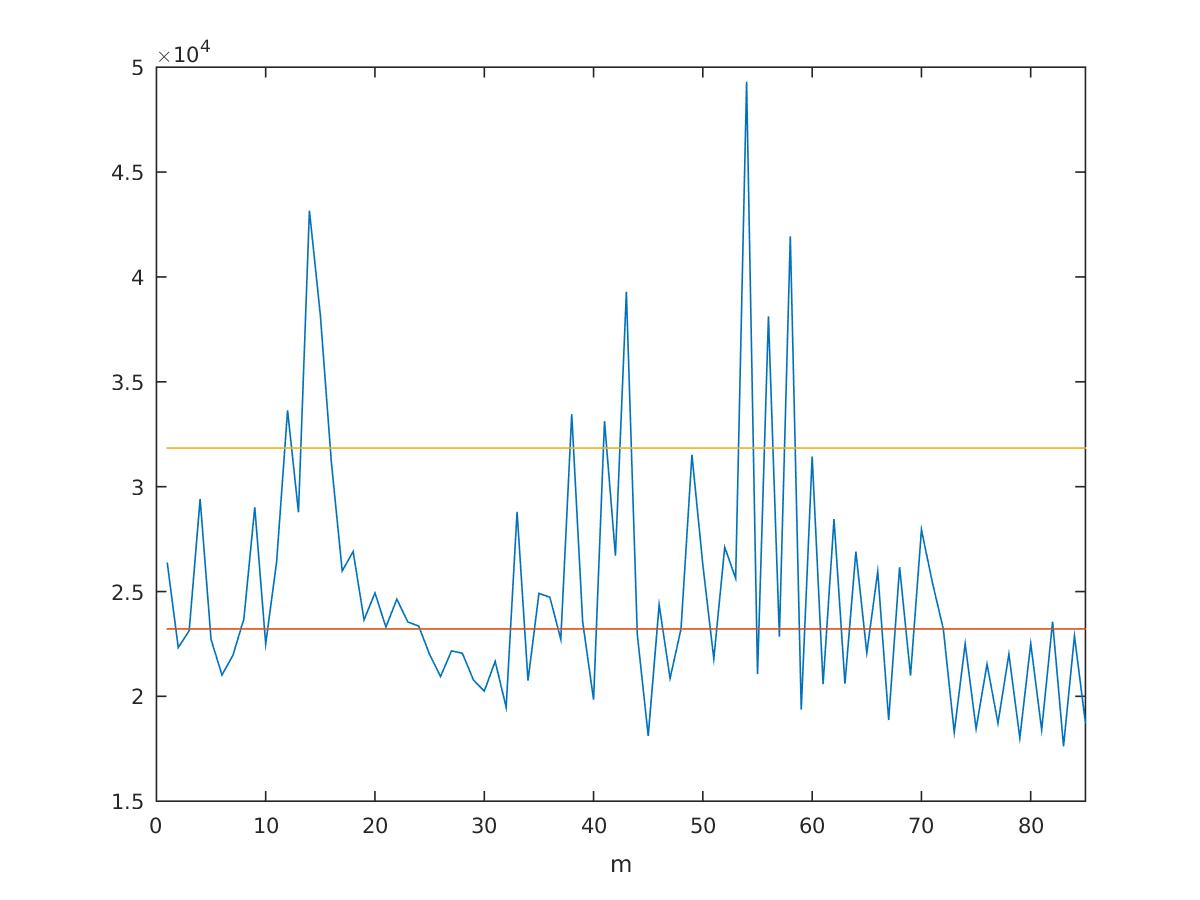}(e)
\includegraphics[width=18pc,height=4pc]{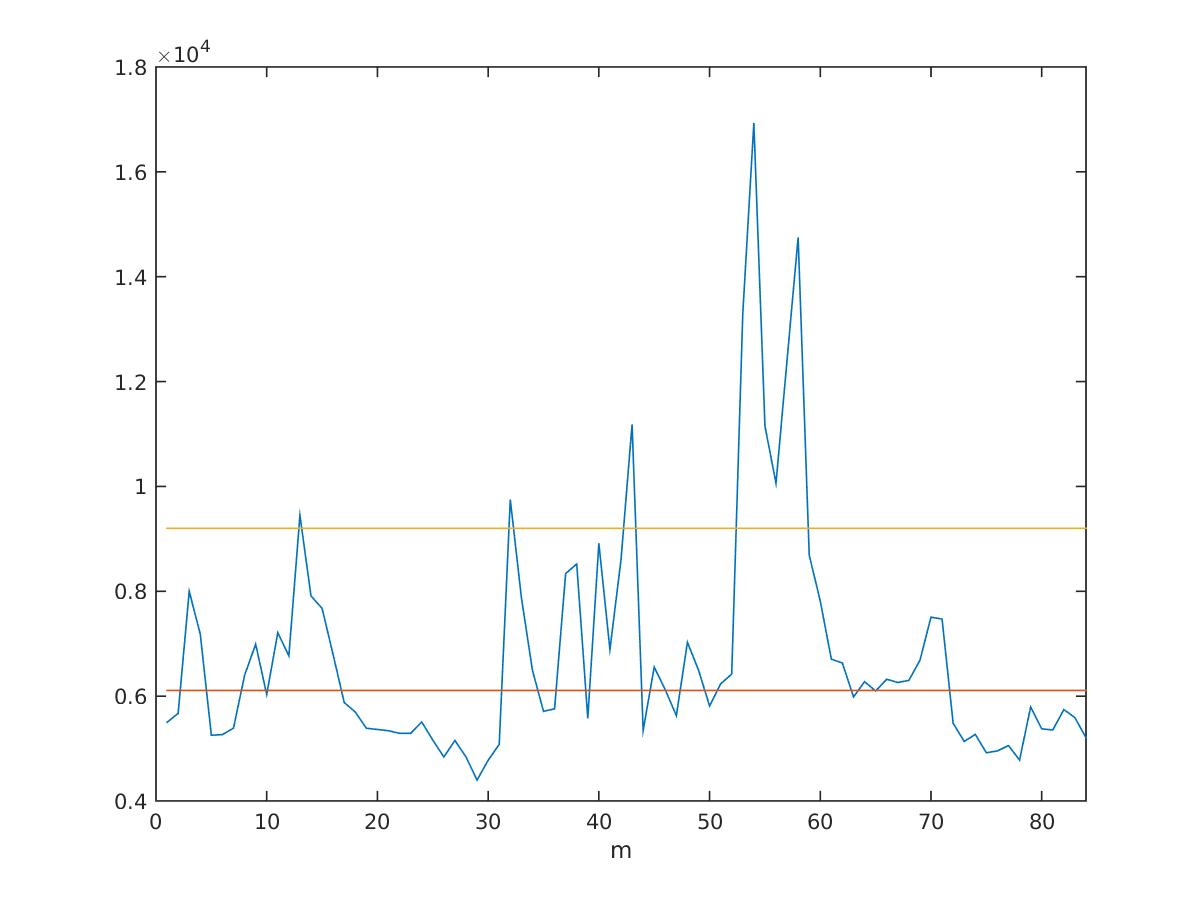}(f)

\includegraphics[width=18pc,height=4pc]{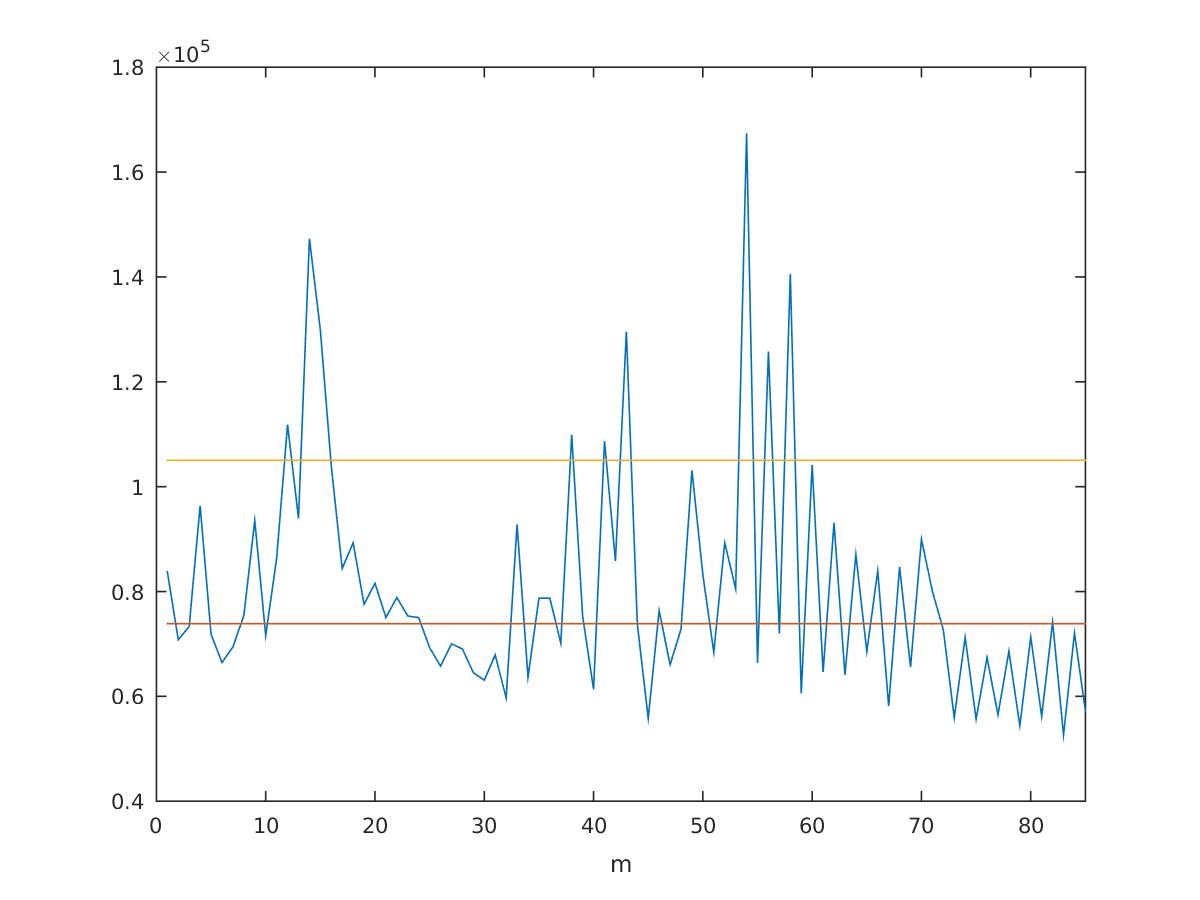}(g)
\includegraphics[width=18pc,height=4pc]{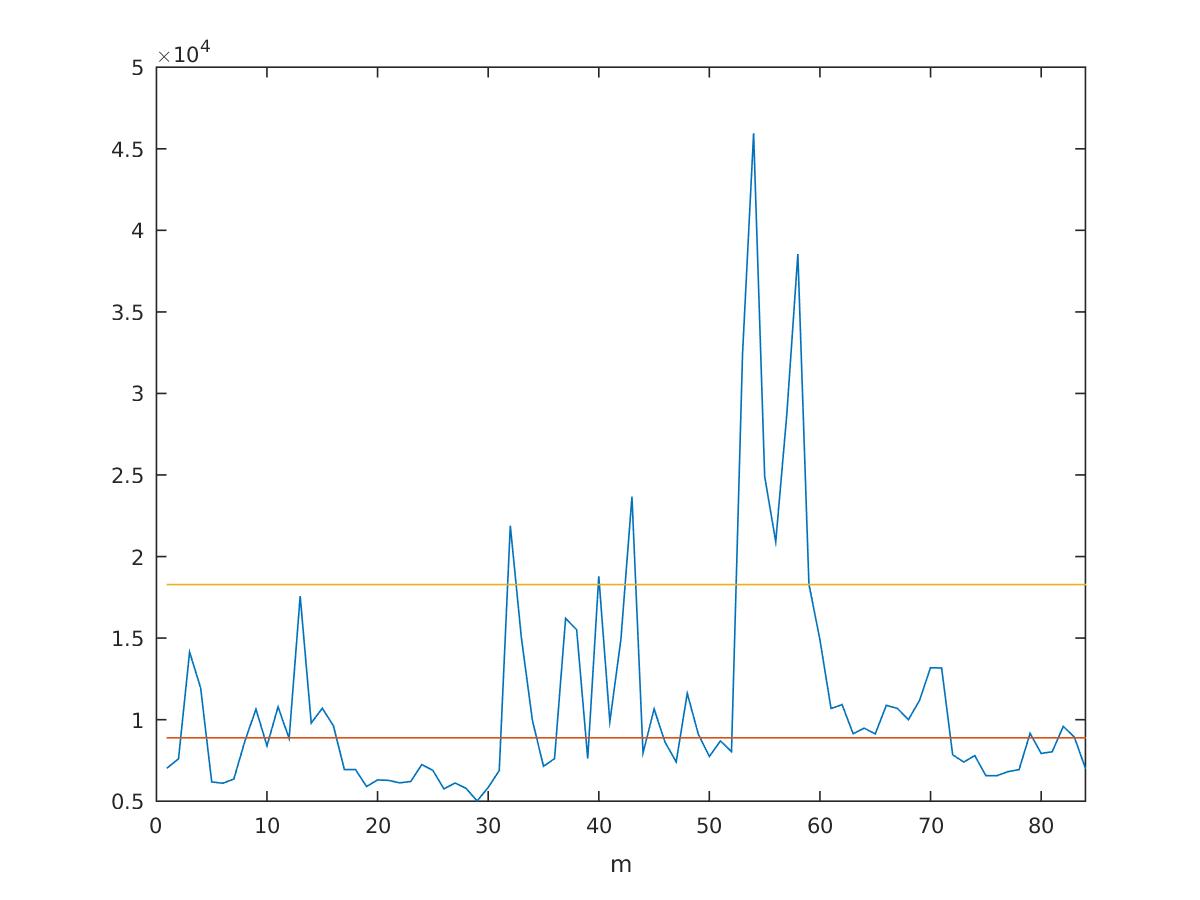}(h)
\caption{{\sc VH Polarization  Channel} Series of squared deviations $d(m)$ and $t(m)$. The red horizontal line represents the median value and the yellow horizontal line represents their median plus two times their absolute median deviation.  $d(m)$ - Approximation Levels:  (a) $J=1$; (c) $J=2$; (e) $J=3$; (g) $J=4$. $t(m)$ - Approximation Levels:  (b) $J=1$; (d) $J=2$; (f) $J=3$; (h) $J=4$. 
}
\label{F:squared_J1-4_VH}
\end{figure}

\begin{figure}[htp!]
\noindent\includegraphics[width=18pc,height=4pc]{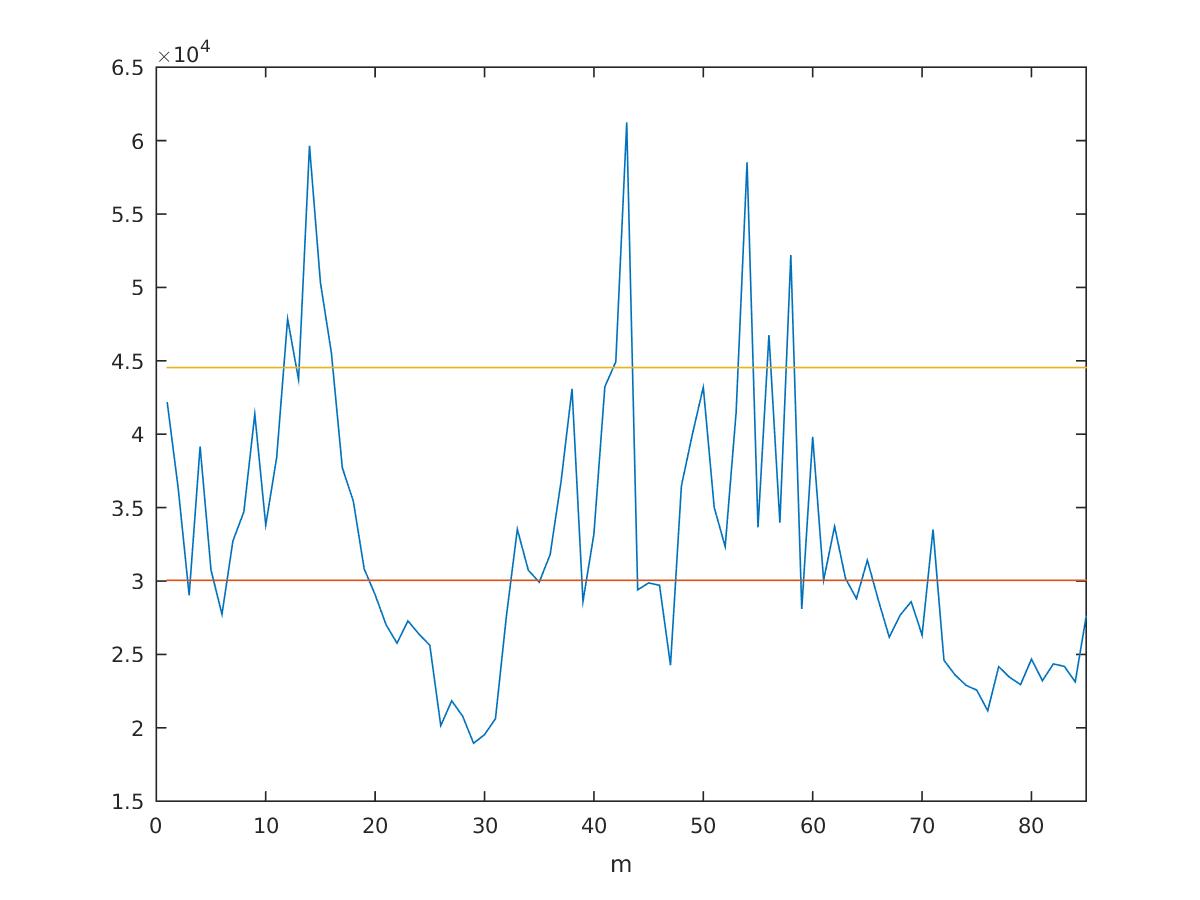}(a)
\noindent\includegraphics[width=18pc,height=4pc]{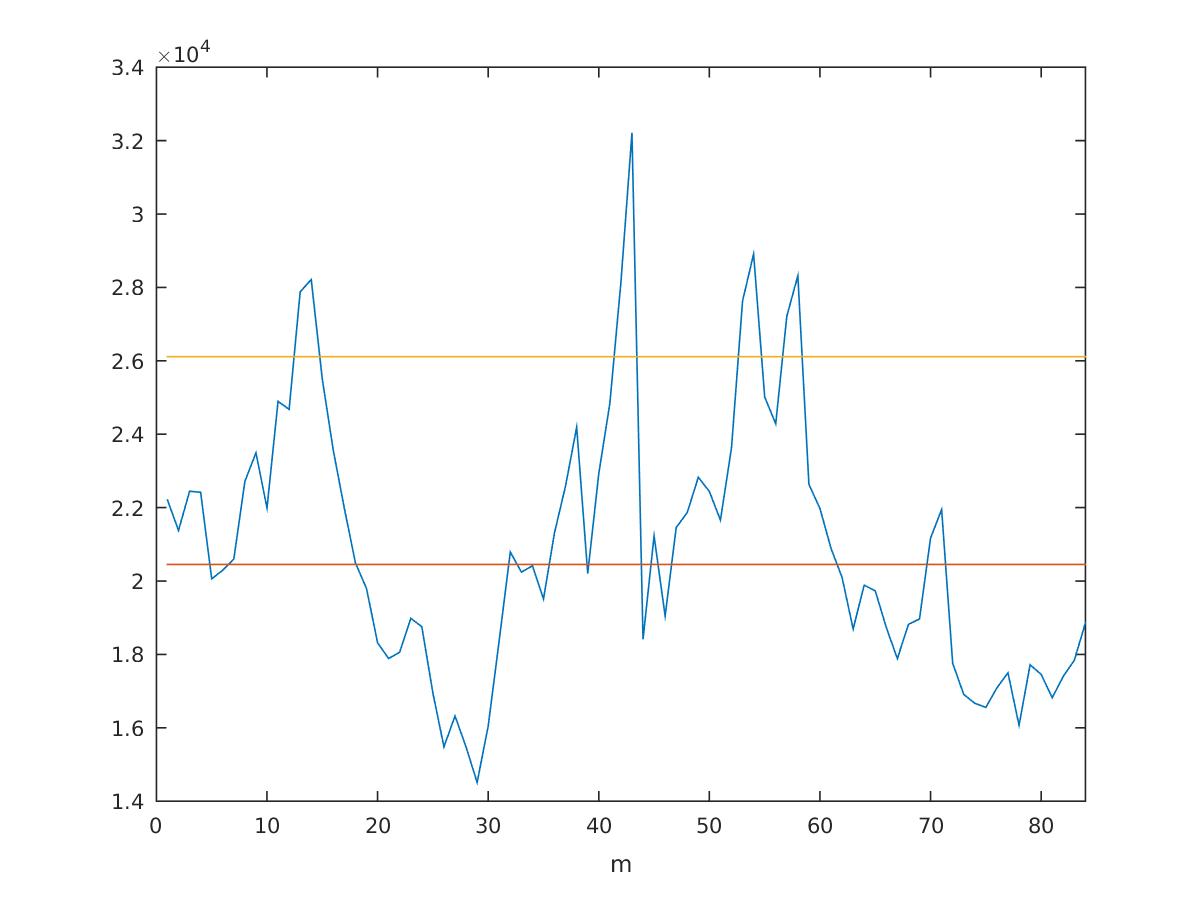}(b)

\includegraphics[width=18pc,height=4pc]{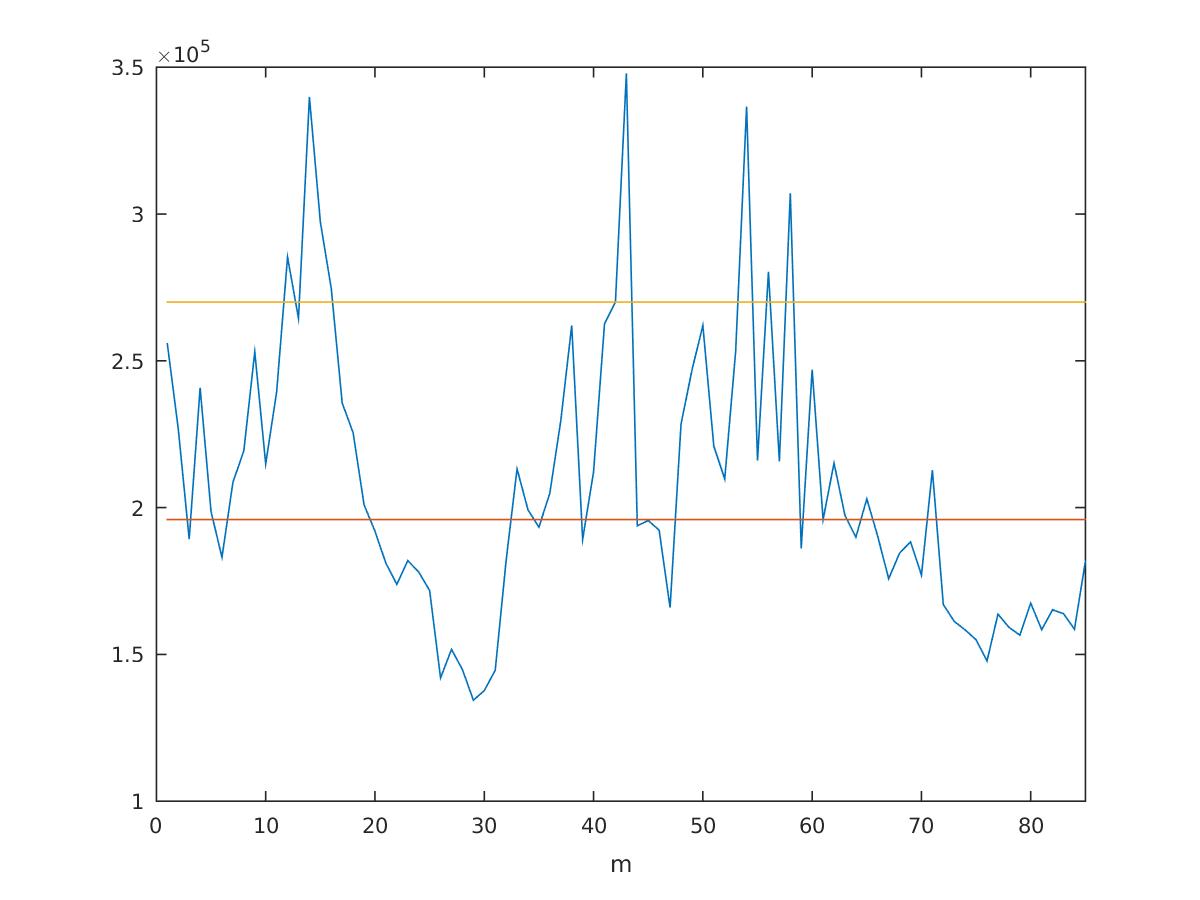}(c)
\includegraphics[width=18pc,height=4pc]{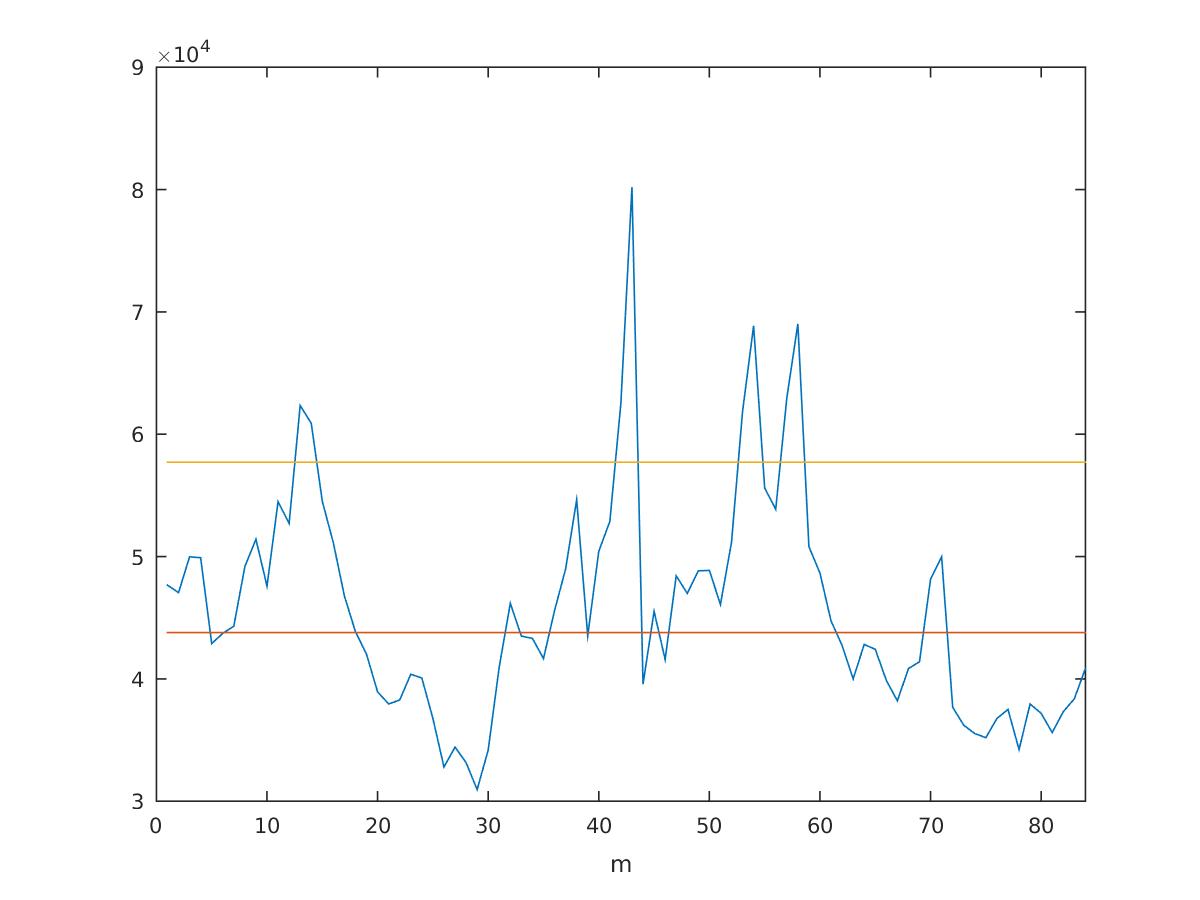}(d)

\includegraphics[width=18pc,height=4pc]{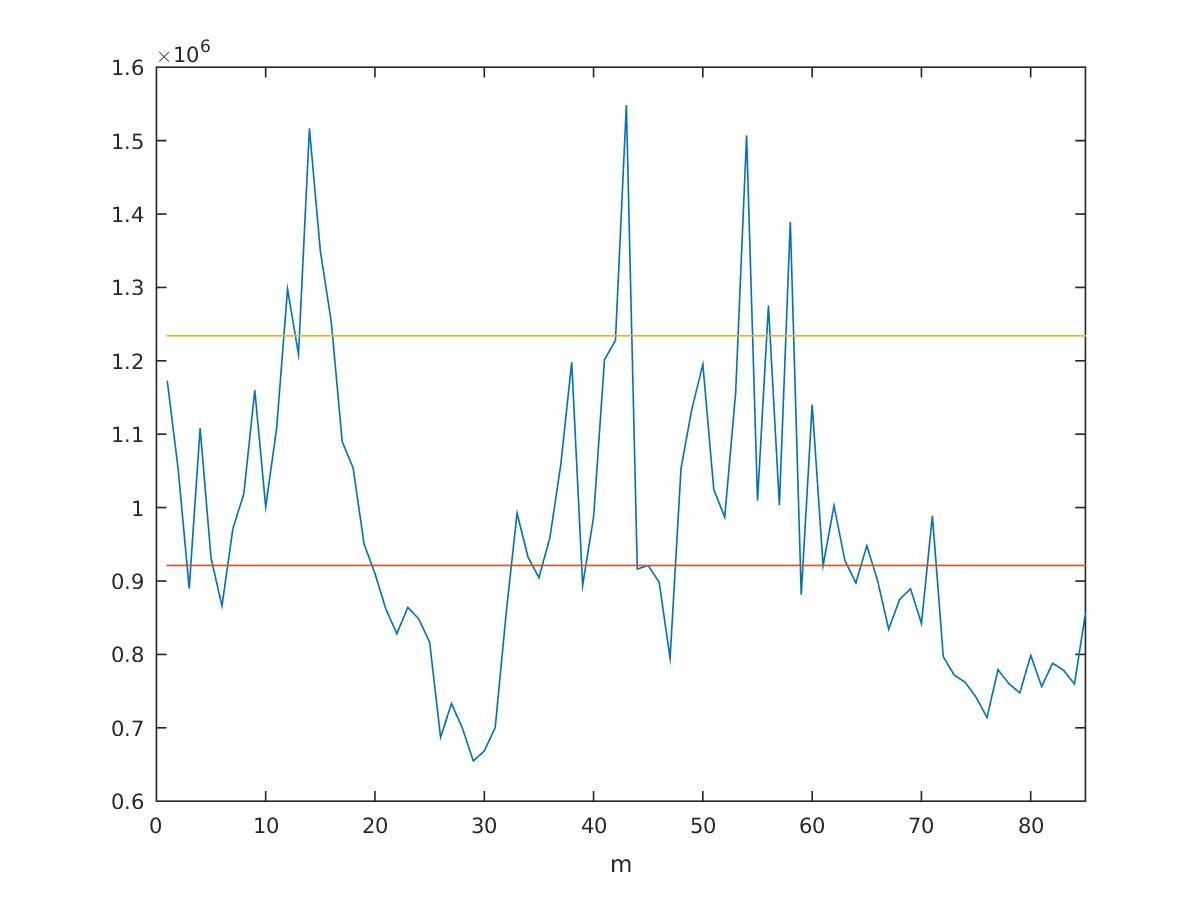}(e)
\includegraphics[width=18pc,height=4pc]{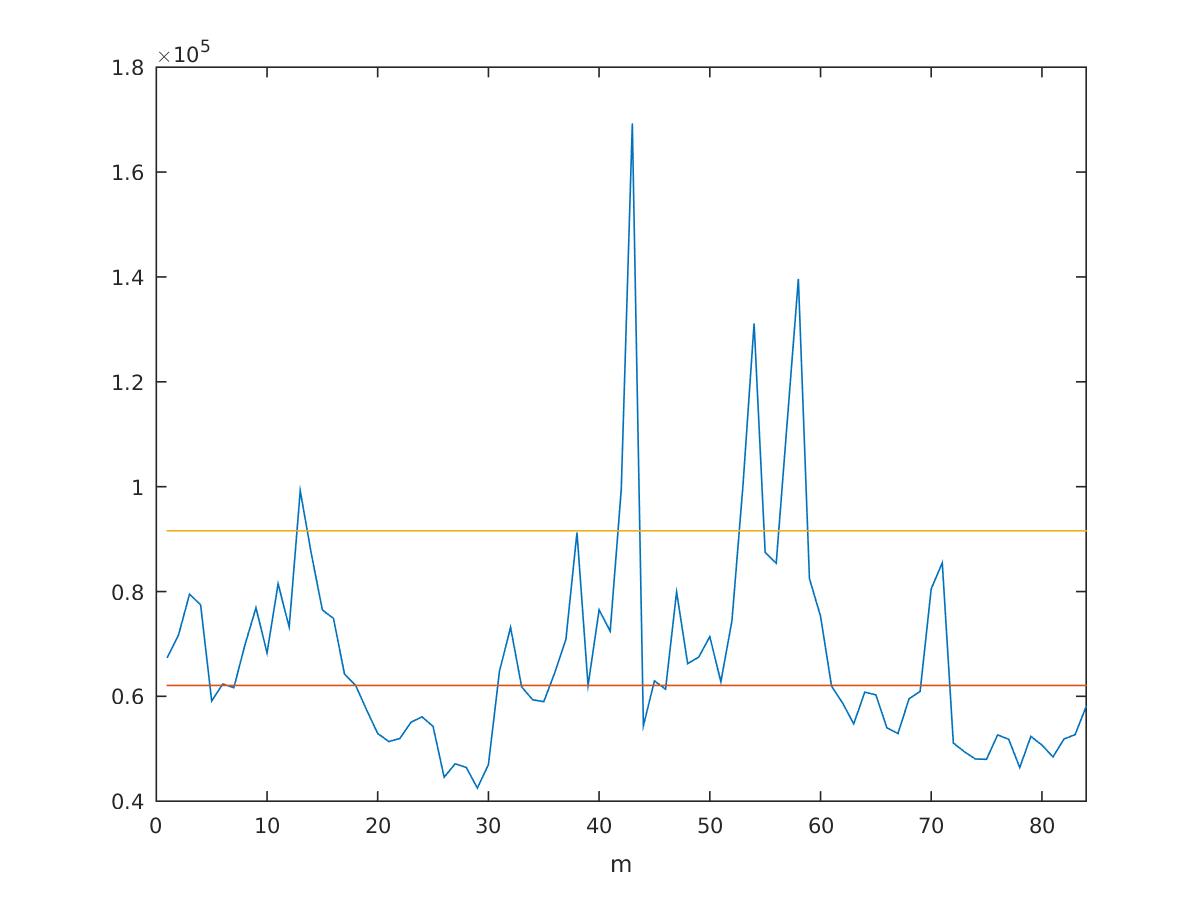}(f)

\includegraphics[width=18pc,height=4pc]{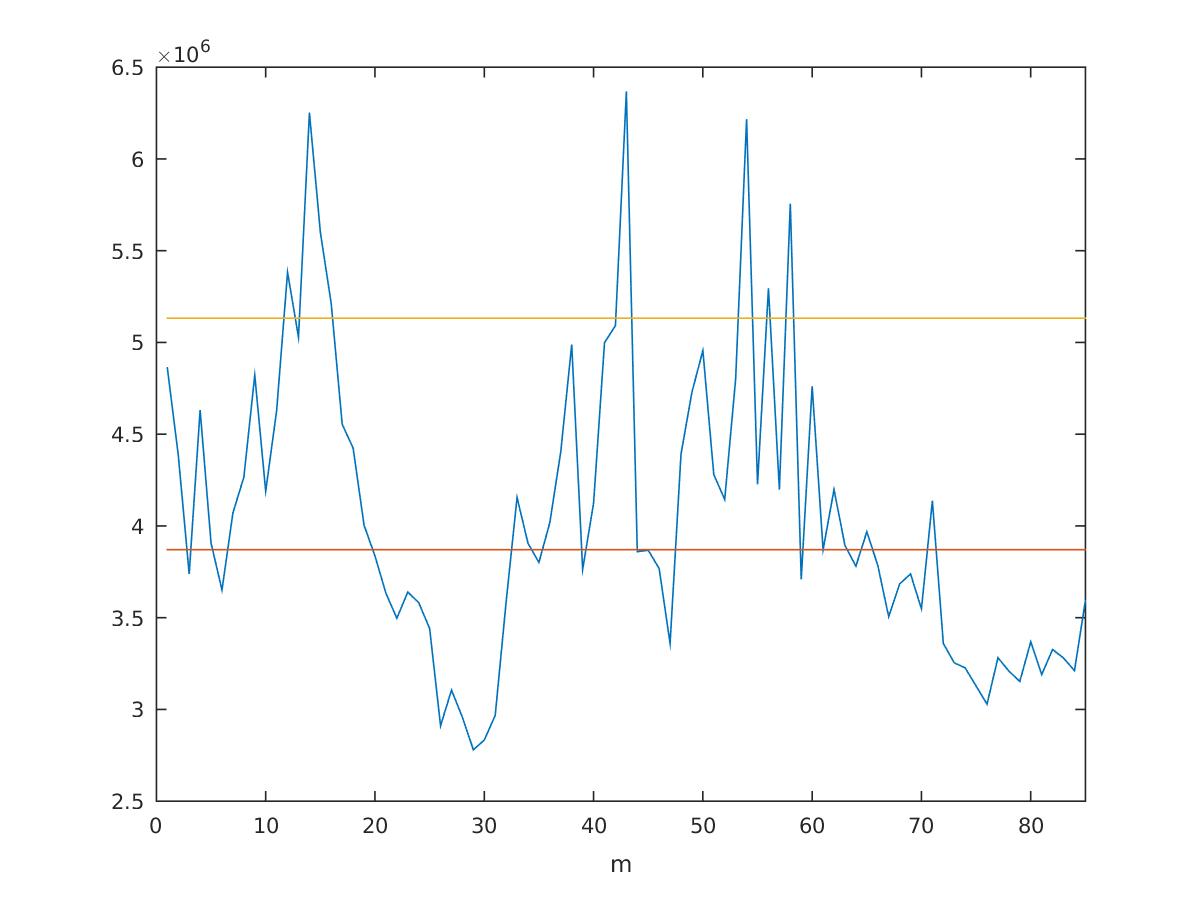}(g)
\includegraphics[width=18pc,height=4pc]{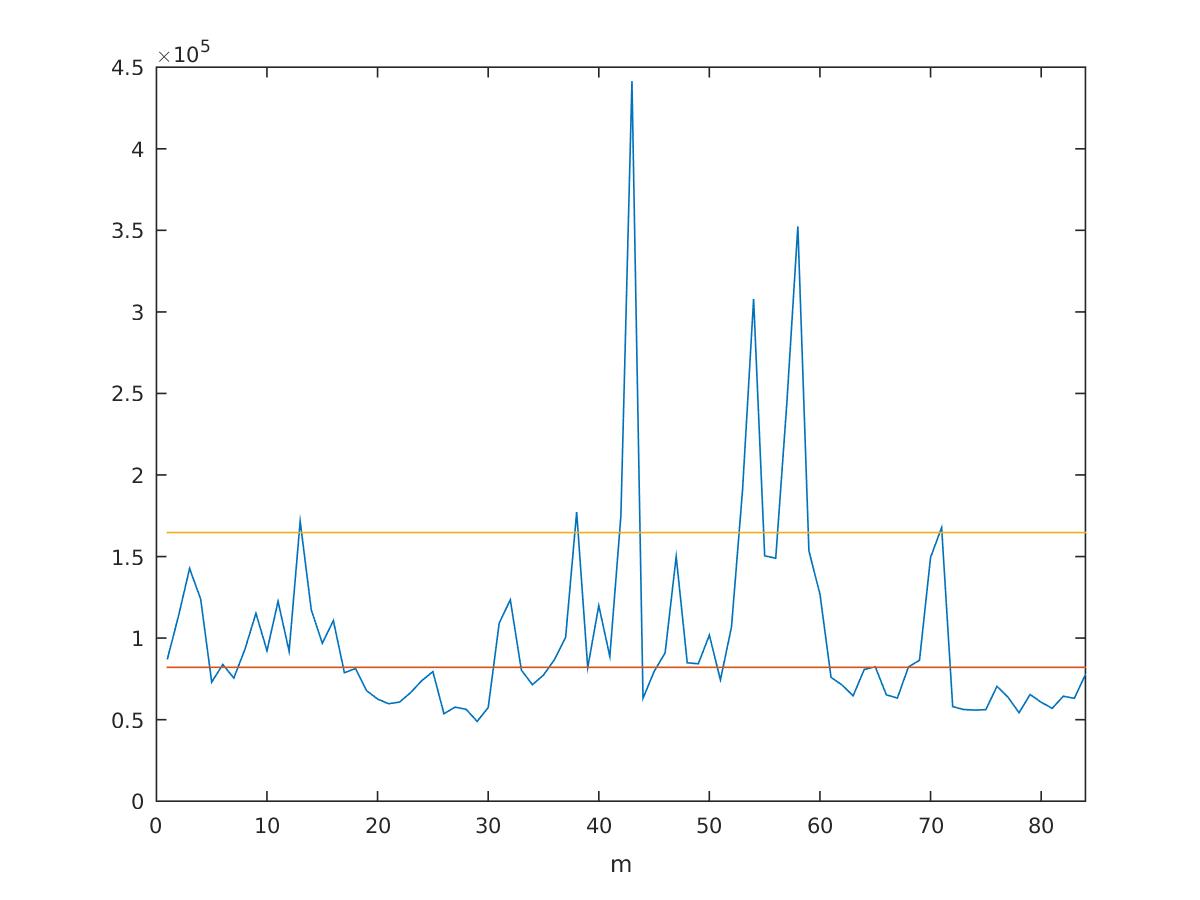}(h)

\caption{{\sc Combined Channels} Series of squared deviations $d(m)$ and $t(m)$. The red horizontal line represents the median value  and the yellow horizontal line represents their median plus two times their absolute median deviation.  $d(m)$ - Approximation Levels:  (a) $J=1$; (c) $J=2$; (e) $J=3$; (g) $J=4$. $d(m)$ - Approximation Levels:  (b) $J=1$; (d) $J=2$; (f) $J=3$; (h) $J=4$. 
}
\label{F:squared_J1-4_euclid}
\end{figure}

Figures \ref{F:squared_J1-4_VV}-\ref{F:squared_J1-4_euclid} show the series of squared deviations $\vd(m)$ and $\vt(m)$, for  the VV channel, VH channel, and Euclidean combination, respectively.  An overall feature on this data is that the VV polarization presents much higher energy than the VH. The amount of energy related to changes  is ten times higher on the former compared to the latter's.

Regarding change time points, in each figure, we can notice a pattern of peaks which are common to all approximation levels. They are time points:
\begin{list}{}{}
\item (a) 14, 43, 54, and 58 by the coefficients' squared deviations on the average VV image;
\item (b) 14, 43, 54, and 58 by the coefficients' squared deviations on the consecutive VV images;
\item (c) 14, 38, 41, 43, 54, 56, and 58 by the coefficients' squared deviations on the average VH image;
\item (d) 14, 38, 41, 43, 54-58 by the coefficients' squared deviations on the consecutive VH images;
\item (e) 14, 43, 54, and 58 by the coefficients' squared deviations on the average combined channels image; and
\item (e) 14, 43, 54, and 58 by the coefficients' squared deviations on the consecutive combined channels images.
\end{list}

\begin{figure}[htp!]
\noindent\includegraphics[width=18pc,height=3pc]{J3_VV_squared_meandev}(a)
\includegraphics[width=18pc,height=3pc]{consecdif_J3_VV_squared_meandev}(b)

\includegraphics[width=18pc,height=3pc]{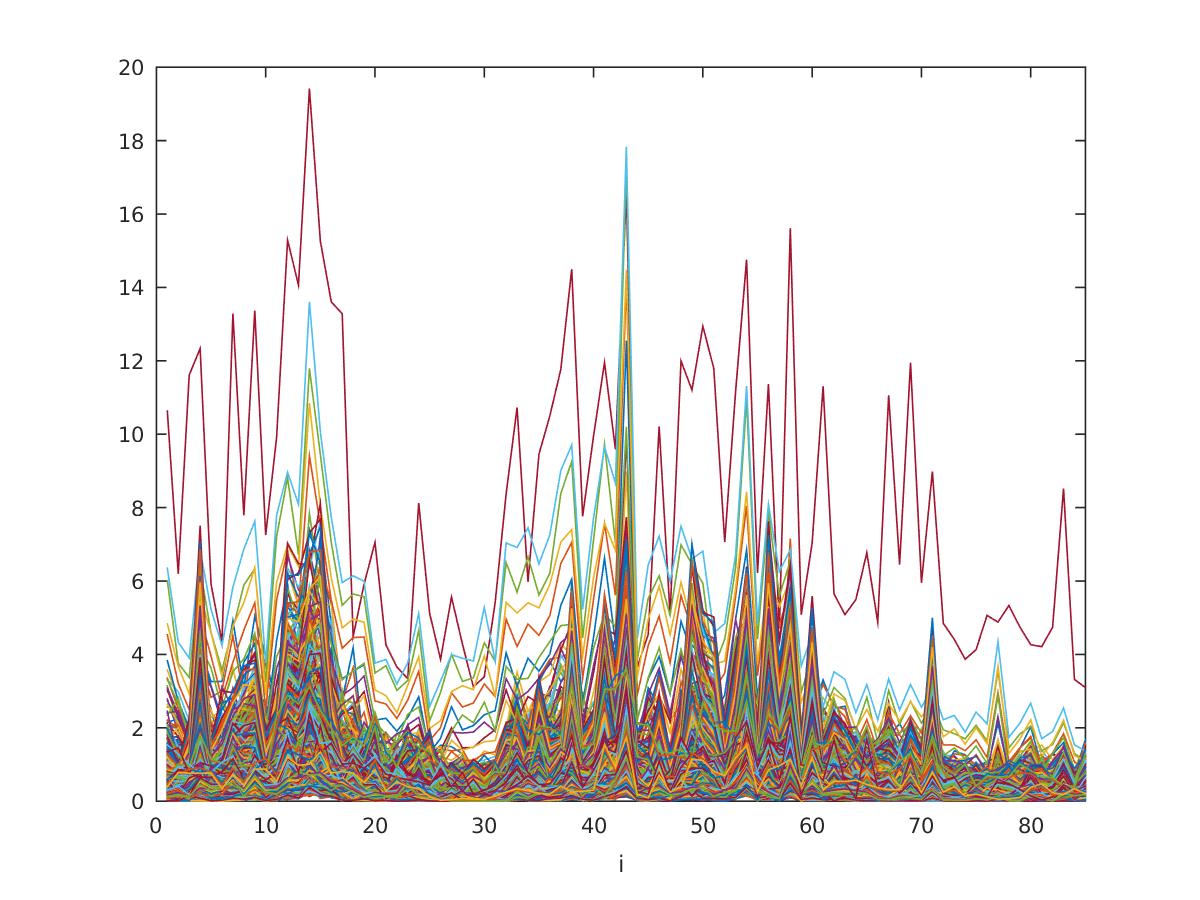}(c)
\includegraphics[width=18pc,height=3pc]{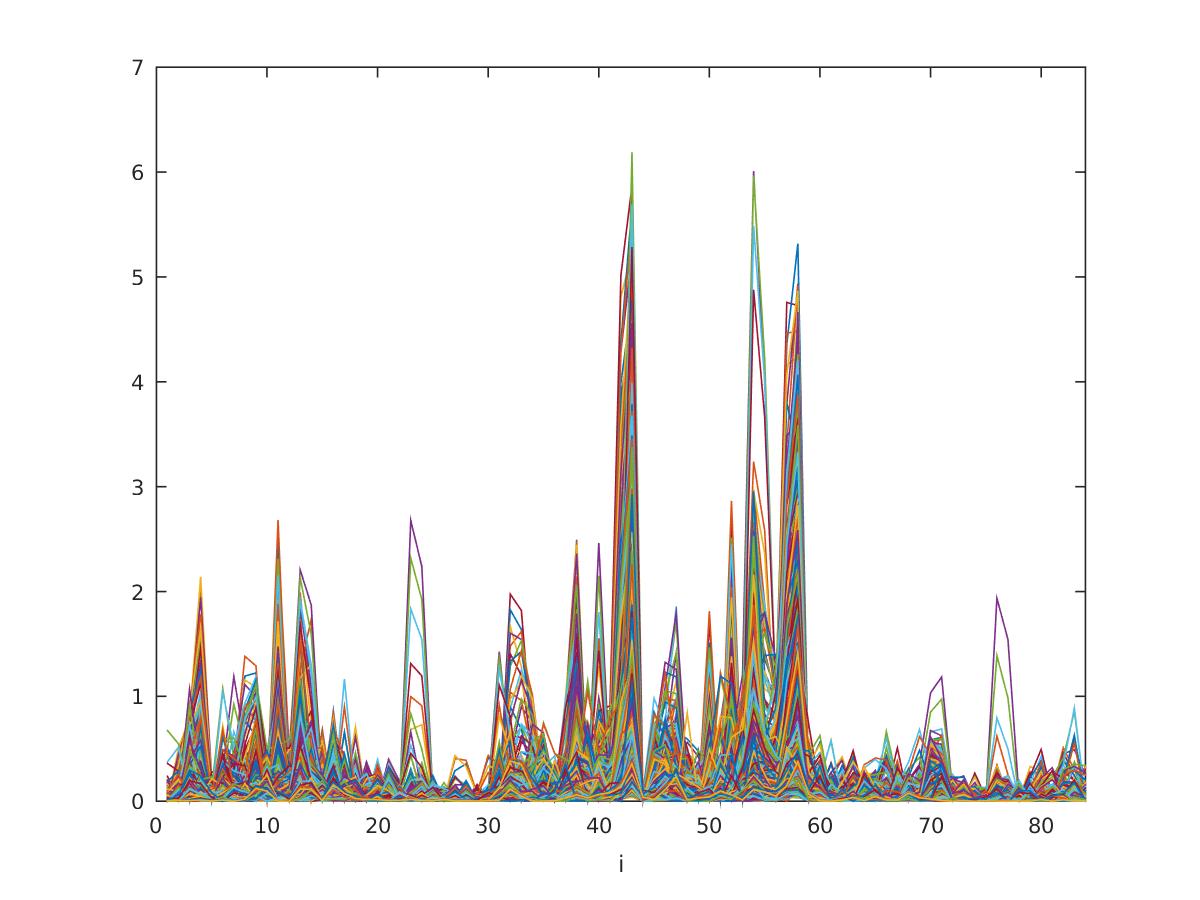}(d)

\includegraphics[width=18pc,height=3pc]{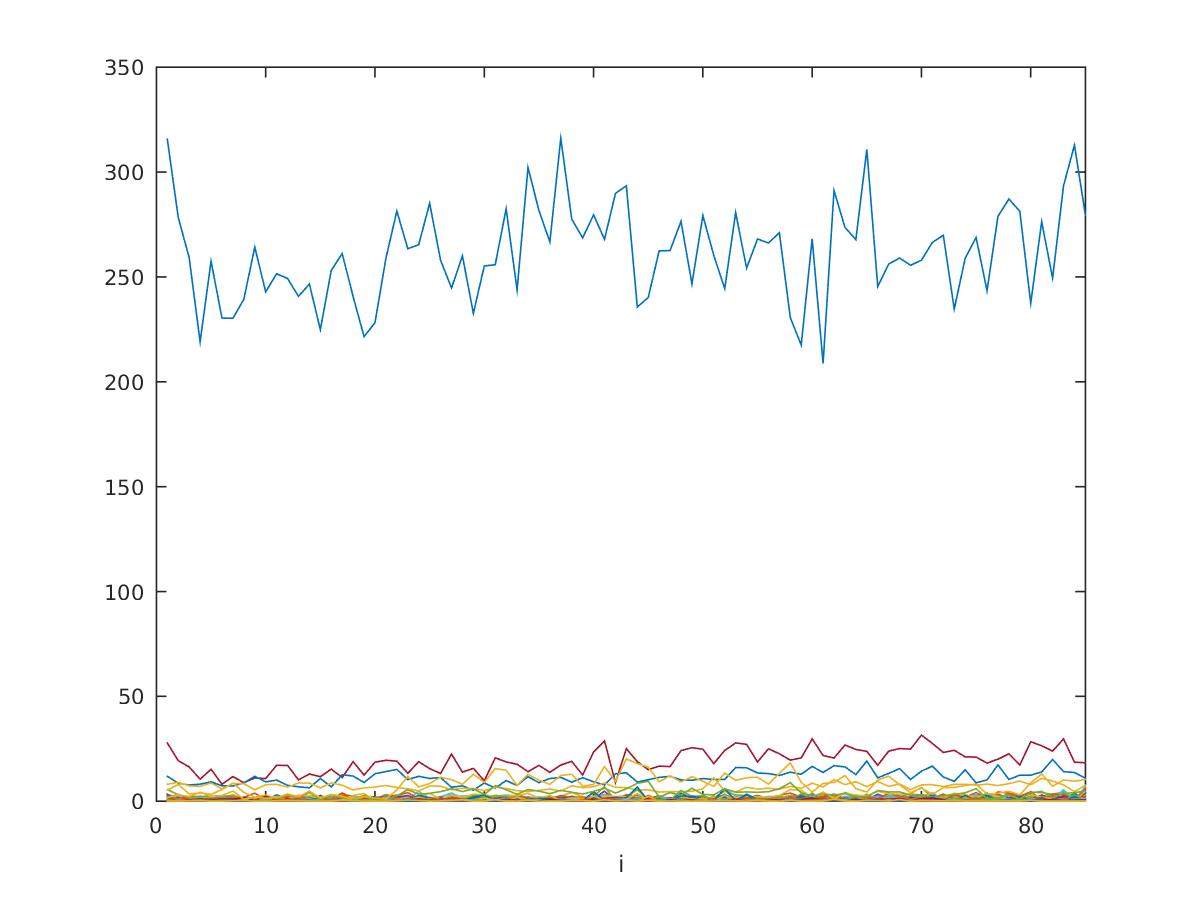}(e)
\includegraphics[width=18pc,height=3pc]{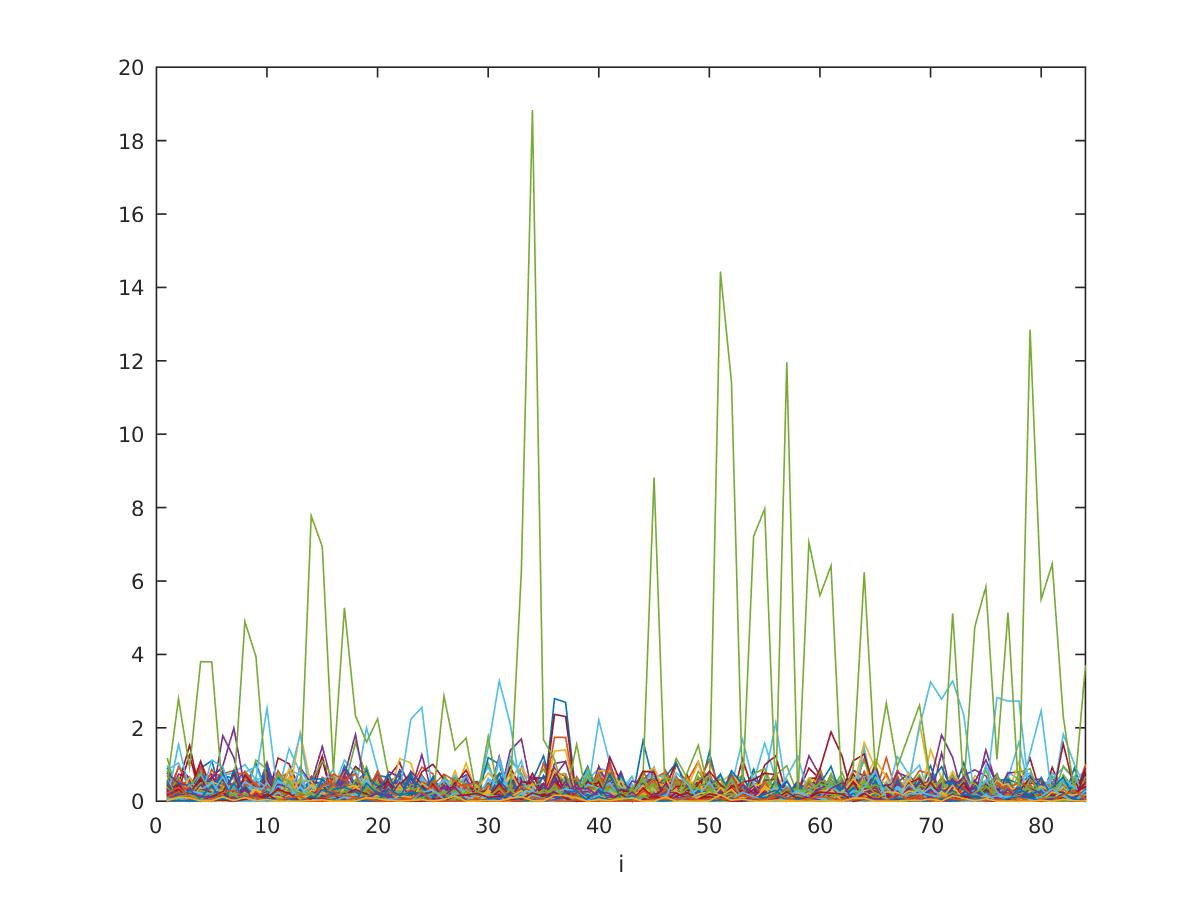}(f)

\includegraphics[width=18pc,height=3pc]{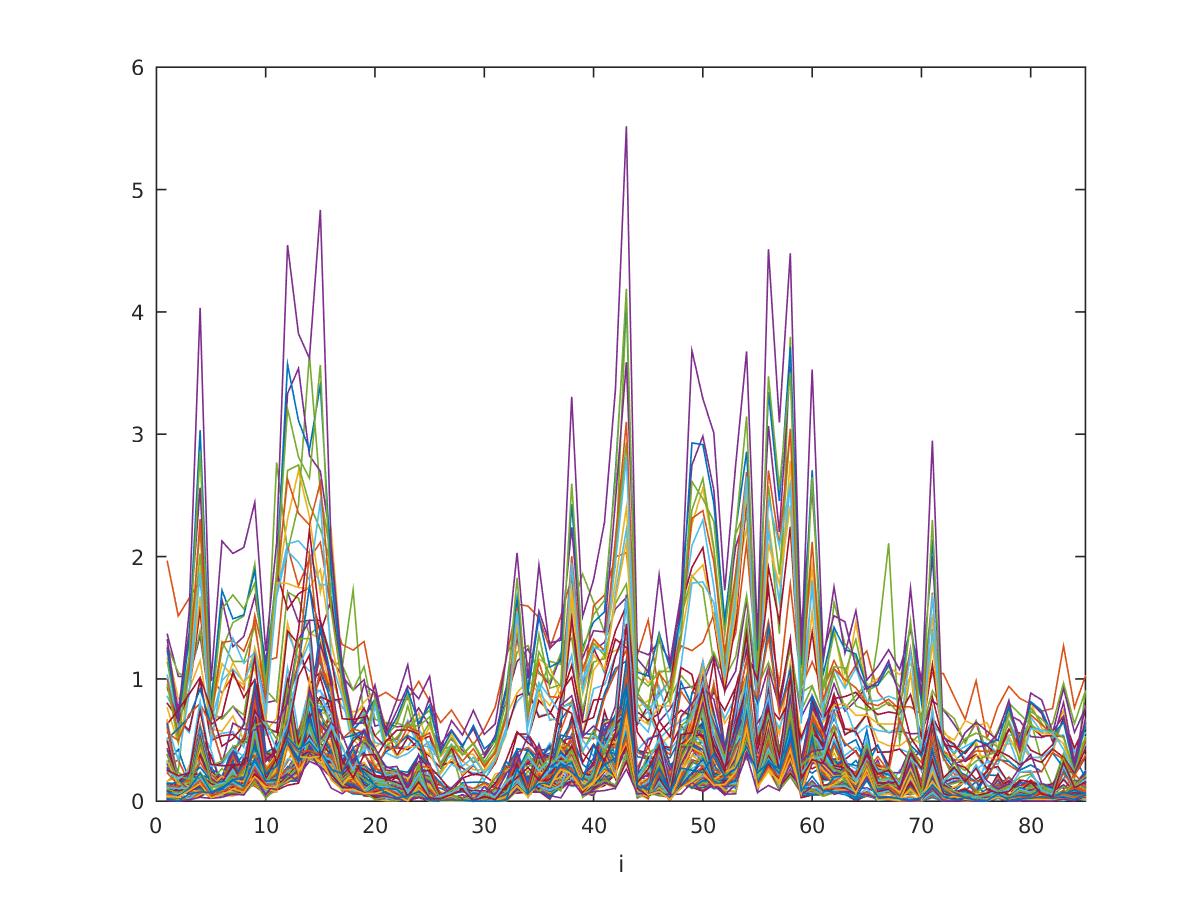}(g)
\includegraphics[width=18pc,height=3pc]{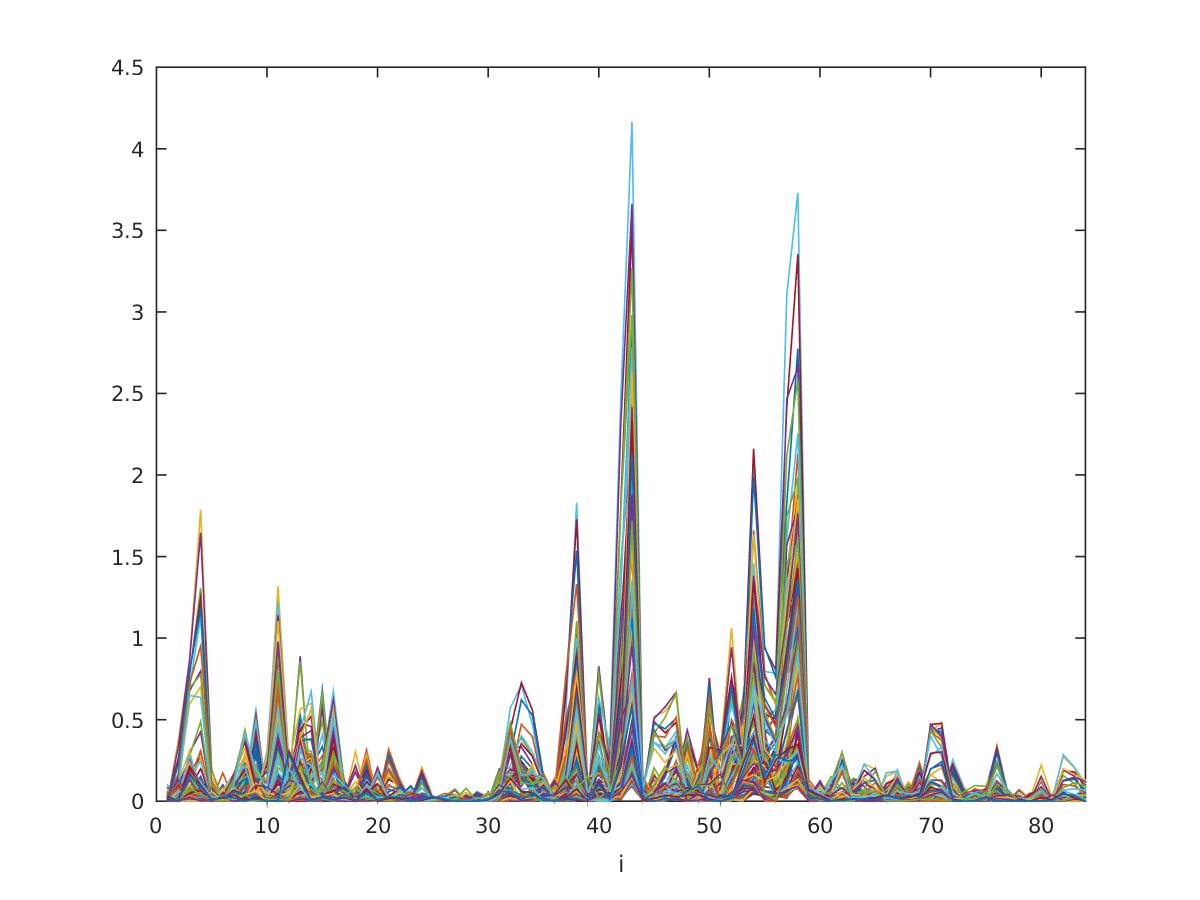}(h)

\includegraphics[width=18pc,height=3pc]{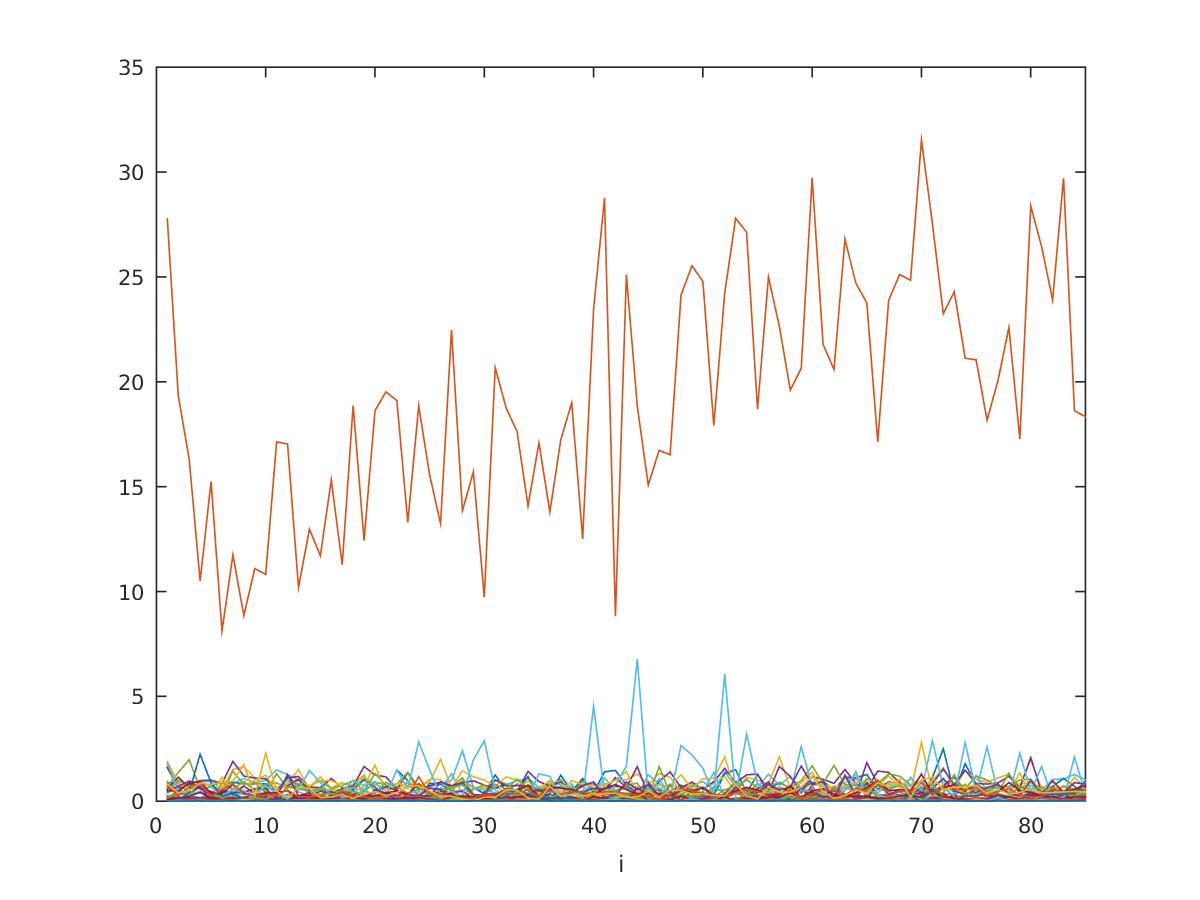}(i)
\includegraphics[width=18pc,height=3pc]{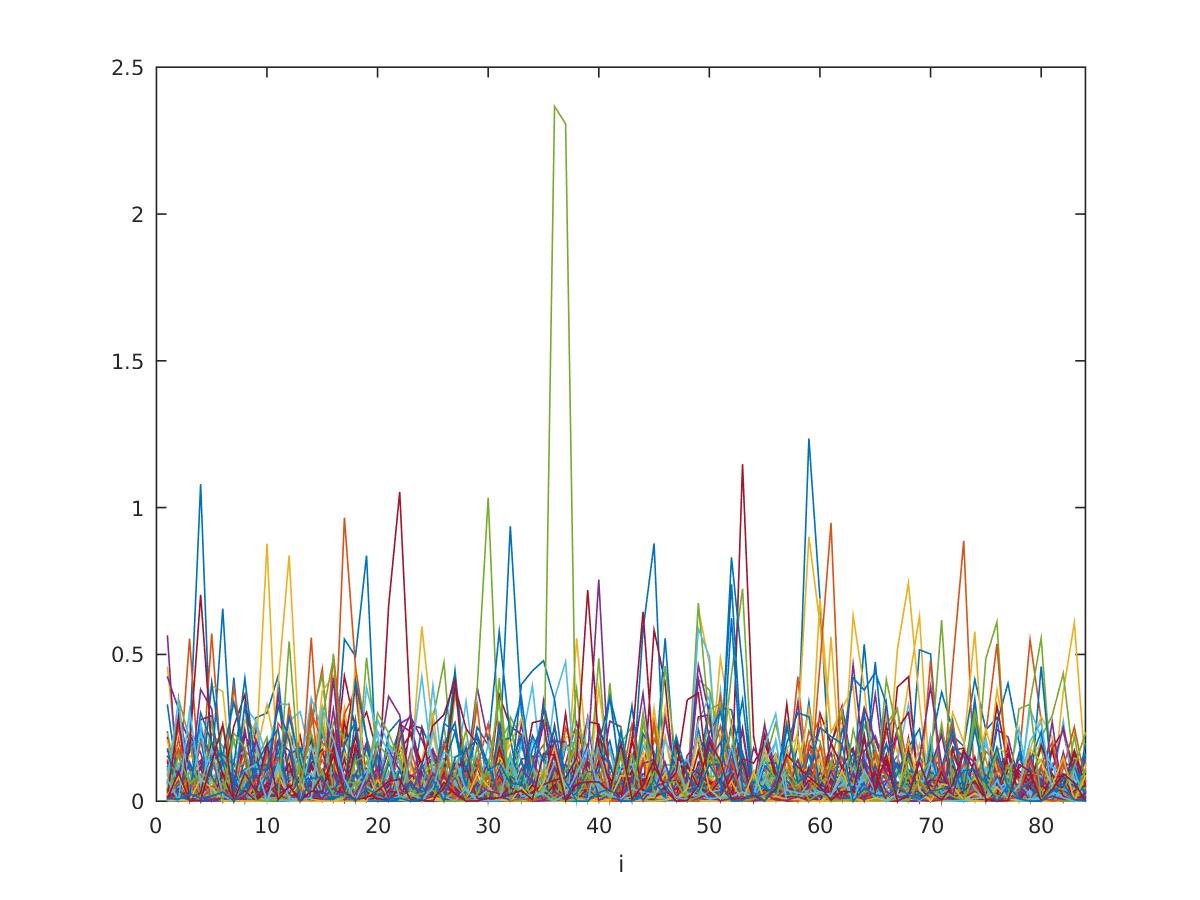}(j)

\caption{{\sc VV Polarization Channel} Level $J=3$ series of squared mean deviations: (a) $\vd(m)$; (b) $\vt(m)$. Red horizontal line represents the median value., Yellow, two absolute median deviations beyond the median. Squared approximation coefficient deviations: 0.1\% highest absolute correlations - (c) $\vd(m)$;  (d) $\vt(m)$; 0.1\% smallest absolute correlations - (e) $\vd(m)$; (f) $\vt(m)$; 0.01\% highest absolute correlations - (g) $\vd(m)$;  (h) $\vt(m)$;  0.01\% smallest absolute correlations - (i) $\vd(m)$; (j)  $\vt(m)$. 
}  
\label{F:squared_meandev_J3_VV}
\end{figure}

\begin{figure}[htp!]
\noindent\includegraphics[width=18pc,height=3pc]{J3_VH_squared_meandev}(a)
\includegraphics[width=18pc,height=3pc]{consecdif_J3_VH_squared_meandev}(b)

\includegraphics[width=18pc,height=3pc]{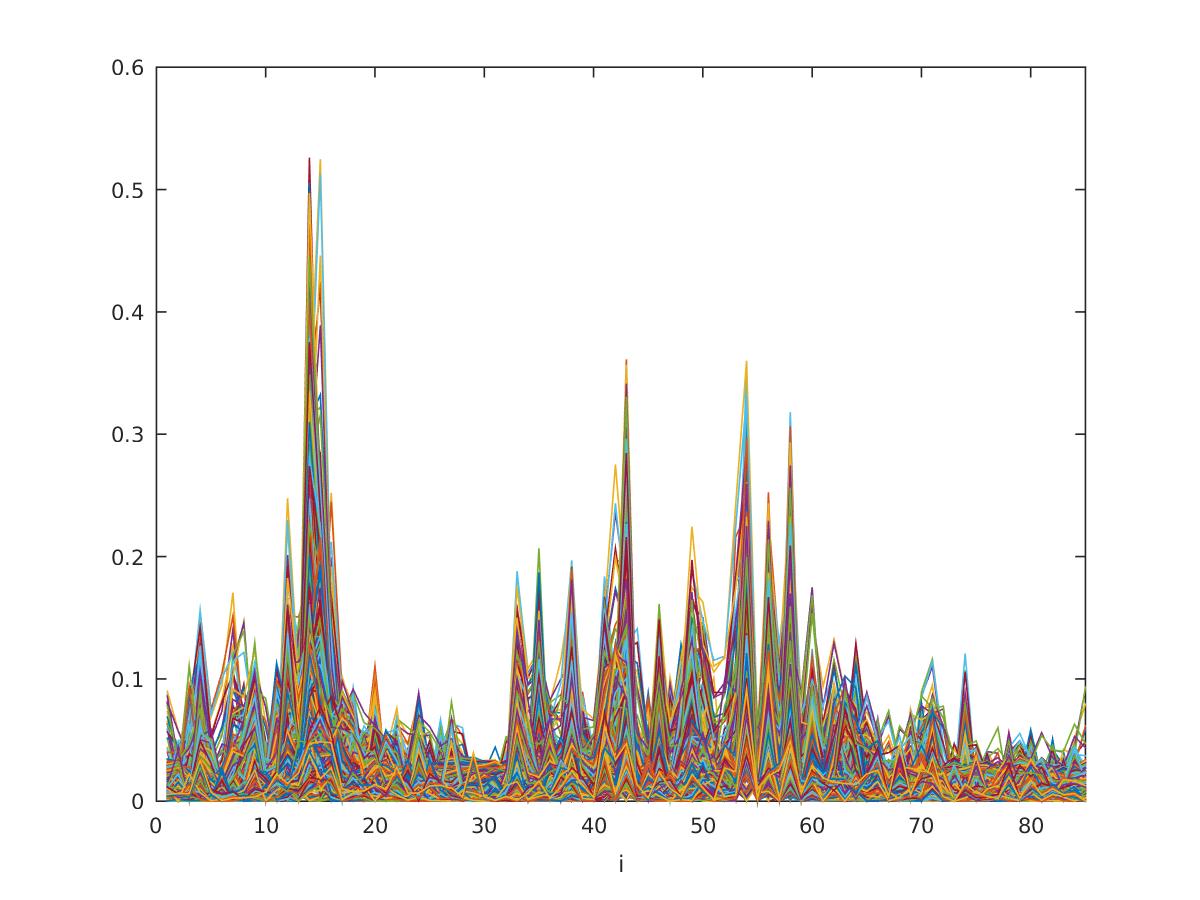}(c)
\includegraphics[width=18pc,height=3pc]{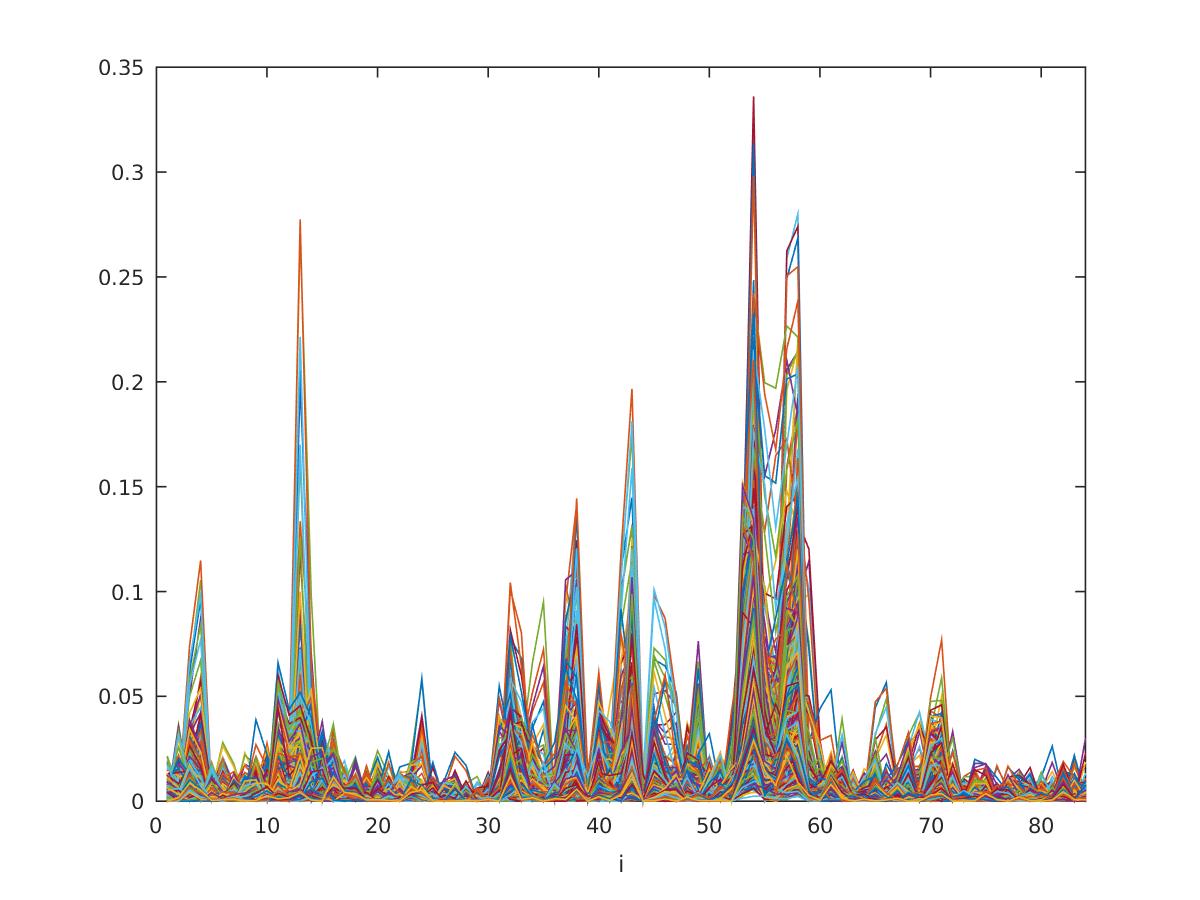}(d)

\includegraphics[width=18pc,height=3pc]{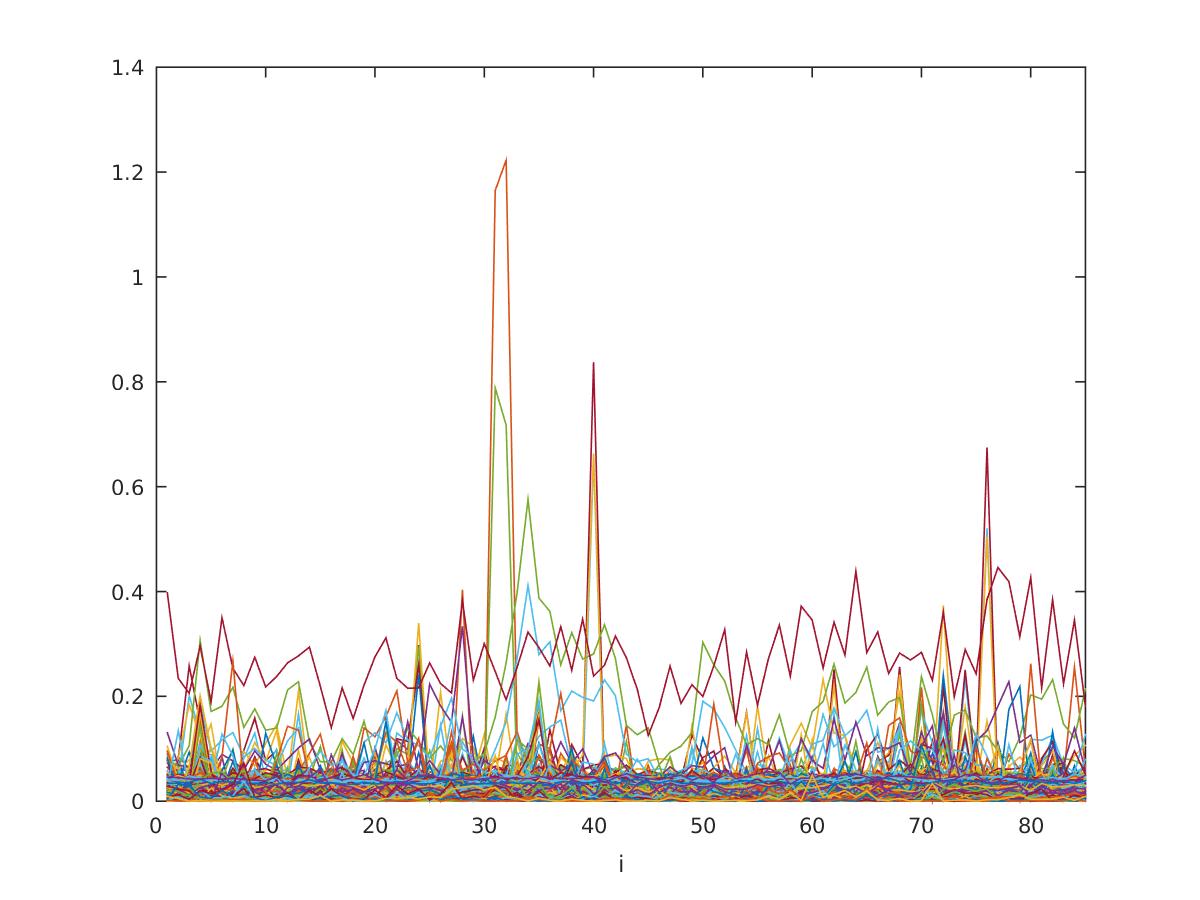}(e)
\includegraphics[width=18pc,height=3pc]{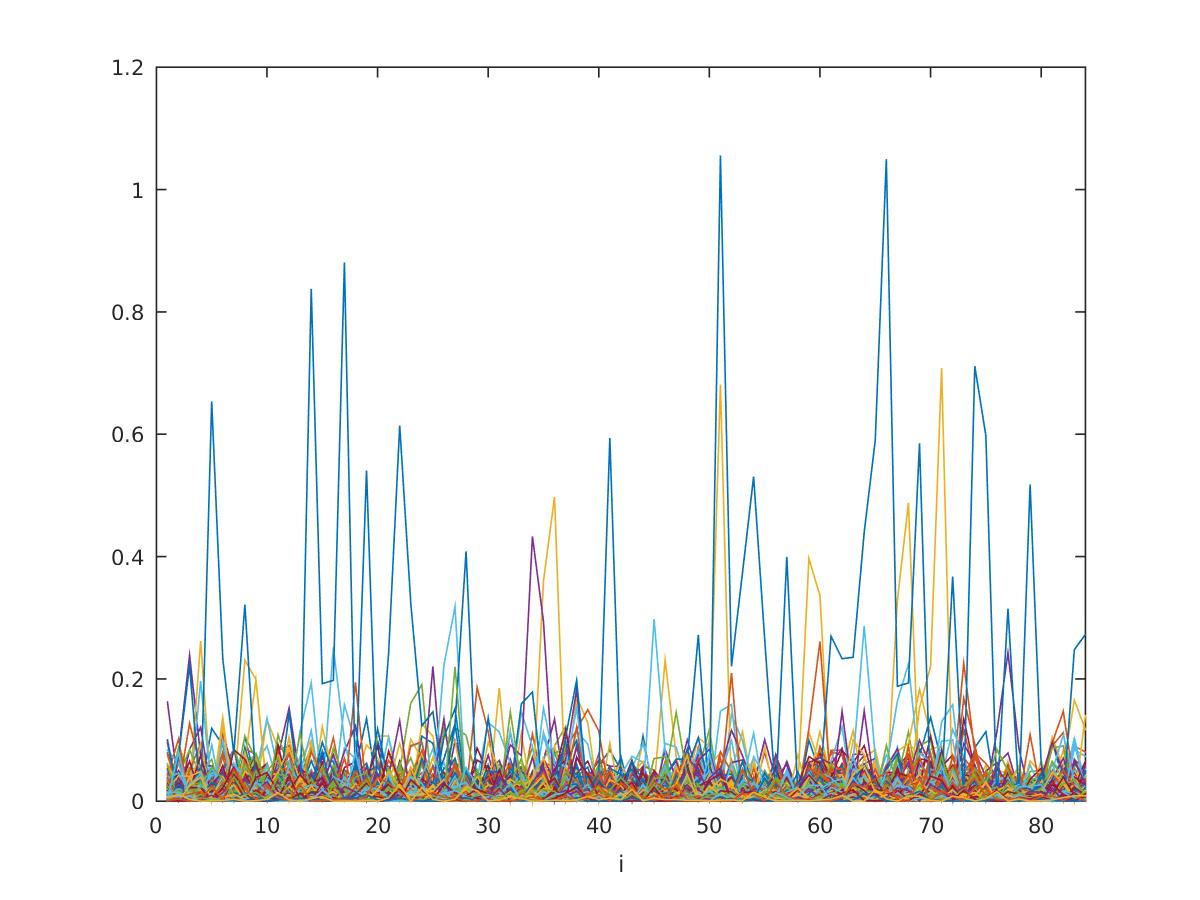}(f)

\includegraphics[width=18pc,height=3pc]{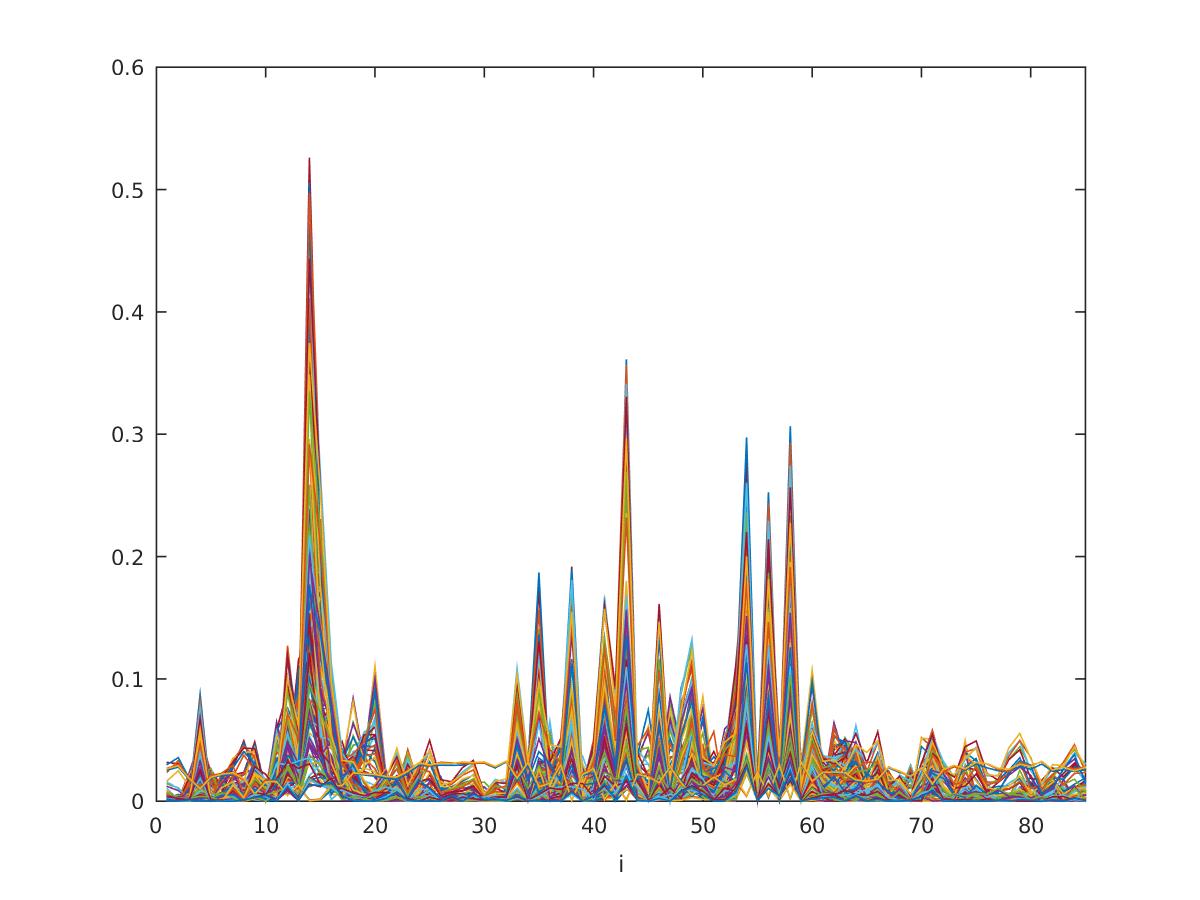}(g)
\includegraphics[width=18pc,height=3pc]{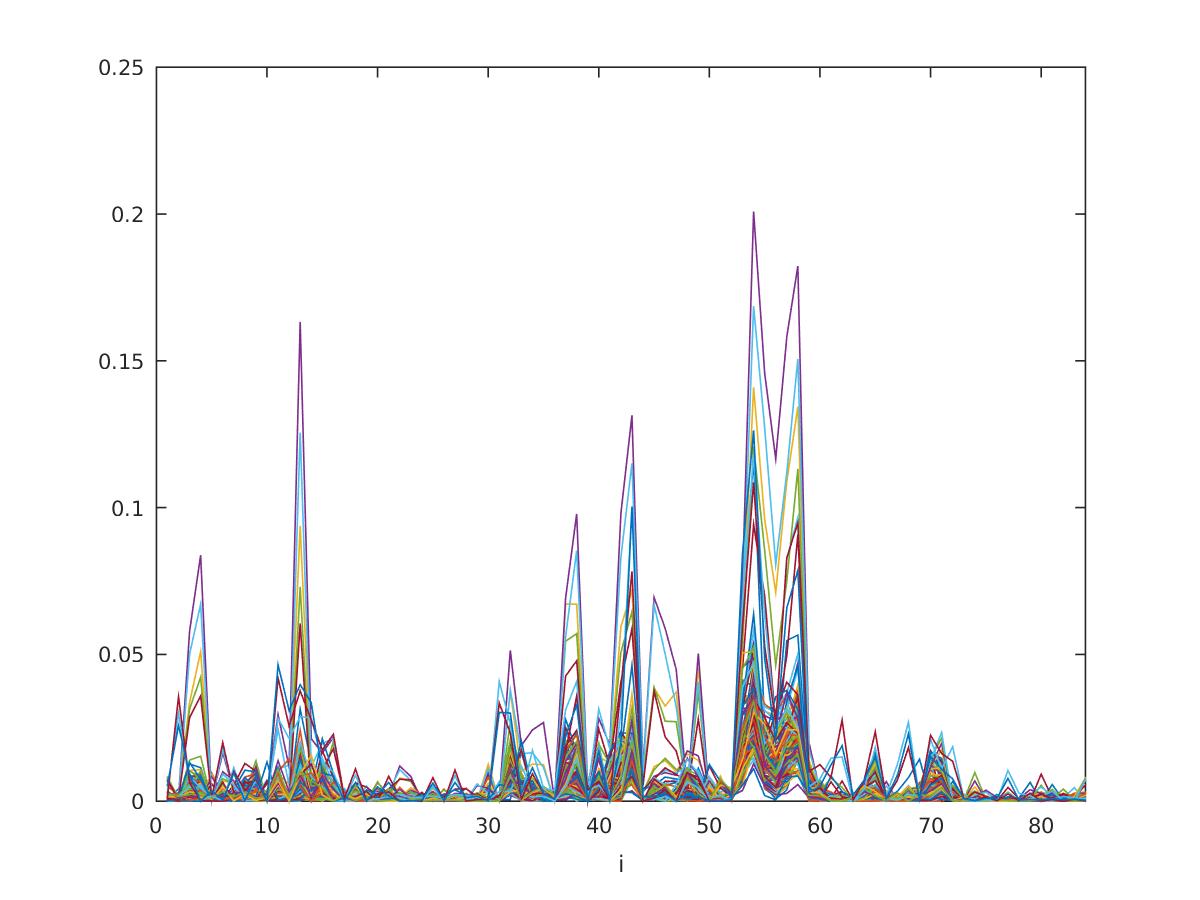}(h)

\includegraphics[width=18pc,height=3pc]{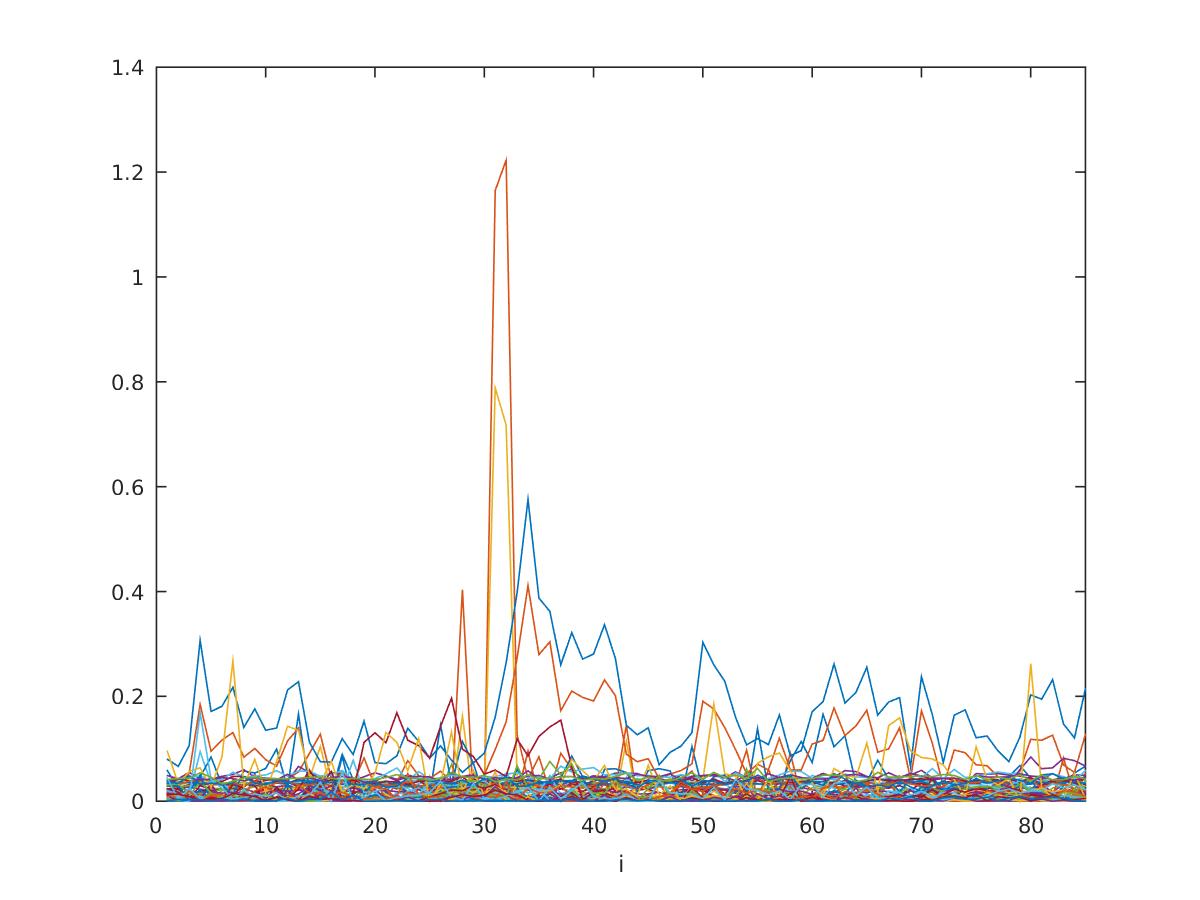}(i)
\includegraphics[width=18pc,height=3pc]{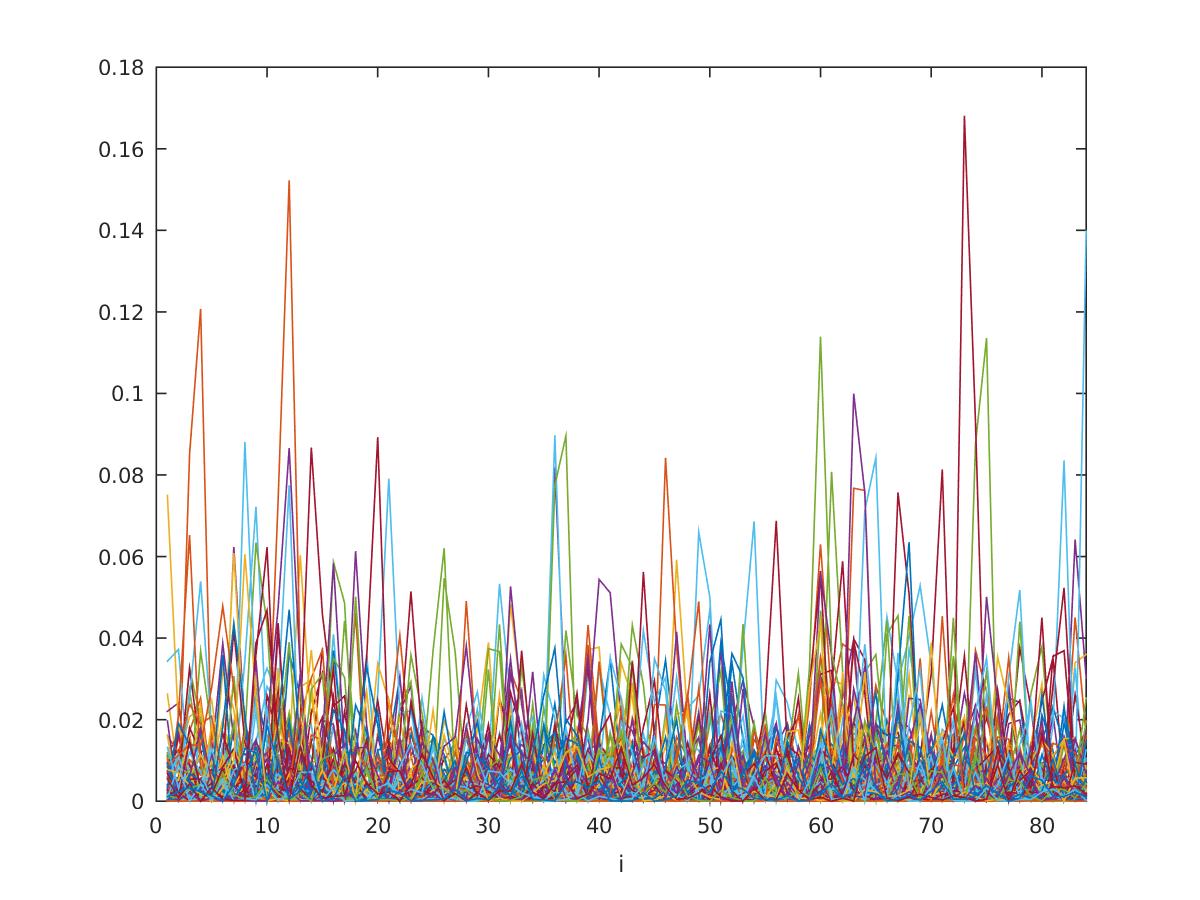}(j)

\caption{{\sc VH Polarization Channel} Level $J=3$ series of squared mean deviations: (a) $\vd(m)$; (b) $\vt(m)$. Red horizontal line represents the median value., Yellow, two absolute median deviations beyond the median. Squared approximation coefficient deviations: 0.1\% highest absolute correlations - (c) $\vd(m)$;  (d) $\vt(m)$; 0.1\% smallest absolute correlations - (e) $\vd(m)$; (f) $\vt(m)$; 0.01\% highest absolute correlations - (g) $\vd(m)$;  (h) $\vt(m)$;  0.01\% smallest absolute correlations - (i) $\vd(m)$; (j)  $\vt(m)$. 
}  
\label{F:squared_meandev_J3_VH}
\end{figure}

\begin{figure}[htp!]
\noindent\includegraphics[width=18pc,height=3pc]{J3_euclid_squared_meandev}(a)
\includegraphics[width=18pc,height=3pc]{consecdif_J3_euclid_squared_meandev}(b)

\includegraphics[width=18pc,height=3pc]{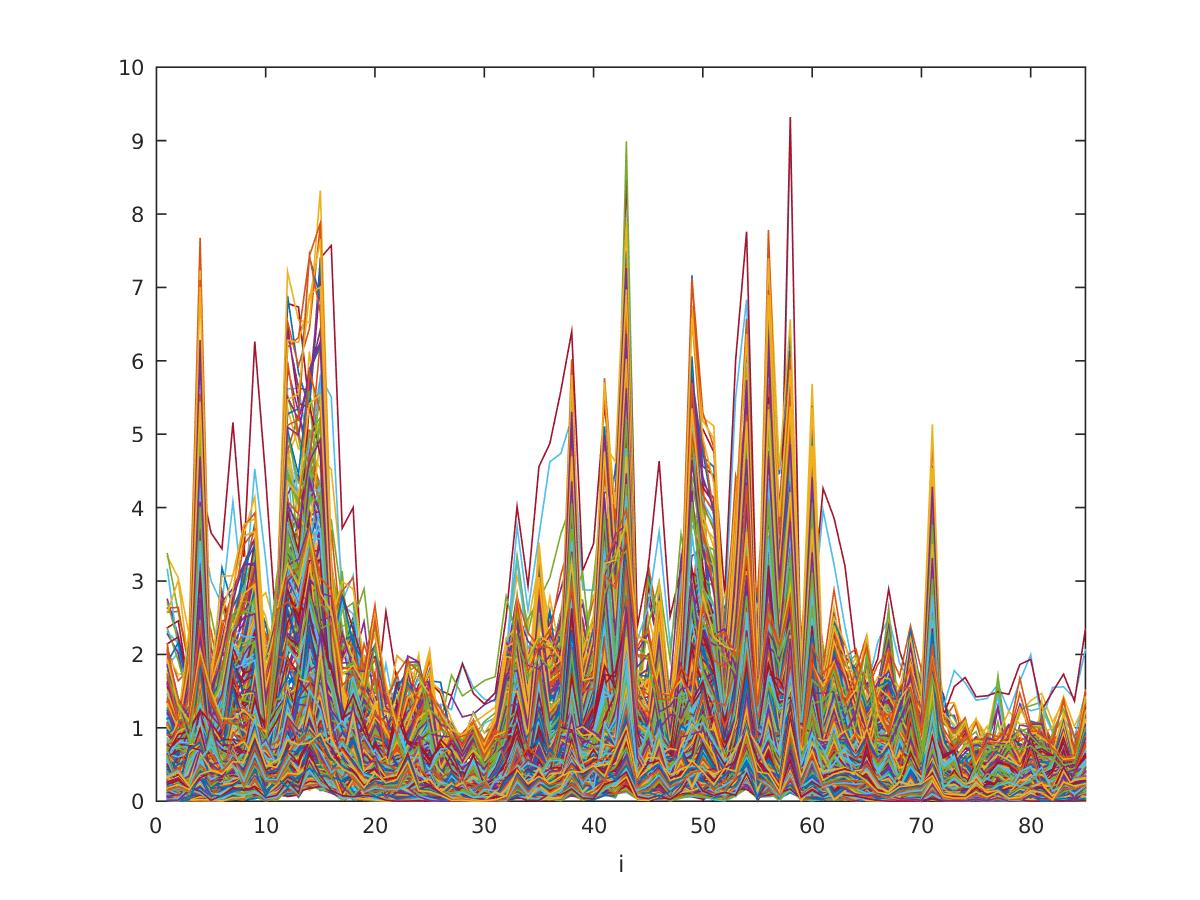}(c)
\includegraphics[width=18pc,height=3pc]{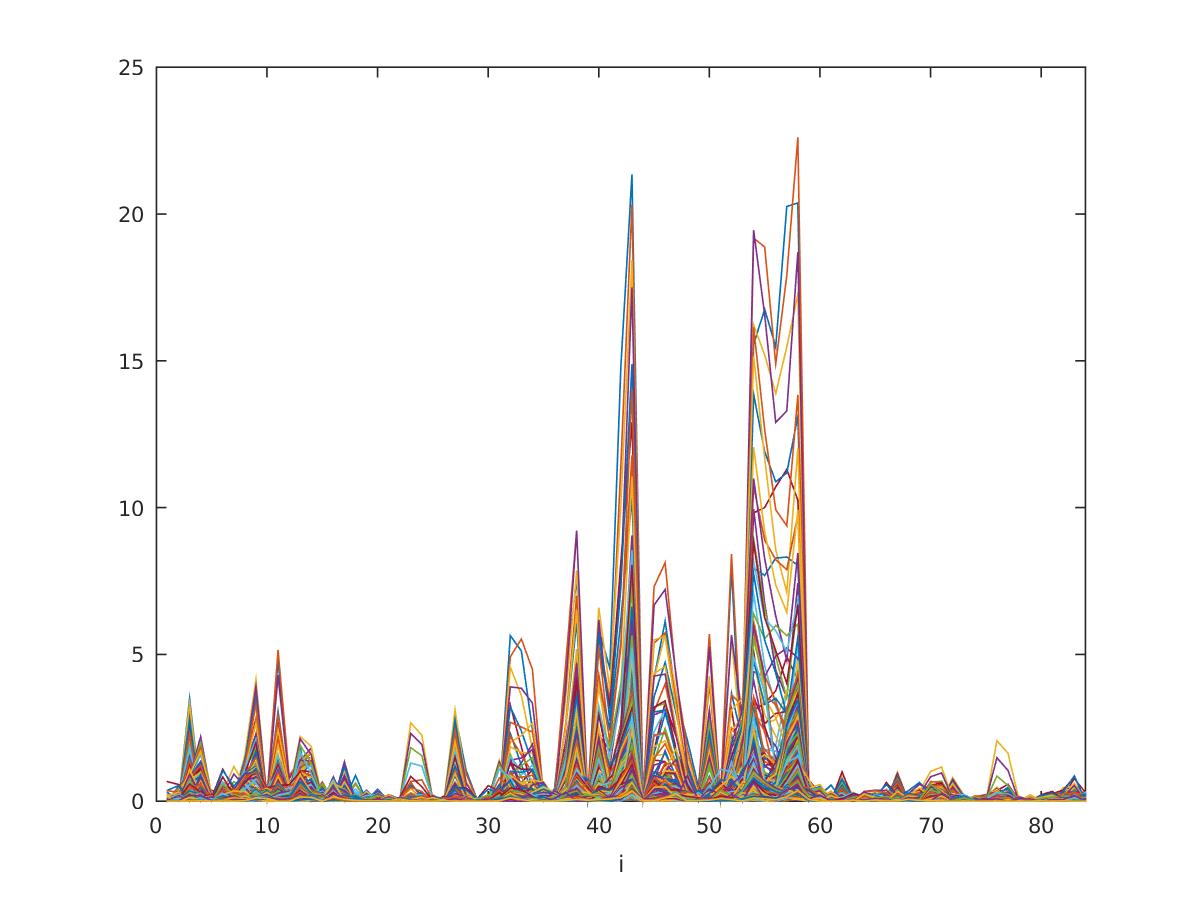}(d)

\includegraphics[width=18pc,height=3pc]{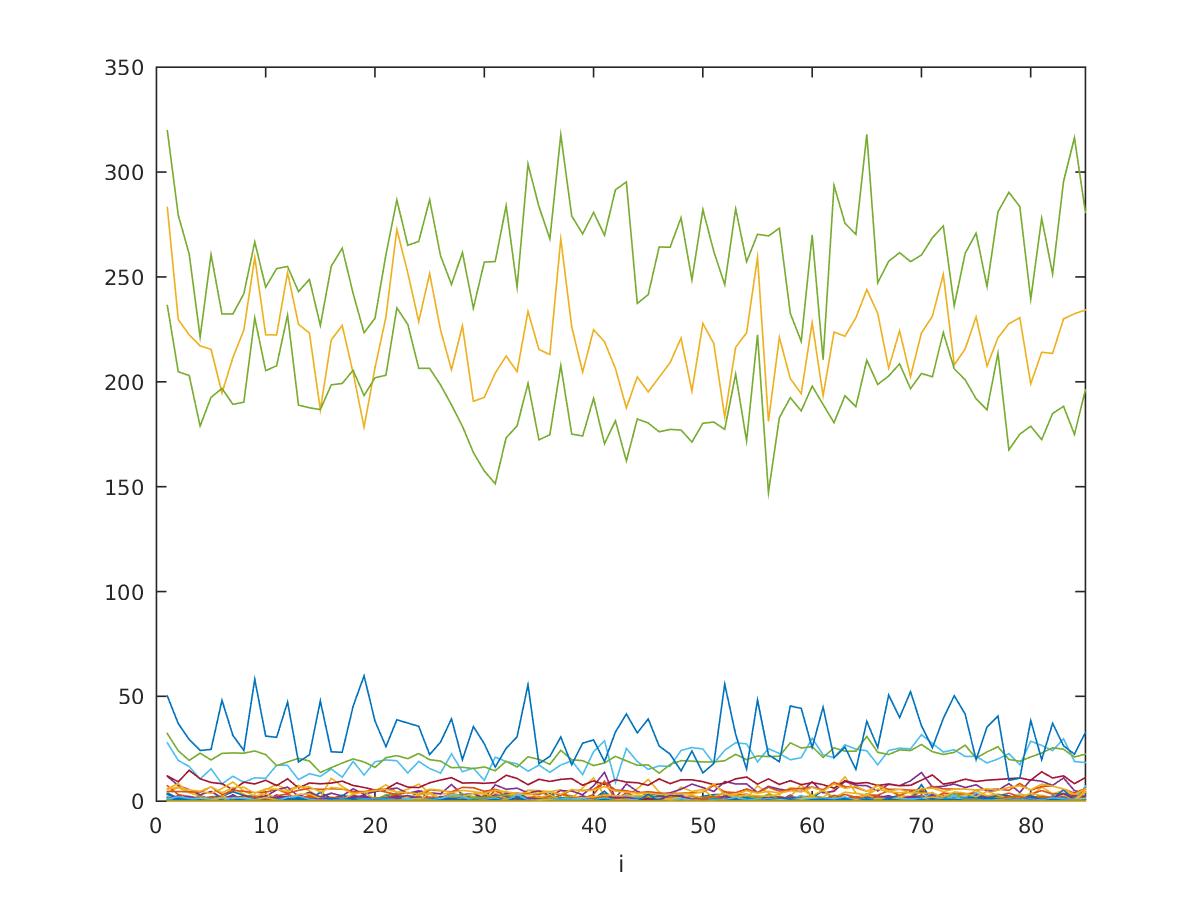}(e)
\includegraphics[width=18pc,height=3pc]{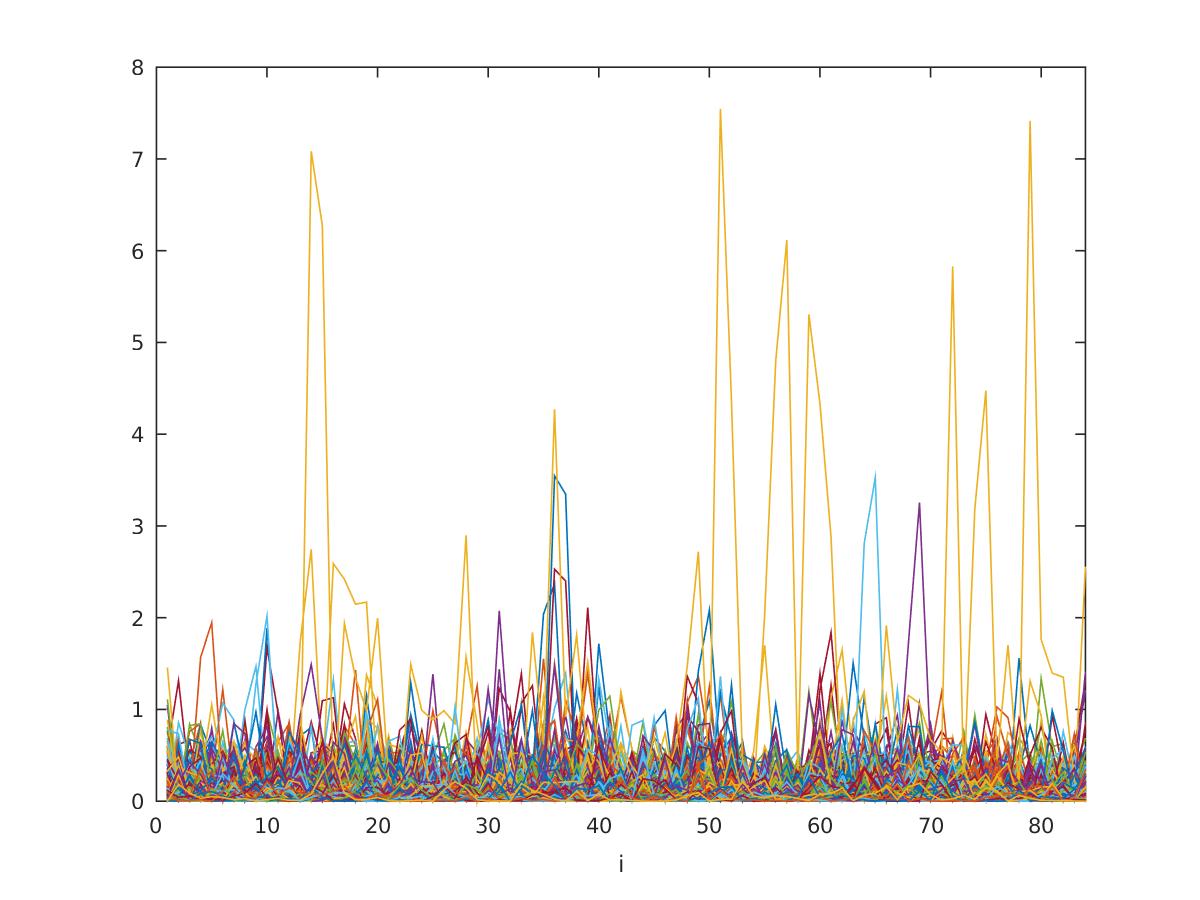}(f)

\includegraphics[width=18pc,height=3pc]{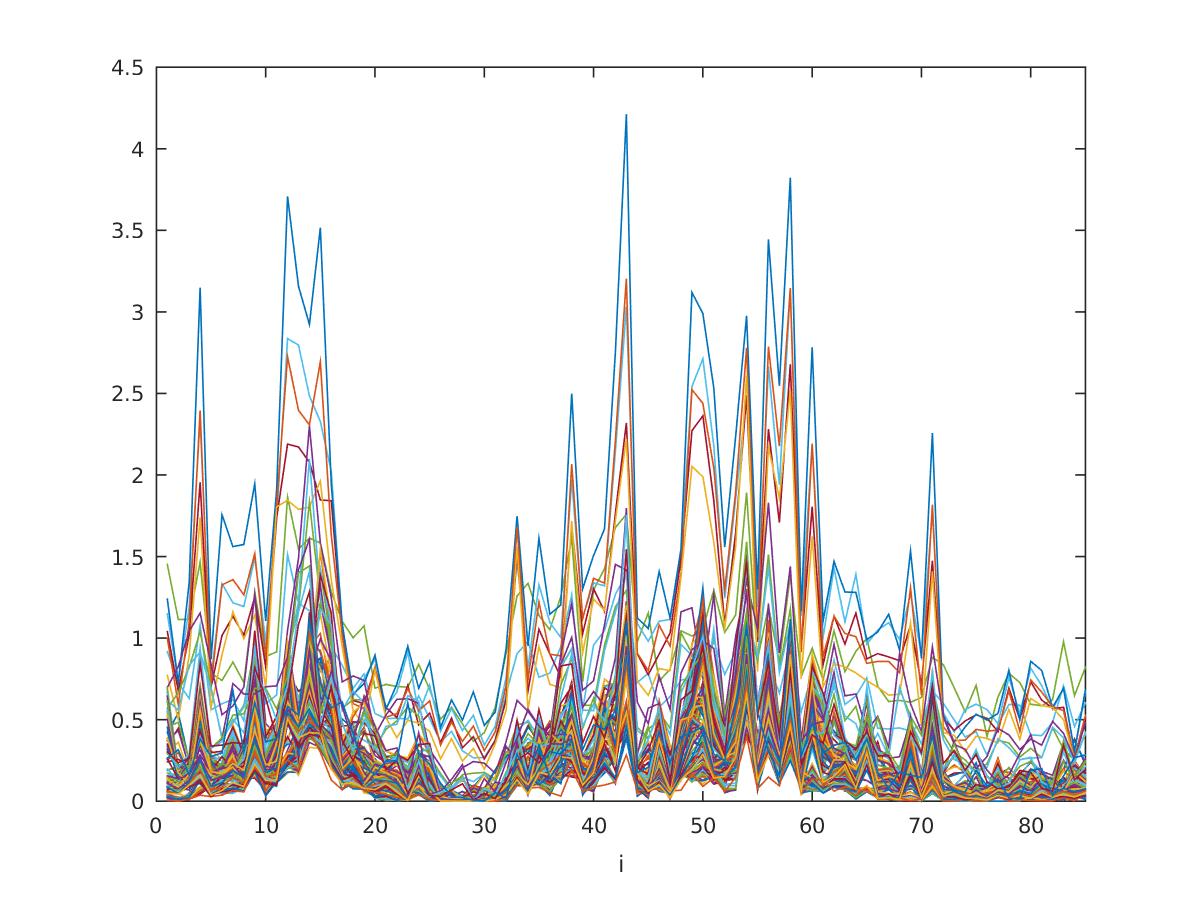}(g)
\includegraphics[width=18pc,height=3pc]{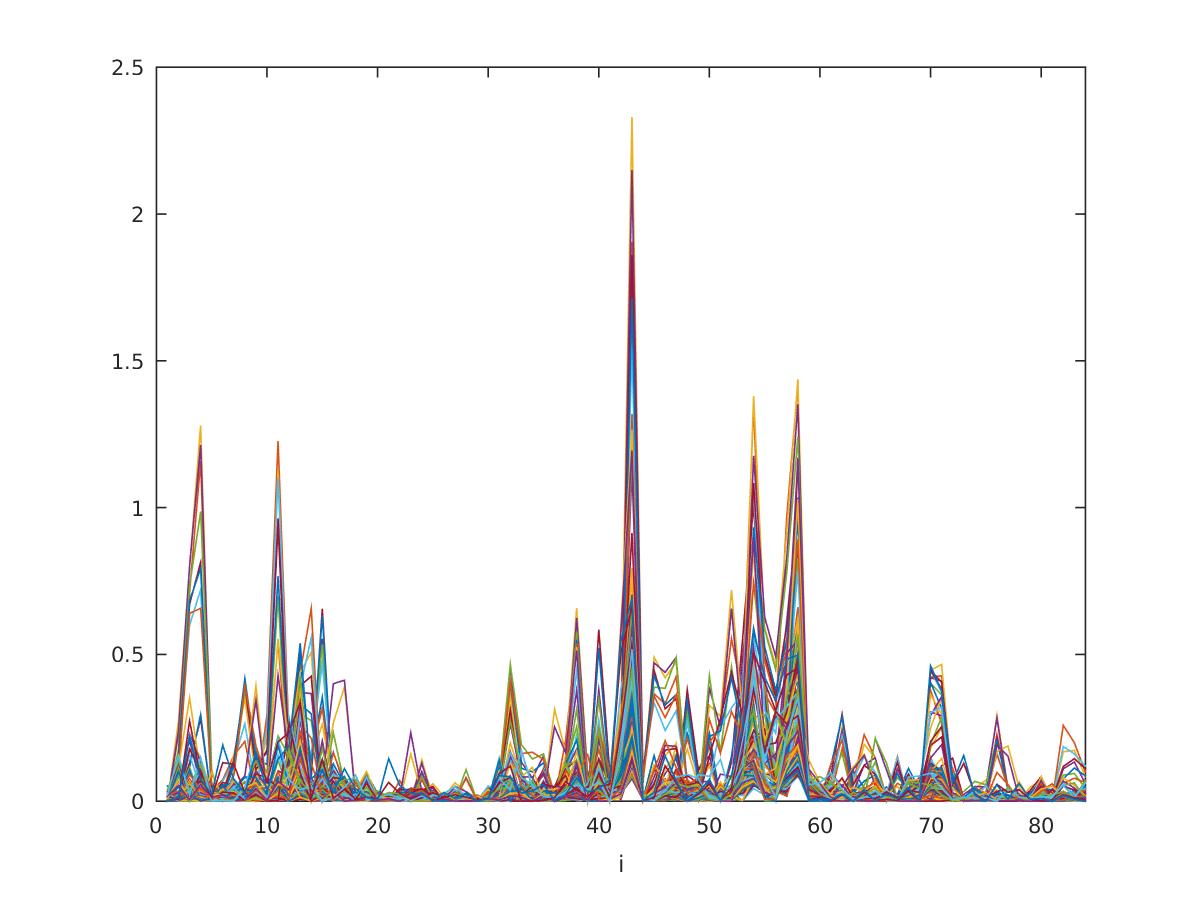}(h)

\includegraphics[width=18pc,height=4pc]{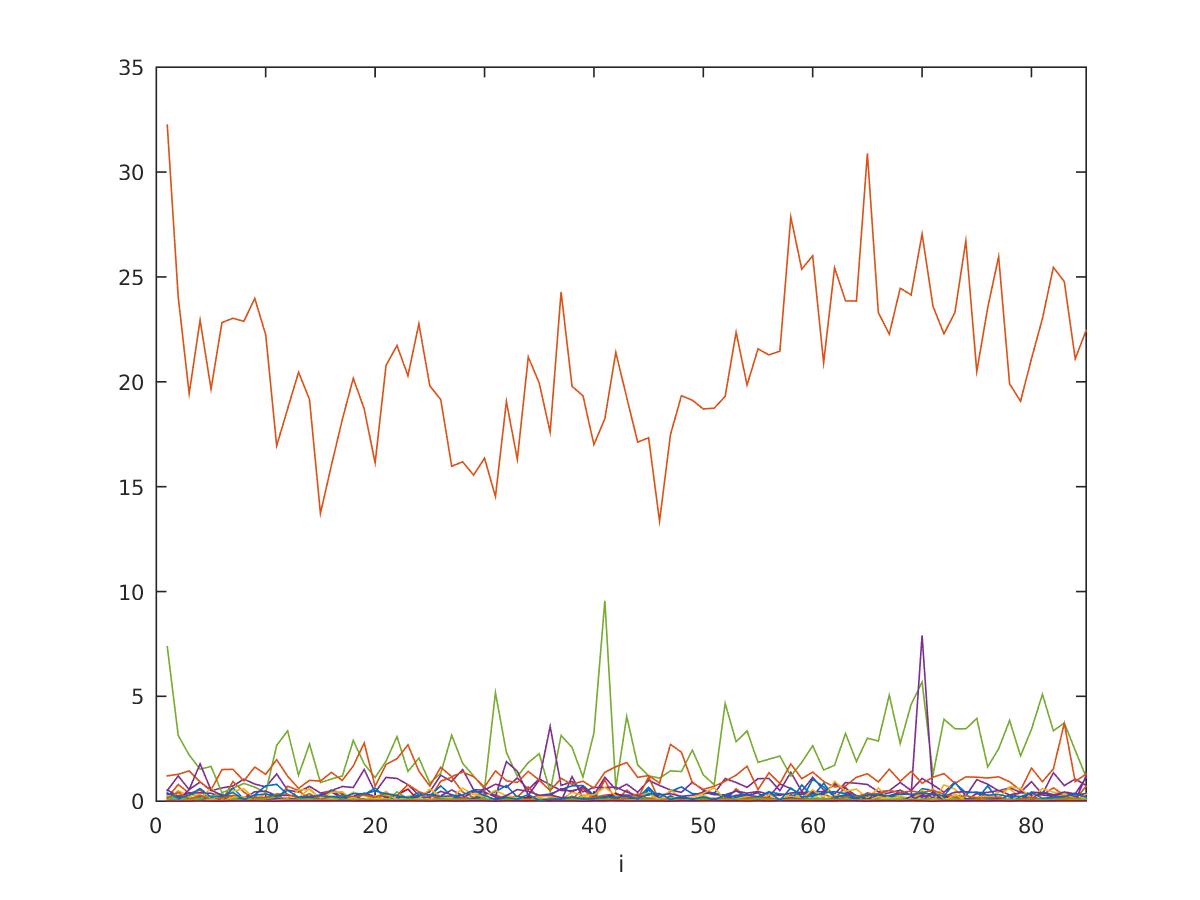}(i)
\includegraphics[width=18pc,height=4pc]{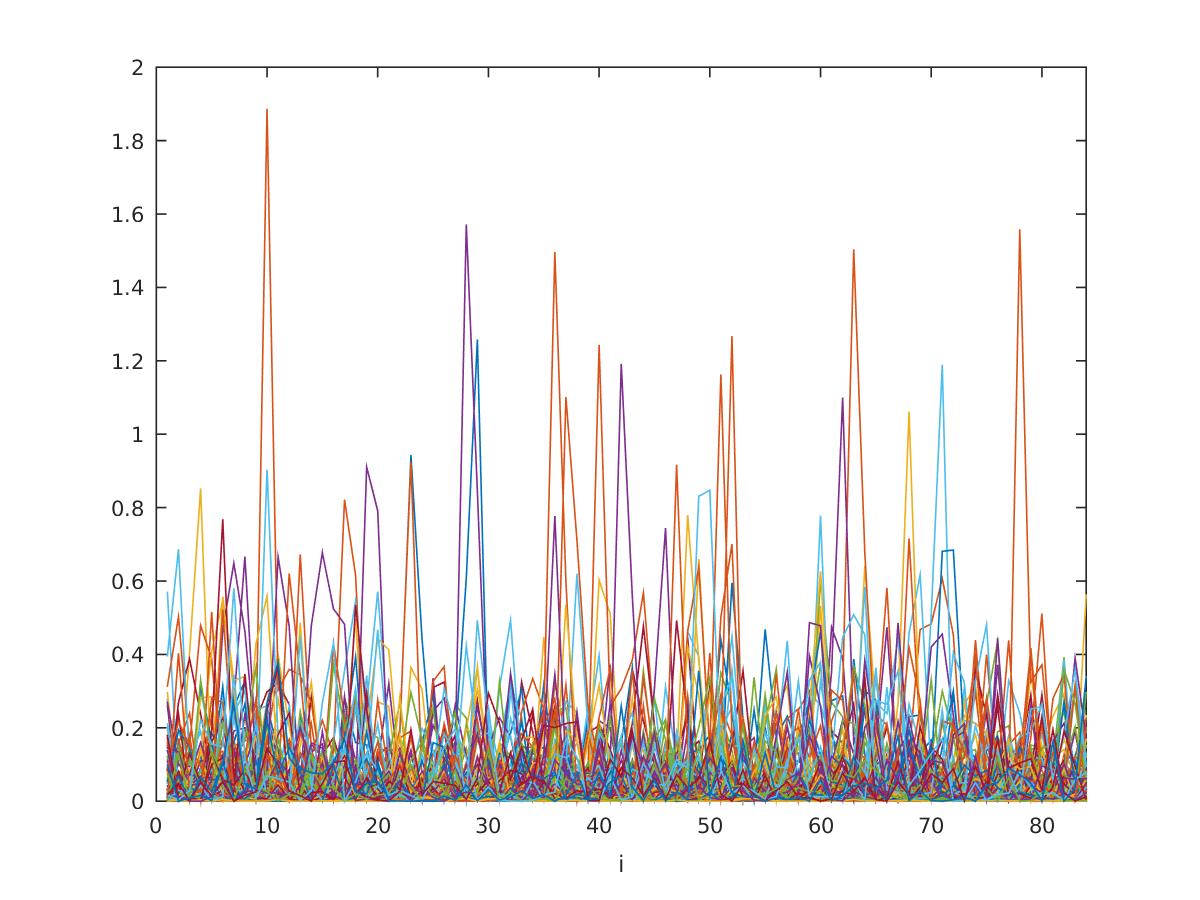}(j)

\caption{{\sc Combined Channels} Level $J=3$ series of squared mean deviations: (a) $\vd(m)$; (b) $\vt(m)$. Red horizontal line represents the median value., Yellow, two absolute median deviations beyond the median. Squared approximation coefficient deviations: 0.1\% highest absolute correlations - (c) $\vd(m)$;  (d) $\vt(m)$; 0.1\% smallest absolute correlations - (e) $\vd(m)$; (f) $\vt(m)$; 0.01\% highest absolute correlations - (g) $\vd(m)$;  (h) $\vt(m)$;  0.01\% smallest absolute correlations - (i) $\vd(m)$; (j)  $\vt(m)$. 
} 
\label{F:squared_meandev_J3_euclid}
\end{figure}

Figures \ref{F:squared_meandev_J3_VV}-\ref{F:squared_meandev_J3_euclid} present a timewise comparison between overall energy variations, $d(m)$ and $t(m)$ and their respective individual coefficients. Each figure has twelve panels. Panels on the left and right deal with the first and second change detection methods respectively. A full description is give in each figure's caption. The overall conclusion from these figures is that as expected there are temporal variations between the methods regarding change detection (Panels (a)-(b)). The correlation screening detects the most relevant indices, as well as the least relevant ones  (Panels (c)-(f)). 
These conclusions hold for analyses based upon VV, VH or combined polarizations, but the VV polarization signal is much stronger than VH's. A slight advantage is perceived for the $\vd(m)$ method as opposed to $\vt(m)$'s. This makes sense, since we are dealing here with data from a forest region over a long time, and seasonal changes will be perceived more easily on the first proposed method.
\FloatBarrier

\begin{table}[ht!]
\caption{Absolute correlation thresholds and number of selected coefficients for $n=85$ multi-temporal images of $1200\times 1000$ . Correlation was computed between approximation coefficients and approximation total energy at level $J=2$ for each image. }\label{tabela_quantis}
\begin{tabular}{c|cccr||c|cccr}
\hline
Qtile &\multicolumn{4}{c||}{\sc Correlation Thresh} &Qtile & \multicolumn{4}{c}{\sc Correlation Thresh}  \\
Level &VV&VH&Comb& Coeffs &Level &VV&VH&Comb& Coeffs \\
\hline
0.50&0.281&0.199&0.292&600000&0.99&0.647&0.582&0.668&12000\\
0.55&0.300&0.218&0.311&540000&0.991&0.651&0.587&0.673&10800\\
0.60&0.319&0.239&0.331&480000& 0.992&0.656&0.594&0.678&9600\\
0.65&0.339&0.260&0.352&420000& 0.993&0.662&0.601&0.684&8400\\
0.70&0.361&0.282&0.375&360000& 0.994&0.669&0.608&0.690&7200\\
0.75&0.385&0.307&0.399&300000& 0.995&0.676&0.617&0.697&6000\\
0.80&0.412&0.335&0.428&240000& 0.996&0.685&0.626&0.705&4800\\
0.85&0.445&0.368&0.462&180000& 0.997&0.696&0.639&0.716&3600\\
0.90&0.487&0.410&0.506&120000& 0.998&0.710&0.654&0.729&2400\\
0.95&0.549&0.472&0.570&60000& 0.999&0.731&0.676&0.748&1200\\
\hline
\end{tabular}
\end{table}

\FloatBarrier

\begin{figure}[htp!]
\noindent\includegraphics[width=11pc]{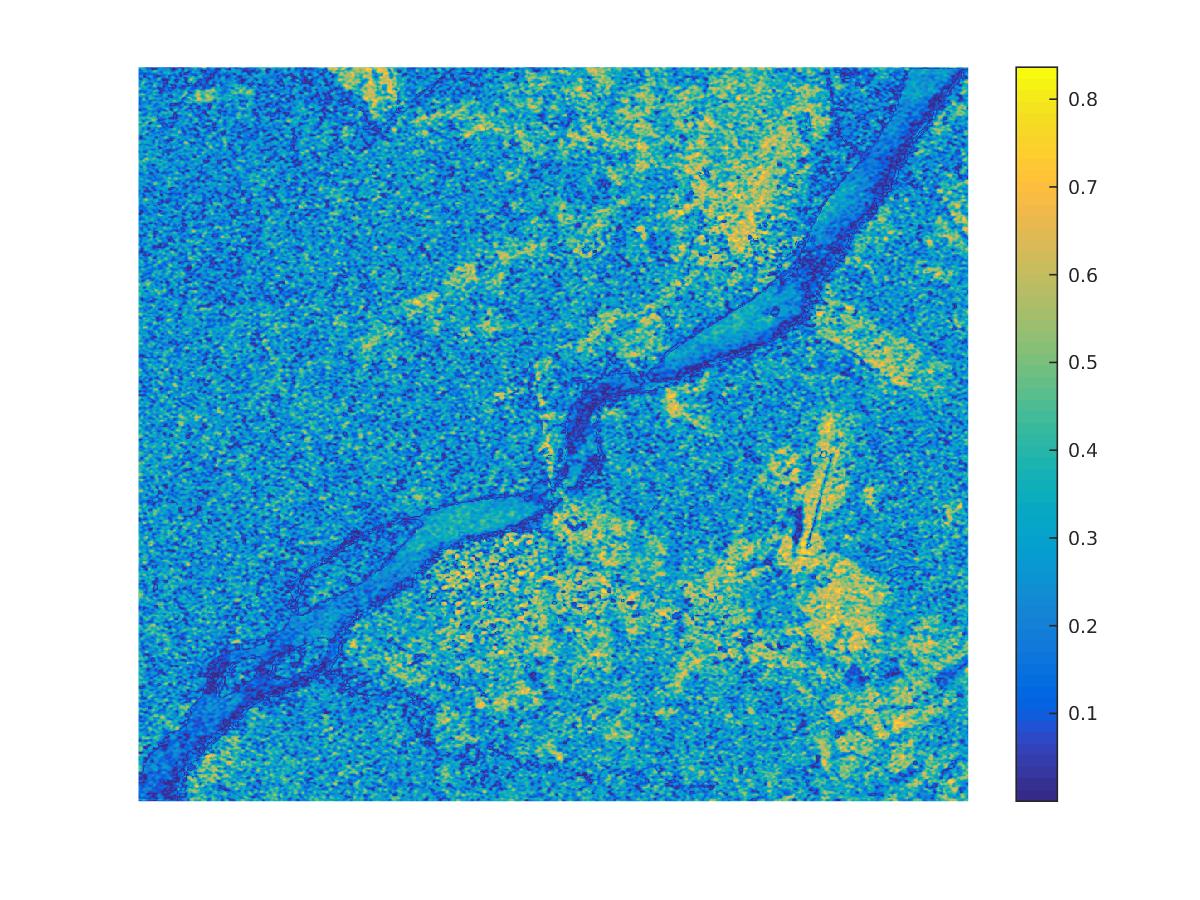}(a)
\noindent\includegraphics[width=11pc]{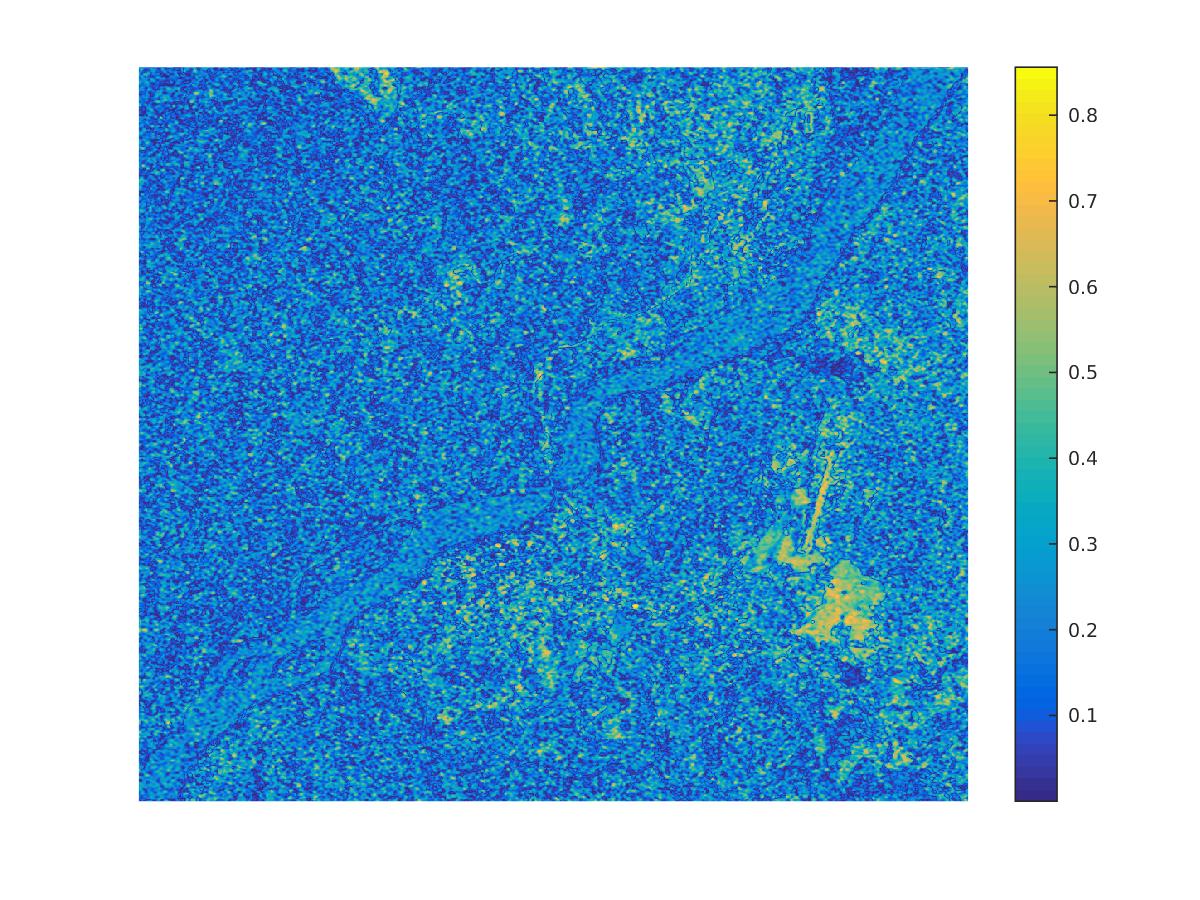}(b)
\noindent\includegraphics[width=11pc]{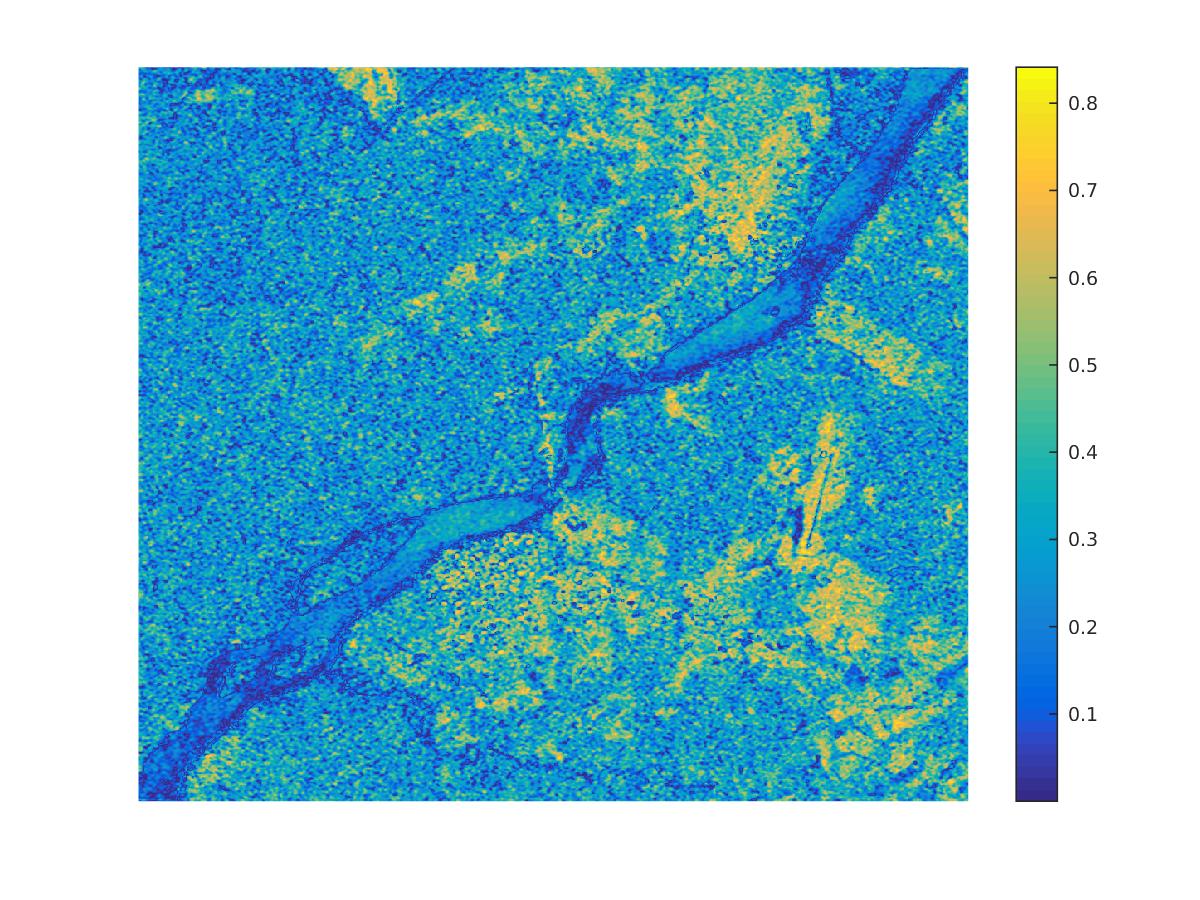}(c)

\includegraphics[width=11pc]{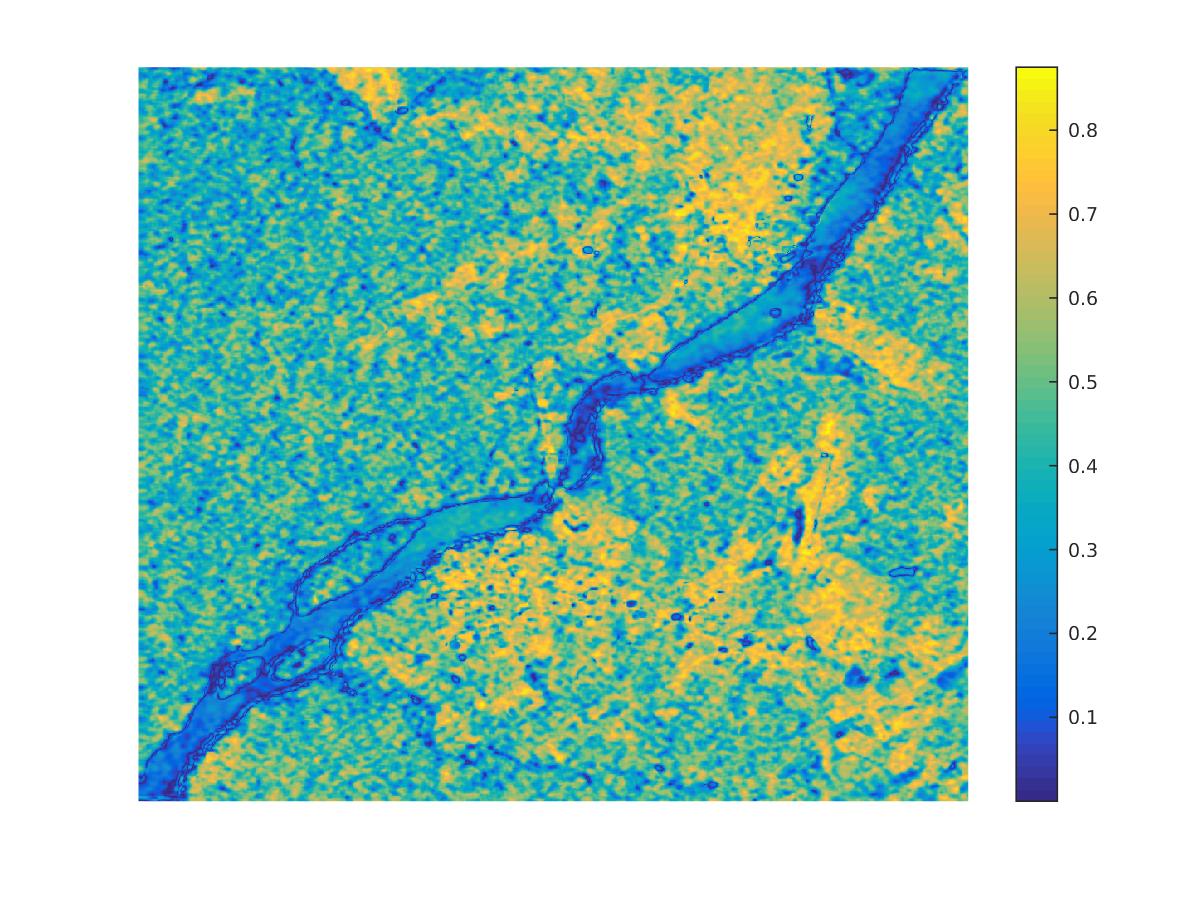}(d) 
\includegraphics[width=11pc]{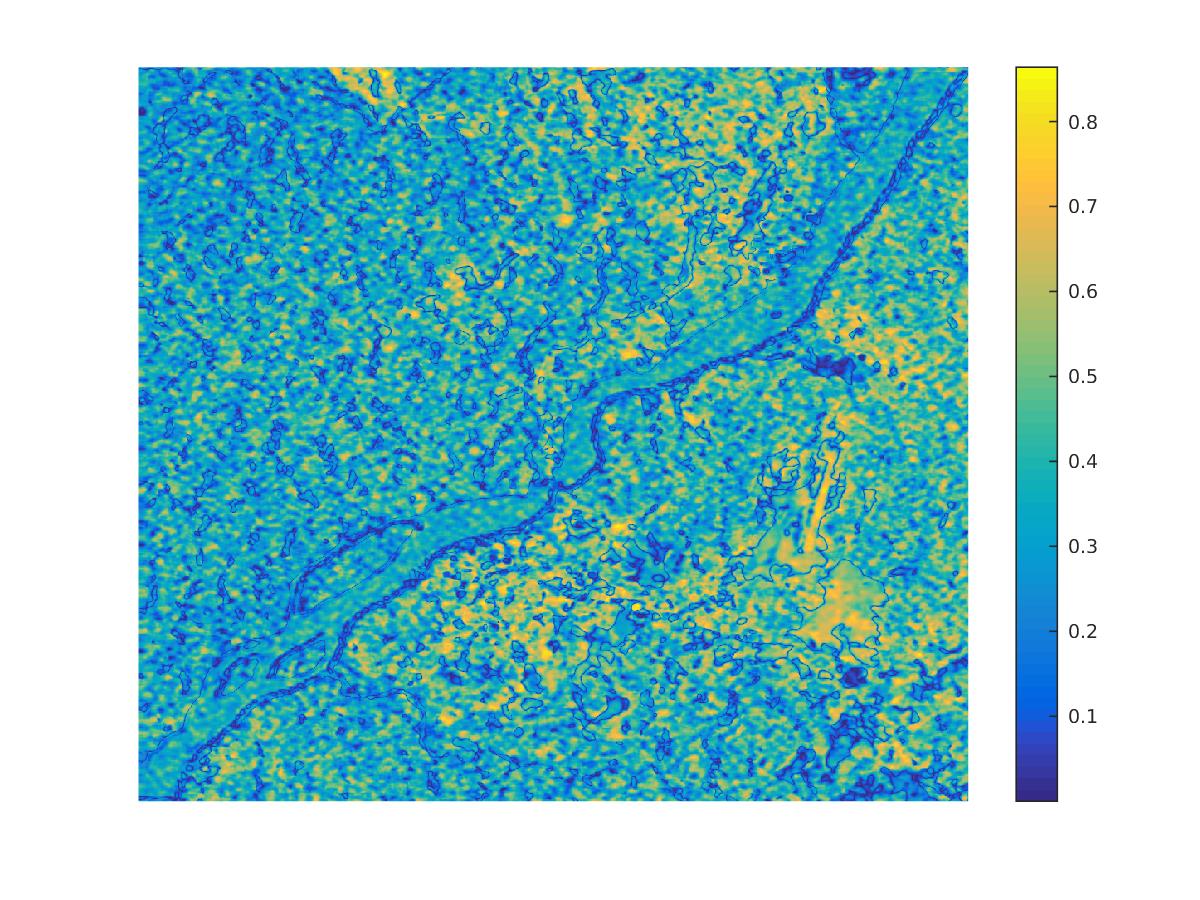}(e)
\includegraphics[width=11pc]{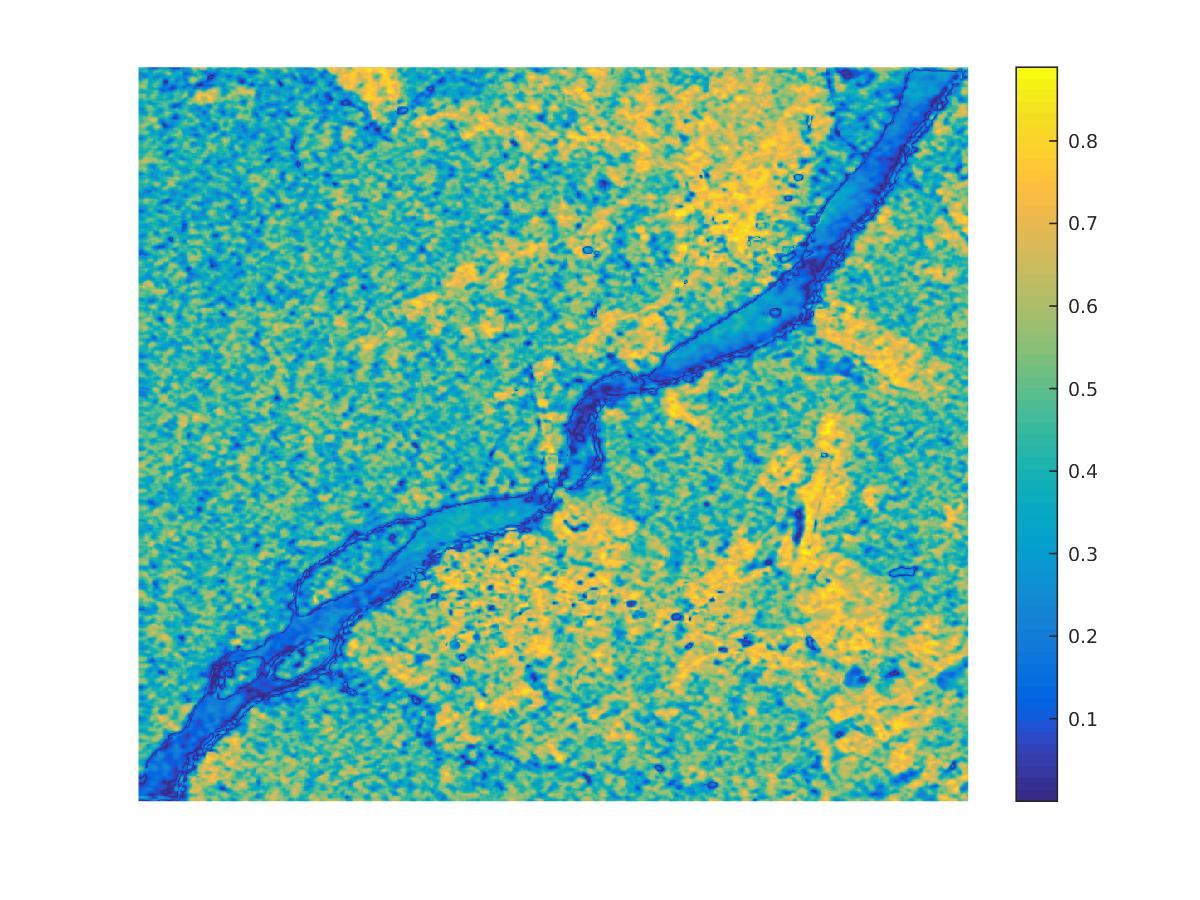}(f)

\caption{ Absolute Correlation matrices $|\vR|$ between mean-corrected squared coeffcients at approximation levels and overall mean-corrected total energy.  $J=2$ (a)-(c);  $J=3$ (d)-(f). $n=85$ multi-temporal images of $1200\times 1000$ ($J=10$). The color bar on the right gives the magnitude of correlation at all positions. (a) {\sc VV  Polarization - Channel} $J=2$. (b) {\sc VH  Polarization - Channel} $J=2$. (c) {\sc Combined Channels} $J=2$.  (d)  {\sc VV  Polarization - Channel} $J=3$. (e) {\sc VH  Polarization - Channel} $J=3$. (f) {\sc Combined Channels} $J=3$} 
\label{F:image_corr_J2J3}
\end{figure}

\begin{figure}[htp!]
\noindent\includegraphics[width=11pc]{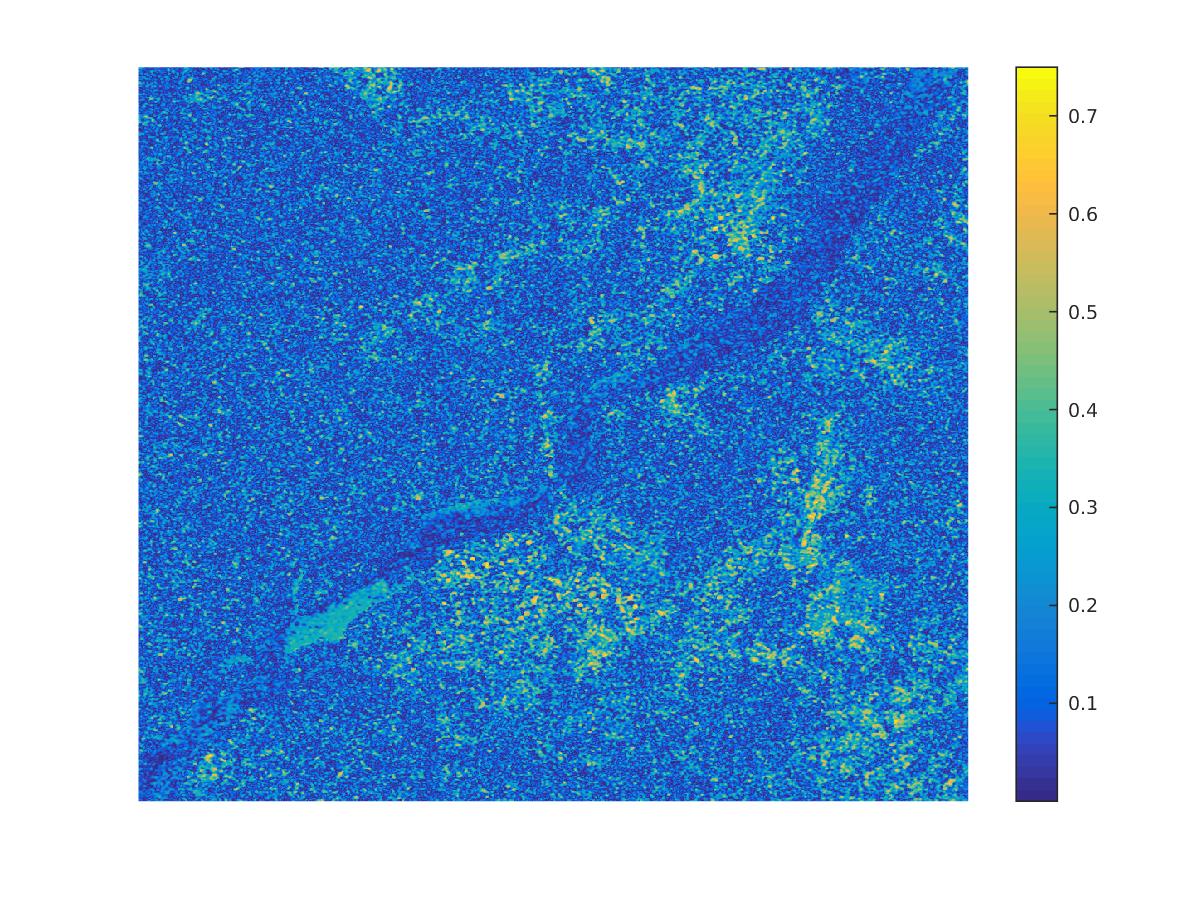}(a)
\noindent\includegraphics[width=11pc]{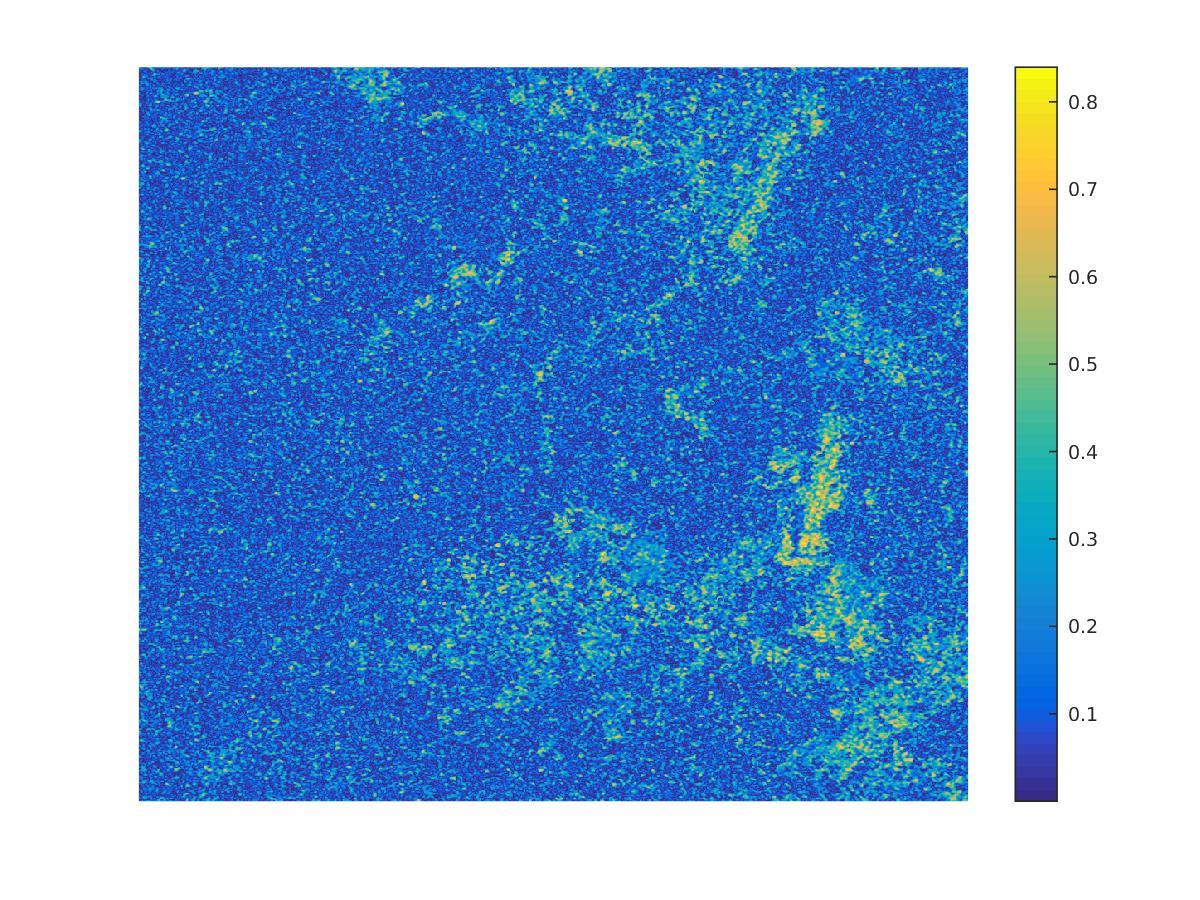}(b)
\noindent\includegraphics[width=11pc]{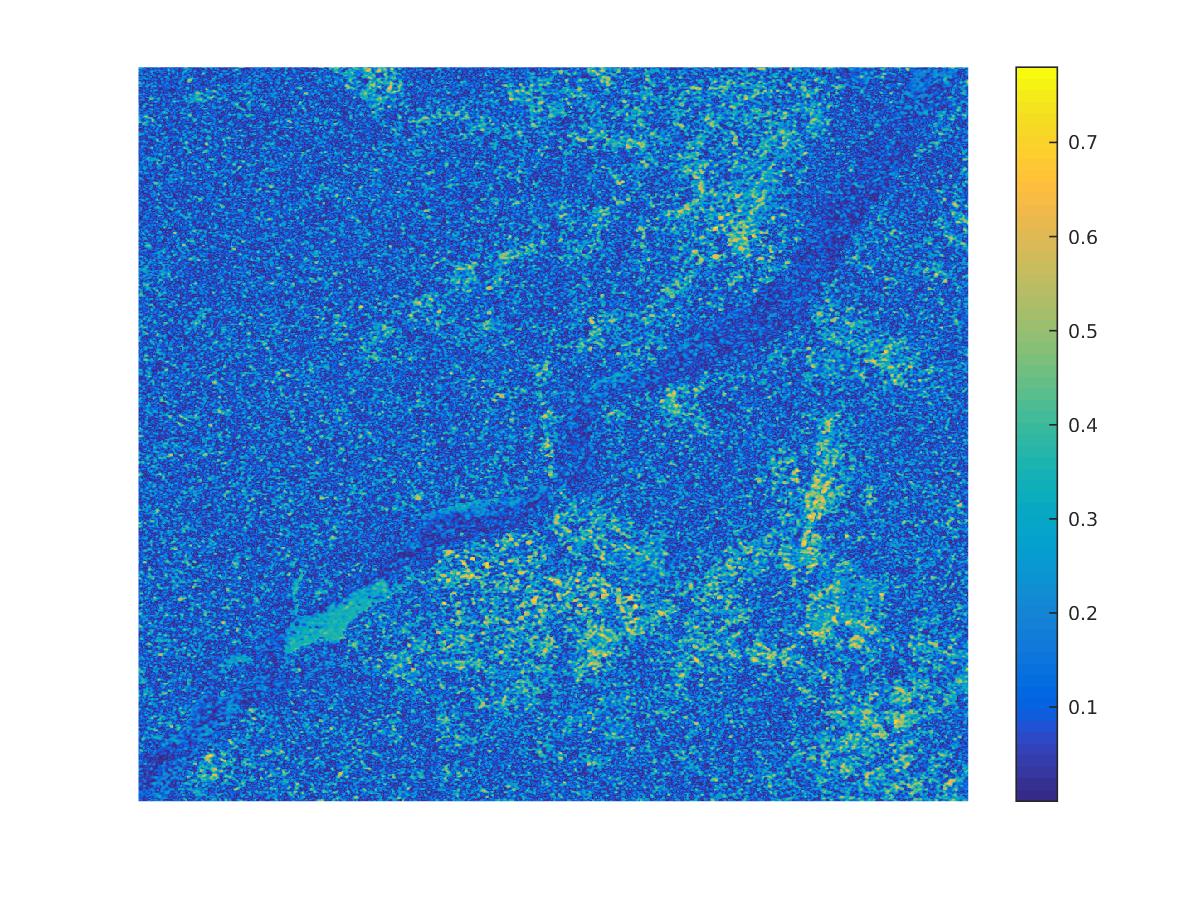}(c)

\includegraphics[width=11pc]{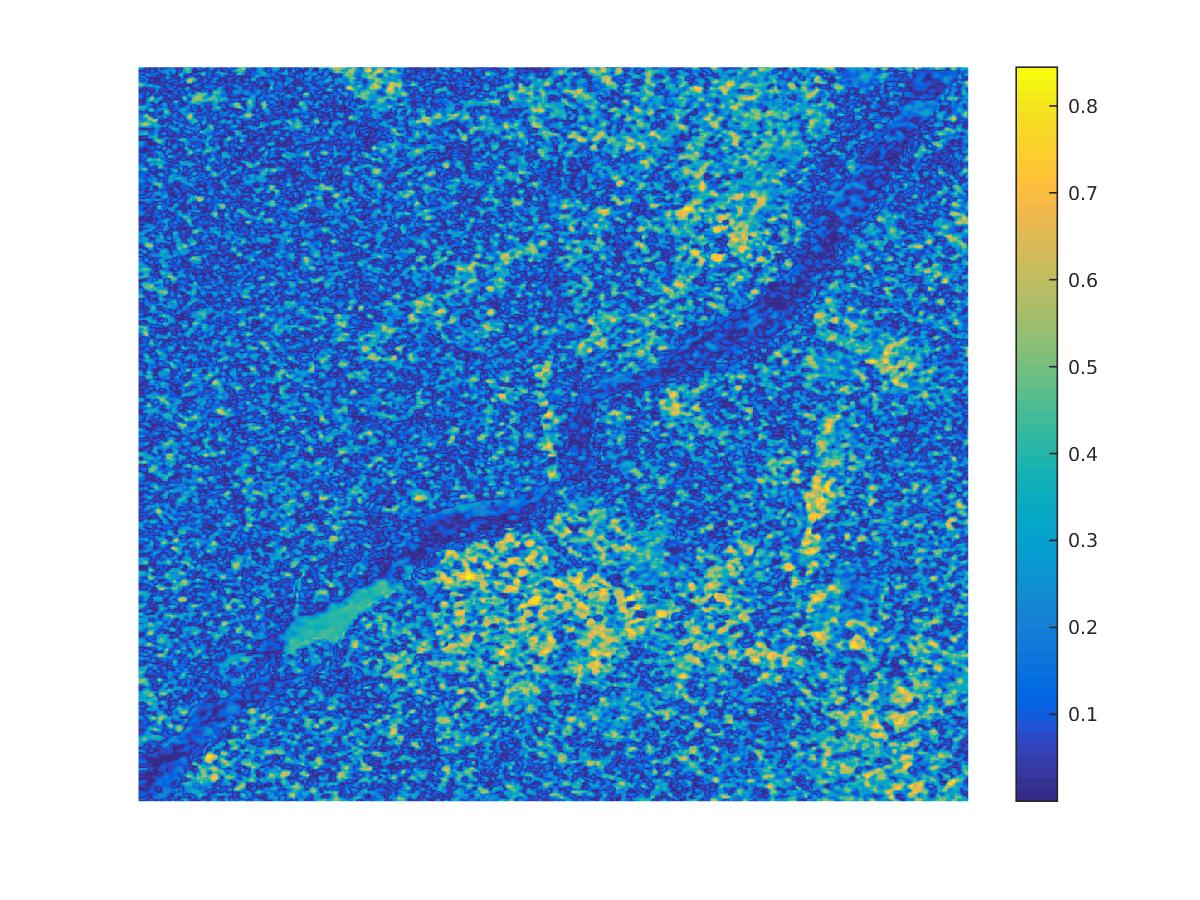}(d) 
\includegraphics[width=11pc]{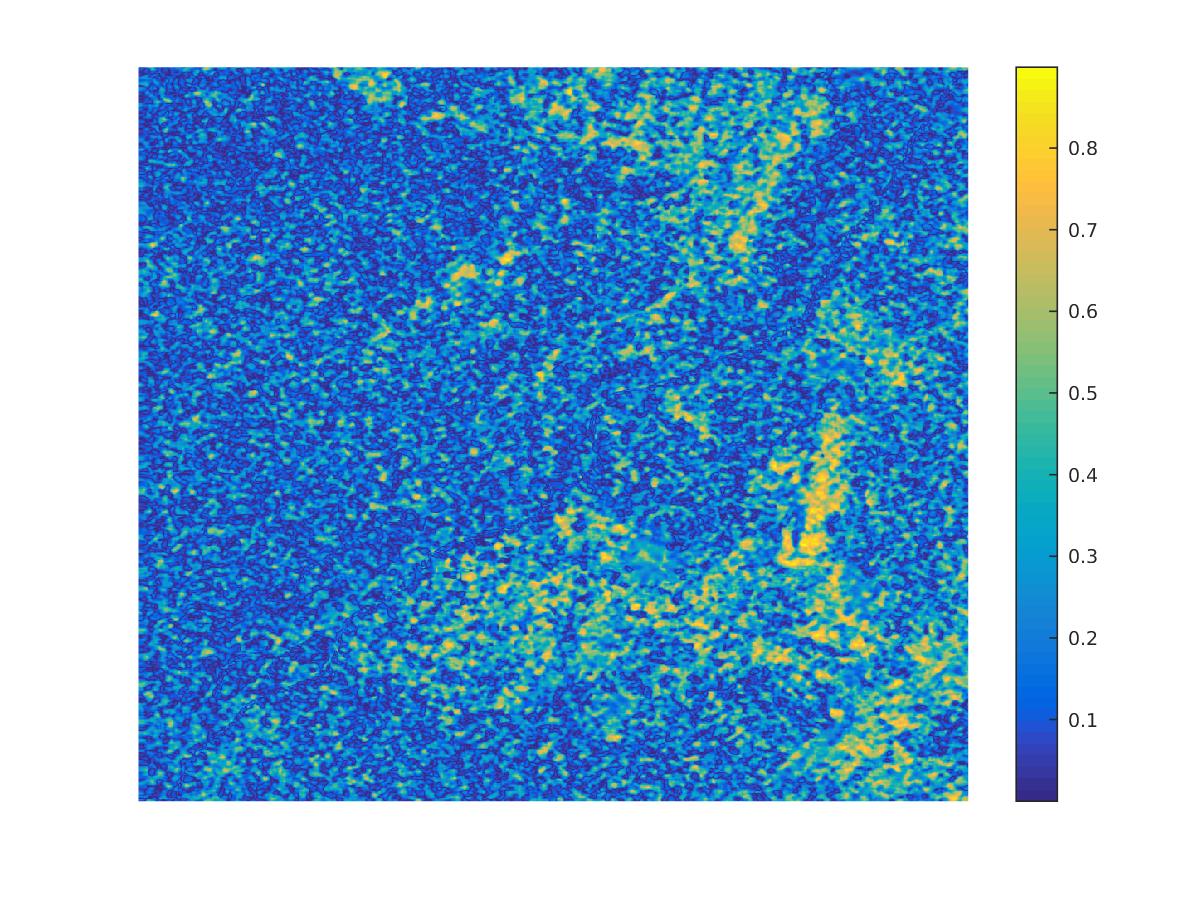}(e)
\includegraphics[width=11pc]{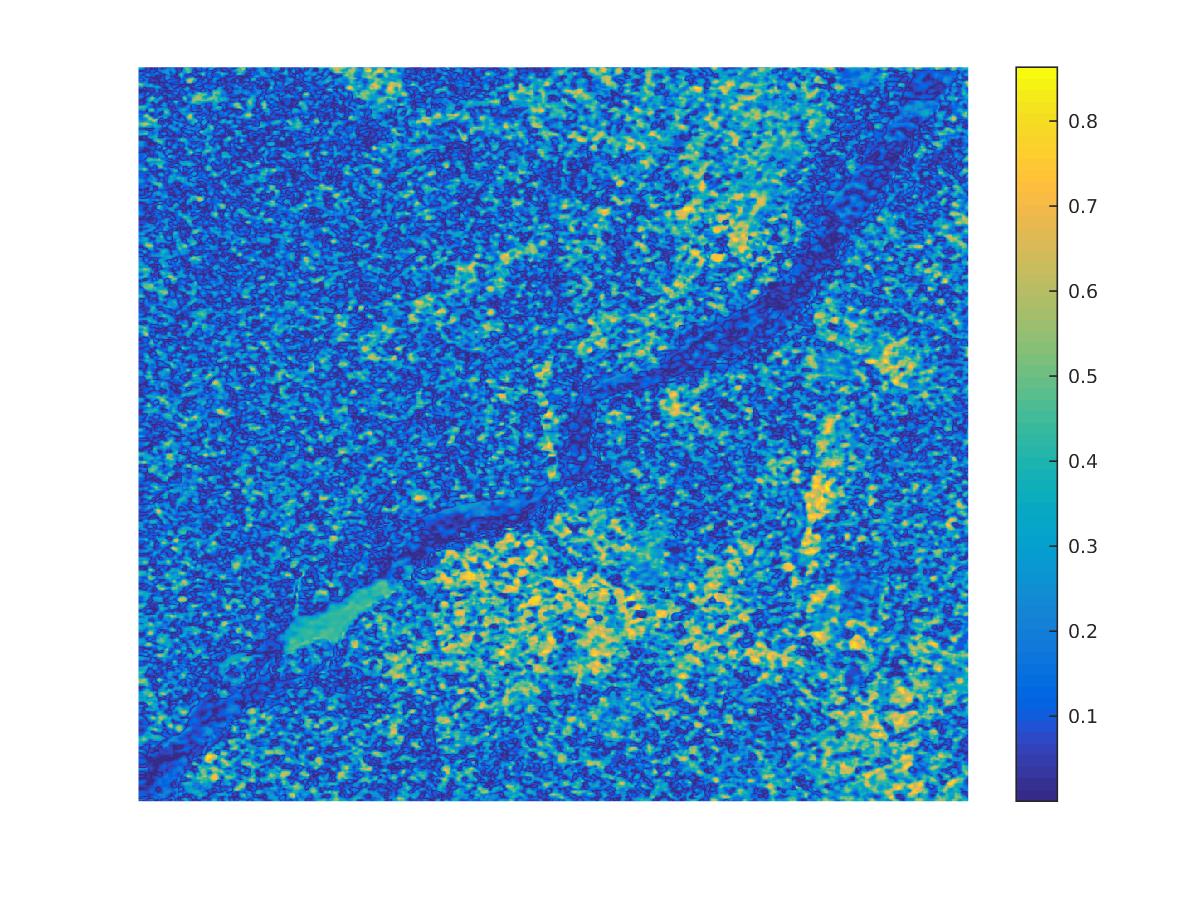}(f)

\caption{ Absolute Correlation matrices $|\vR|$ between consecutive log-images' squared coeffcients at approximation levels and overall mean-corrected total energy . Levels $J=2$ (a)-(c);  $J=3$ (d)-(f). $n=85$ multi-temporal images of $1200\times 1000$. The color bar on the right gives the magnitude of correlation at all positions. (a) {\sc VV  Polarization - Channel} $J=2$. (b) {\sc VH  Polarization - Channel} $J=2$. (c) {\sc Combined Channels} $J=2$. (d)  {\sc VV  Polarization - Channel} $J=3$. (e) {\sc VH  Polarization - Channel} $J=3$. (f) {\sc Combined Channels} $J=3$.} 
\label{F:image_corr_J2J3_consec}
\end{figure}

Figures \ref{F:image_corr_J2J3}-\ref{F:image_corr_J2J3_consec} shows the absolute correlation images for the VV, VH and combined channels for levels $J=2$ and $J=3$. We can notice that high correlation coefficients for $J=2$ are high correlation coefficients for $J=3$ as well, Moreover, there is a clear spatial connection between polarizations and between $d(m)$- and $t(m)$-based  analyses. On the other hand, correlations are more efficiently segregated when we move: from $J=2$ to $J=3$; from VH to VV polarization; or from  $t(m)$ to $d(m)$.


\section{Discussion}\label{section_discussion}

We present a novel way of detecting changes in multi-temporal satellite images, WECS. The procedure is based on wavelet energies from both the estimated individual coefficients as well as the whole image approximation. It makes use of correlation screening for ultra-high dimensional data. The proposed method's performance is shown using both synthetic and real data. The proposed method yields spatio-temporal  change points. Its performance with or without images's pre-treatment is statistically identical, but the computational cost of the proposed method is 180 times smaller than the pre-treatment's cost. Therefore, we may say that this method may be used on untreated images with equivalent performance for a fraction of the computational cost. Because of its reliance on wavelet representation and correlation screening, it is sparse, very fast and scalable. Finally, it is easily adapted to be updatable, so that real-time change detection is feasible even with a portable computer.
    
\section*{Acknowledgement}

RF acknowledges support by FAPESP grant 2016/24469-6. AP acknowledges support by FAPESP grant 2018/04654-9 and CNPq grants 309230/2017-9 and 310991/2020-0.  

\bibliographystyle{agsm}
\bibliography{bibfile}

@article{ansari2020urban,
  title={Urban change detection analysis utilizing multiresolution texture features from polarimetric {SAR} images},
  author={Ansari, Rizwan Ahmed and Buddhiraju, Krishna Mohan and Malhotra, Rakesh},
  journal={Remote Sensing Applications: Society and Environment},
  volume={20},
  pages={100418},
  year={2020},
}

@article{atto2012multidate,
  title={Multidate divergence matrices for the analysis of {SAR} image time series},
  author={Atto, Abdourrahmane Mahamane and Trouv{\'e}, Emmanuel and Berthoumieu, Yannick and Mercier, Gr{\'e}goire},
  journal={IEEE Transactions on Geoscience and Remote Sensing},
  volume={51},
  number={4},
  pages={1922--1938},
  year={2012},
}

@InCollection{ban2016change,
author = {Ban, Yifang and Yousif, Osama},
title = {Change detection techniques: A review},
booktitle = {Multitemporal Remote Sensing},
pages = {19--43},
publisher={Springer},
year = {2016},
editor = {Ban, Yifang},
address = {Cham},
}

@inproceedings{barreto2016deforestation,
  title={Deforestation change detection using high-resolution multi-temporal {X}-Band {SAR} images and supervised learning classification},
  author={Barreto, Thiago LM and Rosa, Rafael AS and Wimmer, Christian and Nogueira, Jo{\~a}o B and Almeida, Jurandy and Cappabianco, F{\'a}bio Augusto Menocci},
  booktitle={2016 IEEE International Geoscience and Remote Sensing Symposium (IGARSS)},
  pages={5201--5204},
  year={2016},
  address = {Beijing, China},
  organization={IEEE}
}

@inproceedings{bouhlel2015multivariate,
  title={Multivariate statistical modeling for multi-temporal {SAR} change detection using wavelet transforms},
  author={Bouhlel, Nizar and Ginolhac, Guillaume and Jolibois, Eric and Atto, Abdourrahmane},
  booktitle={2015 8th International Workshop on the Analysis of Multitemporal Remote Sensing Images (Multi-Temp)},
  pages={1--4},
  year={2015},
  address = {Annecy, France},
  organization={IEEE}
}

@article{bovolo2015time,
  title={The time variable in data fusion: A change detection perspective},
  author={Bovolo, Francesca and Bruzzone, Lorenzo},
  journal={IEEE Geoscience and Remote Sensing Magazine},
  volume={3},
  number={3},
  pages={8--26},
  year={2015},
}

@article{celik2009multiscale,
  title={Multiscale change detection in multitemporal satellite images},
  author={Celik, Turgay},
  journal={IEEE Geoscience and Remote Sensing Letters},
  volume={6},
  number={4},
  pages={820--824},
  year={2009},
}

@article{chen2020change,
  title={Change Detection Algorithm for Multi-Temporal Remote Sensing Images Based on Adaptive Parameter Estimation},
  author={Chen, Yu and Ming, Zutao and Menenti, Massimo},
  journal={IEEE Access},
  volume={8},
  pages={106083--106096},
  year={2020},
}

@article{cui2012statistical,
  title={Statistical wavelet subband modeling for multi-temporal {SAR} change detection},
  author={Cui, Shiyong and Datcu, Mihai},
  journal={IEEE Journal of Selected Topics in Applied Earth Observations and Remote Sensing},
  volume={5},
  number={4},
  pages={1095--1109},
  year={2012},
}

@article{du2019unsupervised,
  title={Unsupervised deep slow feature analysis for change detection in multi-temporal remote sensing images},
  author={Du, Bo and Ru, Lixiang and Wu, Chen and Zhang, Liangpei},
  journal={IEEE Transactions on Geoscience and Remote Sensing},
  volume={57},
  number={12},
  pages={9976--9992},
  year={2019},
}

@book{fan2020statistical,
  title={Statistical Foundations of Data Science},
  author={Fan, Jianqing and Li, Runze and Zhang, Cun-Hui and Zou, Hui},
  year={2020},
  address={Boca Raton},
  publisher={CRC Press}
}

@article{hou2014unsupervised,
  title={Unsupervised change detection in {SAR} image based on {G}auss-log ratio image fusion and compressed projection},
  author={Hou, Biao and Wei, Qian and Zheng, Yaoguo and Wang, Shuang},
  journal={IEEE Journal of Selected Topics in Applied Earth Observations and Remote Sensing},
  volume={7},
  number={8},
  pages={3297--3317},
  year={2014},
}

@article{jia2018novel,
  title={Novel class-relativity non-local means with principal component analysis for multitemporal {SAR} image change detection},
  author={Jia, Meng and Wang, Lei},
  journal={International Journal of Remote Sensing},
  volume={39},
  number={4},
  pages={1068--1091},
  year={2018},
}

@article{johnstone2009statistical,
  title={Statistical challenges of high-dimensional data},
  author={Johnstone, Iain M and Titterington, D Michael},
  year={2009},
  journal={Philosophical Transactions of the Royal Society A},
  volume={367},
  pages={4237--4253},
}

@article{liu2019review,
  title={A review of change detection in multitemporal hyperspectral images: Current techniques, applications, and challenges},
  author={Liu, Sicong and Marinelli, Daniele and Bruzzone, Lorenzo and Bovolo, Francesca},
  journal={IEEE Geoscience and Remote Sensing Magazine},
  volume={7},
  number={2},
  pages={140--158},
  year={2019},
}

@inproceedings{matsunaga2017current,
  title={Current status of hyperspectral imager suite (HISUI) onboard International Space Station (ISS)},
  author={Matsunaga, Tsuneo and Iwasaki, Akira and Tsuchida, Satoshi and Iwao, Koki and Tanii, Jun and Kashimura, Osamu and Nakamura, Ryosuke and Yamamoto, Hirokazu and Kato, Soushi and Obata, Kenta and Mouri, Koichiro and Tachikawa, Tetsushi},
  booktitle={2017 IEEE International Geoscience and Remote Sensing Symposium (IGARSS)},
  pages={443--446},
  year={2017},
  address={Fort Worth, USA},
  organization={IEEE}
}

@book{morettin2017wavelets,
  title={Wavelets in Functional Data Analysis},
  author={Morettin, Pedro A and Pinheiro, Alu{\'\i}sio and Vidakovic, Brani},
  year={2017},
  address={Cham},
  publisher={Springer}
}

@article{ru2021multi,
  title={Multi-Temporal Scene Classification and Scene Change Detection with Correlation based Fusion},
  author={Ru, Lixiang and Du, Bo and Wu, Chen},
  journal={IEEE Transactions on Image Processing},
  volume={30},
  pages={1382--1394},
  year={2021},
}

@article{song2018multi,
  title={Multi-scale feature based land cover change detection in mountainous terrain using multi-temporal and multi-sensor remote sensing images},
  author={Song, Fei and Yang, Zhuoqian and Gao, Xueyan and Dan, Tingting and Yang, Yang and Zhao, Wanjing and Yu, Rui},
  journal={IEEE Access},
  volume={6},
  pages={77494--77508},
  year={2018},
}

@book{vidakovic1999statistical,
  title={Statistical Modeling by Wavelets},
  author={Vidakovic, Brani},
  year={1999},
  address={New York},
  publisher={John Wiley \& Sons}
}

\end{document}